\pdfoutput=1
\documentclass[12pt]{article}

\usepackage{lmodern}

\DeclareFontFamily{T1}{calligra}{}
\DeclareFontShape{T1}{calligra}{m}{n}{<->s*[1.44]callig15}{}
\DeclareMathAlphabet\mathcalligra   {T1}{calligra} {m} {n}
\DeclareMathAlphabet\mathzapf       {T1}{pzc} {mb} {it}
\DeclareMathAlphabet\mathchorus     {T1}{qzc} {m} {n}
\DeclareMathAlphabet\mathrsfso      {U}{rsfso}{m}{n}
\DeclareMathAlphabet\mathfrcal      {T1}{frcursive}{m}{it}
\DeclareFontFamily{T1}{frcursive}{}
\DeclareFontShape{T1}{frcursive}{m}{n}{<->s*[1.44]callig15}{}
\DeclareMathAlphabet\mathfrcal      {T1}{frcursive}{m}{it}

\usepackage{amsmath}
\usepackage{amssymb}
\usepackage{amsthm}
\usepackage{graphicx}
\usepackage[dvipsnames]{xcolor}
\usepackage[english]{babel}
\usepackage{csquotes}
\numberwithin{equation}{section}
\usepackage{array}
\usepackage{mathtools}
\usepackage{dsfont}
\usepackage{mathrsfs}
\usepackage{tikz}\usetikzlibrary{matrix,fit}
\usepackage{varwidth}
\usepackage{enumerate}
 \usepackage{appendix}
\usepackage{ytableau}
\usepackage{simpler-wick}
\usepackage{bm}

\usepackage{caption}

\usepackage[compat=1.1.0]{tikz-feynman}

\usepackage[margin=1in]{geometry}
\usepackage{nicematrix}

\usepackage[
    backend=bibtex,
    style=alphabetic,
maxbibnames=99,
giveninits=true,
minalphanames=1,
maxalphanames=3,url=false
  ]{biblatex}

\usepackage{mathabx}
\usepackage{empheq}

\usepackage{setspace}

\usepackage{fontawesome5}

\usepackage{slashed}
\usepackage{upgreek}

\usepackage{tabstackengine}

\usepackage{wrapfig}
\usepackage[abs]{overpic}

\usepackage{float}

\usepackage{mathtools}

\fixTABwidth{T}

\usepackage{accents}
\newcommand\thickbar[1]{\accentset{\rule{.7em}{.8pt}}{#1}}

\newcommand\smallthickbar[1]{\accentset{\rule{.5em}{.8pt}}{#1}}

\newcommand{\dd}{\partial}

\newcommand{\CP}{\mathds{CP}}
\newcommand{\CC}{\mathds{C}}
\newcommand{\T}{\mathsf{T}}

\renewcommand{\tilde}{\widetilde}

\newcommand{\bea}{\begin{equation}}
\newcommand{\eea}{\end{equation}}
\newcommand{\bear}{\begin{eqnarray}}
\newcommand{\eear}{\end{eqnarray}}
\newcommand{\bearr}{\begin{eqnarray*}}
\newcommand{\eearr}{\end{eqnarray*}}

\newdimen\mytextwidth
\newcommand\rem[2][cyan!40!green]{\noindent\nobreak\hfil\penalty1000\hfilneg
\mytextwidth=\linewidth\advance\mytextwidth by 2mm
\begin{tikzpicture}[baseline=-\the\dimexpr\fontdimen22\textfont2\relax]\node[outer sep=0pt,draw=black,fill=#1,fill opacity=1,text opacity=1,rectangle,rounded corners]{\begin{varwidth}{\mytextwidth}\textcolor{white}{#2}\end{varwidth}};
\end{tikzpicture}\allowbreak
}

\newcommand\whiterem[2][white!]{\noindent\nobreak\hfil\penalty1000\hfilneg
\mytextwidth=\linewidth\advance\mytextwidth by 2mm
\begin{tikzpicture}[baseline=-\the\dimexpr\fontdimen22\textfont2\relax]\node[outer sep=0pt,draw=black,fill=#1,fill opacity=1,text opacity=1,rectangle,rounded corners,line width=1.5pt]{\begin{varwidth}{\mytextwidth}\textcolor{black}{#2}\end{varwidth}};
\end{tikzpicture}\allowbreak
}

\newcommand{\SU}{\mathsf{SU}}

\renewbibmacro{in:}{}

\usepackage{mdframed}

\newbibmacro{string+doiurlisbn}[1]{%
  \iffieldundef{doi}{%
    \iffieldundef{url}{%
      \iffieldundef{isbn}{%
        \iffieldundef{issn}{%
          #1%
        }{%
          \href{http://books.google.com/books?vid=ISSN\thefield{issn}}{#1}%
        }%
      }{%
        \href{http://books.google.com/books?vid=ISBN\thefield{isbn}}{#1}%
      }%
    }{%
      \href{\thefield{url}}{#1}%
    }%
  }{%
    \href{http://dx.doi.org/\thefield{doi}}{#1}%
  }%
}

\DeclareFieldFormat{title}{\usebibmacro{string+doiurlisbn}{\mkbibemph{#1}}}
\DeclareFieldFormat[article,incollection]{title}%
    {\usebibmacro{string+doiurlisbn}{\mkbibquote{#1}}}

\newcommand\BbbGamma{\reflectbox{\rotatebox[origin=c]{180}{$\mathds L$}}}

\newmdenv[
  topline=false,
  bottomline=false,
  rightline=false,
  linewidth=2pt,
  skipabove=\topsep,
  skipbelow=\topsep
]{siderules}

\newmdenv[
  topline=false,
  bottomline=false,
  linewidth=2pt,
  skipabove=\topsep,
  skipbelow=\topsep
]{siderulesright}

\makeatletter
\renewcommand{\@seccntformat}[1]{\csname the#1\endcsname.\quad}
\makeatother

\usepackage{setspace}
\usepackage{xpatch}

\makeatletter
\renewcommand{\@chap@pppage}{
  \clear@ppage
  \thispagestyle{plain}
  \if@twocolumn\onecolumn\@tempswatrue\else\@tempswafalse\fi
  \null\vfil
  \markboth{}{}
  {\centering
   \interlinepenalty \@M
   \normalfont
   \MakeUppercase \appendixpagename\par}
  \if@dotoc@pp
    \addappheadtotoc
  \fi
  \vfil\newpage
  \if@twoside
    \if@openright
      \null
      \thispagestyle{empty}
      \newpage
    \fi
  \fi
  \if@tempswa
    \twocolumn
  \fi
}
\makeatother

\definecolor{navycol}{RGB}{100,150,160}
   \definecolor{pinkcol}{RGB}{242,55,55}
   \definecolor{greencol}{RGB}{50,205,50}

   \definecolor{bluecol}{RGB}{30,144,255}

\usepackage{tocloft}

\usepackage{titlesec}

\titleformat*{\section}{\large\bfseries}
\titleformat*{\subsection}{\normalsize\bfseries}
\titleformat*{\subsubsection}{\normalsize\bfseries}
\titleformat*{\paragraph}{\large\bfseries}
\titleformat*{\subparagraph}{\large\bfseries}
\titlespacing{\author}{-5pt}{-5pt}{-5pt}[-5pt]

\makeatletter
\renewcommand\subsubsection{\@startsection{subsubsection}{3}{\z@}
                                     {-3.25ex\@plus -1ex \@minus -.2ex}
                                     {-1.5ex \@plus -.2ex}
                                     {\normalfont\normalsize\bfseries}}
\renewcommand\subsection{\@startsection{subsection}{3}{\z@}
                                     {-3.25ex\@plus -1ex \@minus -.2ex}
                                     {-1.5ex \@plus -.2ex}
                                     {\normalfont\normalsize\bfseries}}                                     
\makeatother

\usepackage{soul}

\usepackage{scalerel}[2016/12/29]
\newcommand{\scale}[2]{\scaleobj{#2}{#1}}   
\newcommand\N{\raisebox{-0.2em}{\scale{\boldsymbol{:}}{1.9}}}

\DeclareFontFamily{U}{solomos}{}
\DeclareErrorFont{U}{solomos}{m}{n}{10}
\DeclareFontShape{U}{solomos}{m}{n}{
  <-> s*[1.1]  gsolomos8r
}{}

\newcommand{\vkappa}{\text{\usefont{U}{solomos}{m}{n}\symbol{'153}}}

   \usepackage{stmaryrd}

\usepackage{tikz}
\usetikzlibrary{arrows.meta}

\usepackage{indentfirst}

\let \savenumberline \numberline
\def \numberline#1{\savenumberline{#1.}}

\usepackage{xurl}
\usepackage{scrextend}

\usepackage{upgreek}

\usepackage{etoolbox}
\patchcmd{\tableofcontents}{\@starttoc}{\vspace{-0.3cm}\@starttoc}{}{}

\usepackage{nicematrix}

\newcounter{Chapcounter}

\newcommand{\chapter}[1] 
{ {\centering          
  \addtocounter{Chapcounter}{1} \Large \underline{\sffamily \texorpdfstring{\textbf{  Chapter \theChapcounter: ~#1}}{Lg}} }   
  \addcontentsline{toc}{section}{ \color{blue} \texorpdfstring{Chapter ~}{Lg}\theChapcounter.\texorpdfstring{~~}{Lg} #1 }    
}

\newcommand{\appendixbig}[1] 
{ {\centering          
   \Large \underline{\sffamily \textbf{  Appendices}} }   
  \addcontentsline{toc}{section}{ \color{blue} Appendices}    
}

\newcommand{\introbig}[1] 
{ {\centering          
   \Large \underline{\sffamily \textbf{  #1}} }   
  \addcontentsline{toc}{section}{ \color{blue} #1} 
}

\newcommand{\conclbig}[1] 
{ {\centering          
   \Large \underline{\sffamily \textbf{  #1}} }   
  \addcontentsline{toc}{section}{ \color{blue} #1} 
}

\usepackage[hyperfootnotes=]{hyperref}
\hypersetup{
    colorlinks=true,
    linkcolor=blue,
    filecolor=magenta,      
    urlcolor=cyan,
    breaklinks=true
    }

\begin{document}

\title{\textbf{Oscillator Calculus on Coadjoint Orbits} \\ \textbf{and Index Theorems} \vspace{1cm}}

\author{Dmitri Bykov$^{\,a,\,b,\,c}$\footnote{Emails:
 bykov@mi-ras.ru, dmitri.v.bykov@gmail.com} \,, Viacheslav Krivorol$^{\,a,\,b,\,}$\footnote{Emails:
 vkrivorol@itmp.msu.ru, v.a.krivorol@gmail.com} \,, Andrew Kuzovchikov$^{\,a,\,b\,}$\footnote{Emails: kuzovchikov@mi-ras.ru, 
 andrkuzovchikov@mail.ru}
\\  \vspace{0cm}  \\
{\small $a)$ \emph{Steklov
Mathematical Institute of Russian Academy of Sciences,}} \\{\small \emph{Gubkina str. 8, 119991 Moscow, Russia} }\\
{\small $b)$ \emph{Institute for Theoretical and Mathematical Physics,}} \\{\small \emph{Lomonosov Moscow State University, 119991 Moscow, Russia}} \\
{\small $c)$ \emph{Beijing Institute of Mathematical Sciences and Applications (BIMSA),}} \\{\small \emph{Huairou District, Beijing
101408, China}} 
}

\date{}

{\let\newpage\relax\maketitle}

\maketitle

\ytableausetup{centertableaux}

\vspace{0cm}
\textbf{Abstract.} We consider quantum mechanical systems of spin chain type, with finite-dimensional Hilbert spaces and  $\mathcal{N}=2$ or $\mathcal{N}=4$ supersymmetry, described in $\mathcal{N}=2$ superspace in terms of nonlinear chiral multiplets. We prove that they are natural truncations of 1D sigma models, whose target spaces are $\SU(n)$ (co)adjoint orbits. 
As a first application, we compute the Witten indices of these finite-dimensional models showing that they reproduce the Dolbeault and de Rham indices of the target space. The problem of finding the exact spectra of generalized Laplace operators on such orbits is shown to be equivalent to the diagonalization of spin chain Hamiltonians.

\newpage
\tableofcontents

\vspace{0.5cm}

\newpage
\introbig{Introduction}

\vspace{1cm}
Dynamical systems on homogeneous spaces are a classical subject in both differential geometry and mathematical physics. For example, the study of tops has been foundational for the theory of integrable systems (cf.~\cite{Babelon}). The study of geodesic flows on homogeneous spaces naturally leads to 1D sigma models. Leaving aside the case of symmetric spaces, the complete description of geodesics on such spaces is far from being fully understood\footnote{As we shall see below in the example of flag manifolds, typically invariant metrics on homogeneous spaces come in families.}~\cite{Thimm, Paternain, Bolsinov}. The quantum mechanical counterpart of this story is in diagonalizing Laplace operators on such homogeneous spaces and is also of considerable interest (cf.~\cite{Gurarie} for classic results).

\subsubsection{1D sigma models via `spin chains'.} 
In the present paper we will consider a special class of homogeneous spaces with particularly nice properties: these are the (co)adjoint orbits of classical simple compact Lie groups\footnote{In this case there is no difference between adjoint and coadjoint orbits, since the adjoint and codjoint representations may be identified by means of the Killing metric on the Lie algebra.} (see~\cite{KirillovReview} for an introduction).  We believe our methods are applicable to all classical groups, although below we will exclusively deal with the case of $\SU(n)$ (see~\cite{Bordemann} or~\cite{Affleck_2022} for a review of $\SU(n)$ orbits). In this case the orbits, in general, are the so-called \emph{flag manifolds}:  they include the complex projective space $\CP^{n-1}$ as well as the Grassmannians $\mathsf{Gr}(m, n)$ and, in full generality, are defined  below.

Our strategy in studying the corresponding quantum 1D sigma models is to construct finite-dimensional approximations, which are rudimentary spin chains of a kind~\cite{Bykov_2024}. Each site of the spin chain contains a single spin, i.e.~a (finite-dimensional) representation of $\SU(n)$ that may be described by its Young diagram. Throughout the paper we will only need the representations corresponding to rectangular Young diagrams, of some height\footnote{For the most part we will have $k=1$;  generalizations to partial flag manifolds will require~${k>1}$.} $k$ and width $p$. The remarkable property of these spin chains is that their spectra give the \emph{exact spectra} of the Laplacians on the corresponding flag manifolds, albeit truncated to a finite-dimensional subset of all harmonics. Increasing the value of $p$, we enlarge the corresponding subspace and in the limit $p\to \infty$ the full Hilbert space and hence the full spectrum are~recovered.

\subsubsection{Supersymmetric extensions of sigma models.} 
Although this method is applicable to purely bosonic models as well, the most interesting results are obtained for their supersymmetric extensions. 

We consider two large classes of supersymmetric sigma models in~1D. In short, these are the models whose supercharges correspond to the differentials in the Dolbeault and K\"ahler-de Rham complexes (cf.~\cite{IvanovSmilga, SmilgaDiffGeom} for a detailed treatment). Although the former has $\mathcal{N}=2$ supersymmetry, whereas the latter has $\mathcal{N}=4$, both are most easily explained in $\mathcal{N}=2$ language. In 1D there are two fundamental types of $\mathcal{N}=2$ sigma models, which have been called $2a$ and $2b$ in~\cite{Hull}. The $2a$ model may be obtained by dimensional reduction from the $\mathcal{N}=(1,1)$ model in 2D and describes the de Rham complex. In turn, the $2b$ model may be obtained from the (`chiral')  $\mathcal{N}=(0,2)$ model in 2D and describes the Dolbeault complex. If, in addition, the metric of the target space is K\"ahler, supersymmetry of the $2a$ model is automatically upgraded to $\mathcal{N}=4$ and in this case it can be called the $4a$ model\footnote{This is parallel to the way $\mathcal{N}=(1, 1)$ supersymmetry of the 2D sigma model is upgraded to $\mathcal{N}=(2, 2)$ for K\"ahler metrics. We note also that there are interesting 1D $\mathcal{N}=4$ models with hyper-K\"ahler target spaces that cannot be obtained by dimensional reduction~\cite{Delduc, Fedoruk}.}. On the other hand, supersymmetry of the $2b$ model is not upgraded, but here a different coincidence occurs in the K\"ahler case: the $2b$ model then coincides with the \emph{minimally coupled} $\mathcal{N}=1$ model, whose supercharge is the Dirac operator. This is one way of seeing the well-known relation between the Dolbeault and Dirac operators on K\"ahler manifolds, which will be discussed in detail below. In the $2b$ case one may as well couple the models to gauge fields of magnetic monopole type (technically they are connections in non-trivial line bundles $\mathcal{L}$ over the target space). We will widely use this additional freedom, assuming that all allowed magnetic fluxes are turned on.

It is worth noting that all coadjoint orbits of simple compact Lie groups are K\"ahler manifolds~\cite{Bordemann} (this is one of their `nice' properties referred to earlier). Nevertheless, unless we are dealing with a symmetric space, not all invariant metrics on these orbits are K\"ahler. It turns out that our construction naturally works for $2b$ (either K\"ahler or not) and $4a$ (K\"ahler) sigma models.

\subsubsection{Nonlinear chiral multiplets and `spin chains'.} 
So far we have reviewed how supersymmetry is applied to 1D sigma models. What about the `spin chain' truncations thereof? From the perspective of supersymmetry, these seem to be novel models, which are most conveniently described in $\mathcal{N}=2$ superspace in terms of \emph{nonlinear chiral multiplets} -- a generalization of the multiplet considered in~\cite{IvanovToppan}. The nonlinear constraint is really a generalized chirality condition w.r.t.~a connection $\mathscr{A}$ in $\mathcal{N} = 2$ superspace (the trivial connection corresponds to the usual chirality condition).  Just as in the sigma model case, there are two types of models:
\begin{itemize}
    \item \textbf{D-model:} superconnection $\mathscr{A}$ valued in upper-triangular $n\times n$ matrices,
    \item \textbf{K-model:} superconnection $\mathscr{A}$ valued in general $n\times n$ matrices.  
\end{itemize}
It turns out that these are the natural spin chain truncations of the 2b and 4a sigma models, respectively (\textbf{D}~is an abbreviation for Dolbeault and \textbf{K} for K\"ahler-de Rham). In particular, supersymmetry of the K-model is automatically upgraded to $\mathcal{N}=4$, in analogy with the $4a$ model. Sigma models are restored in the limit $p\to\infty$, where $p$ is the truncation level discussed above (related to the number of allowed harmonics on the sigma model side, or the value of `spin' on the spin chain side).

Rather strikingly, for both types of models the Lagrangians  can be taken as `free Lagrangians' for the nonlinear multiplets in $\mathcal{N} = 2$ superspace, and the interactions are \emph{entirely} encoded in the deformed chirality constraints\footnote{There are $\mathcal{N}=4$ sigma models with hyper-K\"ahler target spaces, whose superspace formulation also has this property~\cite{Delduc}.}. Moreover, quantization of such theories leads to oscillator representations for the Hamiltonian and supercharges (supercharges are cubic in the oscillators, resembling the models\footnote{Related lattice models were constructed in~\cite{Fendley1, Fendley2}. Spin chain models with $\mathcal{N}=1$ supersymmetry have as well appeared in the literature~\cite{Sannomiya, Totsuka}. The crucial difference of our models is that they involve bosonic oscillators together with fermionic ones. Finite-dimensionality of the Hilbert space is achieved by imposing supersymmetry-invariant constraints. }  of~\cite{Nicolai76, Nicolai77}). The construction also naturally leads to constraints on the Hilbert space, which ensure that it is a \emph{finite-dimensional} truncation of the oscillator Fock space. The structure of the Hilbert space may be encoded in the so-called \textit{framed quiver diagrams} of a special kind. Moreover, the classical phase spaces are the corresponding  \textit{quiver varieties}, whereas the relevant Hilbert spaces are obtained by the geometric quantization thereof. 

\subsubsection{The Witten index for `spin chains' and index theorems.}  
Studying the full spectral problem of the resulting Hamiltonians is an important goal which might shed light on the integrability of these systems and could be addressed using our methods. However, it is beyond the scope of the present paper. Here instead we will concentrate on the computation of the Witten index\footnote{Here $\mathscr{H}$ is the Hilbert space, $\mathcal{H}$ the Hamiltonian, $F$  the fermion number and $\beta\geq 0$ a parameter.} (see~\cite{SmilgaWitten} for a comprehensive and pedagogical exposition)
\bea\label{Windexdef}
W = \mathsf{Tr}_{\mathscr{H}} (-1)^F e^{-\beta \mathcal{H}}
\eea
as well as its refined version, the `equivariant' Witten index $\widetilde{W}$, which takes into account the $\SU(n)$ symmetry of the models. By standard arguments, the index is independent of $\beta$, so letting $\beta\to \infty$ amounts to projecting on the zero-energy eigenstates of the Hamiltonian, i.e.~the states annihilated by the supercharges. Thus, in fact these are the only states contributing to the index. On the other hand, in our spin chain systems the Hilbert space is finite-dimensional, so no harm is done by setting $\beta=0$, in which case the index~(\ref{Windexdef}) simply becomes the superdimension $\mathsf{Sdim}(\mathscr{H})$ of the Hilbert space or its supercharacter in the equivariant case. We use this approach to calculate the Witten index, finding that miraculously it is independent of the truncation~$p$ (although changing the value of $p$ means changing the Hilbert space, which is not a smooth deformation of the system). It therefore coincides with the Witten index of the respective sigma models that appear in the limit $p\to \infty$, the latter being equal to the index of the Dolbeault and de Rham operators (in the $2b$ and $4a$ cases, respectively). We are thus able to reproduce the index theorems for coadjoint orbits in a novel way, using finite-dimensional truncations, or spin chains.

\subsubsection{Classical index theorems for coadjoint orbits.} 
Finally, let us briefly recall the developments around the index theorem, with emphasis on applications to coadjoint orbits. An index theorem expresses the index of an elliptic operator on a compact manifold $\mathcal{M}$ in terms of an integral of some top form (characteristic class) over that manifold. First index formulas were obtained in~\cite{Index1}; more detailed proofs were given in a subsequent series of papers (cf.~\cite{Index3}, which contains the most explicit expressions). For example, the index of the Dolbeault operator $\smallthickbar{\dd}$ acting on forms of type $(0, \bullet)$ with values in some holomorphic vector bundle $\mathcal{V}$ reads:
\bea\label{Dindex1}
\mathrm{Ind}\big(\,\smallthickbar{\dd}, \mathcal{V}\big)=\int_{\mathcal{M}}\,\mathrm{ch}(\mathcal{V})\wedge\mathrm{Td}(\mathcal{M})\,.
\eea
Here $\mathrm{ch}(\mathcal{V})$ is the Chern character of the vector bundle $\mathcal{V}$ and $\mathrm{Td}(\mathcal{M})$ is the Todd genus, which is a polynomial in the curvature tensor. In our applications $\mathcal{V}=\mathcal{L}$ will always be a line bundle corresponding to magnetic monopole fluxes referred to earlier, in which case  $\mathrm{ch}(\mathcal{L})=e^{c_1(\mathcal{L})}$. There is also a similar formula for the index of the Dirac operator, where $\mathrm{Td}(\mathcal{M})$ is replaced by the so-called $\hat{A}$-genus. 

The proof of the index theorem was elucidated with the advent of supersymmetric quantum mechanics~\cite{WittenSUSYQM}. It was shown in~\cite{Alvarez-Gaume} \cite{FriedanWindey} that index formulas like the one in~(\ref{Dindex1}) could be obtained by a simple evaluation of the Witten index for various 1D sigma models, whose target space is the relevant manifold $\mathcal{M}$. The path integral for the Witten index may, in these cases, be dimensionally reduced to a finite-dimensional integral, which reproduces~(\ref{Dindex1}) (see, for example,~\cite{Alvarez} or \cite{IvanovSmilga, SmilgaDiffGeom}). The literature on the index theorem, where one can find details, is vast; for a brief account see~\cite{Gilkey}, or the lectures~\cite{FreedDirac} for a more informal discussion.

Either way, the above formula~(\ref{Dindex1}) is too general for our purposes. Throughout this paper $\mathcal{M}$ is a very special manifold -- the coadjoint orbit --, where drastic simplifications occur. First of all, in case $\mathcal{L}$ is an ample line bundle over the coadjoint orbit, the Borel-Weil-Bott theorem~\cite{Bott}, \cite[\S{23.3}]{FultonHarris} states that the vector space of its holomorphic sections ${H^{0}(\mathcal{M}, \mathcal{L})=\mathcal{R}_{\mathcal{L}}}$ constitutes a representation $\mathcal{R}_{\mathcal{L}}$ of the symmetry group $G$. Moreover, all higher cohomology groups vanish, and the equivariant index is simply the character of the corresponding representation:
\bea\label{BWBindex}
\mathrm{Ind}_G\big(\,\smallthickbar{\dd}, \mathcal{L}\,\big)=\upchi(\mathcal{R}_{\mathcal{L}})\,.
\eea
This result 
can be obtained from an integral of the type~(\ref{Dindex1}) (its equivariant generalization was given in~\cite{BerlineVergne}; see also the book~\cite{Berline}) in at least two ways: either by recognizing that the integral is an instance of the Kirillov character formula for~$\mathcal{R}_{\mathcal{L}}$~\cite{KirillovCharacter} (an integral over the orbit), or by reducing the integral to the fixed-point set of a torus action (cf.~\cite{Swiggen} for a review of methods for calculating such integrals). The torus in question is the Cartan subgroup of $G$, whose fixed point set is a finite collection of points on~$\mathcal{M}$. The result is the Weyl  formula for the character of $\mathcal{R}_{\mathcal{L}}$, or the character of a virtual representation in the non-ample setting.

It is formulas of the type~(\ref{BWBindex}) that are most relevant in our setup of finite-dimensional supersymmetric quantum mechanics. For example, our calculation of the Witten index for the D-model will result in the character $\upchi(\mathcal{R}_{\mathcal{L}})$ featuring in~(\ref{BWBindex}).

\subsection{Structure of the paper.}

Since the paper has become rather voluminous in the process of writing, we have decided to split it into three chapters. This is meant to help the reader follow the logic of the exposition more easily.

Chapter 1 is dedicated entirely to `spin chains' describing truncations of the supersymmetric $\CP^1$ sigma models.  We start in Section~\ref{N=2 CP1section} by introducing the D-system for the $\CP^1$ sigma model, showing that its Hamiltonian  reproduces the truncated spectrum of the Laplacian on $\CP^1$. We also discuss the quiver formulation of this theory. We then compute its (equivariant) Witten index, proving that it coincides with the index of the Dolbeault operator on $\CP^1$ and is  independent of the truncation.  In Section~\ref{N4CP1} we introduce the second class of models that we study, the K-model. We prove that its Witten index is the Euler characteristic.

In Chapter 2 we formulate the models of Chapter 1 in superspace and describe the `large-spin' ($p\to \infty$) limit, which leads from spin chains to sigma models.  In Section~\ref{superspacesec} we describe our models in superspace: first using $\mathcal{N}=2$ fields, then reducing to $\mathcal{N}=1$ superspace and, finally, to component form. Up to this point the models that we consider are defined somewhat abstractly as oscillator-type models with finite-dimensional Hilbert spaces. The goal of Section~\ref{Derive of CP1} is to provide a path integral-based proof that in a certain limit, $p\to \infty$, these systems reproduce 1D sigma models with target space $\CP^1$.

Finally, Chapter 3 is dedicated to generalizations to other coadjoint orbits of $\SU(n)$. 
Starting from $\CP^1$, there are two natural generalizations, either to higher-dimensional projective spaces $\CP^{n-1}$ or to complete flag manifolds $\mathcal{F}_n$. The reason is that $\CP^1$ is also the manifold of complete flags in $\CC^2$ and hence could be thought of as being a representative of both classes of models. This is summarized by the following diagram:
\begin{center}
\begin{tabular}{ccccc}
     &  &$\CP^1$  & & \\
     &{\large$\swarrow$} & & {\large $\searrow$} &\\
   $\mathcal{F}_n$  & & & & $\CP^{n-1}$ \\
&{\large$\searrow$} & & {\large $\swarrow$} & \\
&  &$ \mathcal{F}_{n_1, \ldots,n_k}$  & & 
\end{tabular}
\end{center}
The lower entry indicates that, subsequently, one can generalize all of these examples to the (most general) class of partial flag manifolds. We also note that, from a gauge-theoretic perspective, complete flag manifolds correspond to \emph{Abelian} gauge theories, whereas partial flag manifolds lead to \emph{non-Abelian} extensions.

In Section~\ref{completeflagsec} we generalize the oscillator models to the case of complete flag manifolds. We observe that the K\"ahler condition on the metric arises naturally from the SUSY algebra in the case of K-models, which correspond to $\mathcal{N}=4$ SUSY. We compute the indices for both D- and K-models, reproducing the corresponding characters and Euler characteristics.  Sections~\ref{CPnsec} and \ref{partialflagsec} are dedicated to the non-Abelian generalizations: first to the case of $\CP^{n-1}$ and then to generic partial flag manifolds, thus exhausting all (co)adjoint orbits of $\SU(n)$.

Finally, in Section~\ref{DDsec} we recall the well-known isomorphism between the Dolbeault and Dirac operators on K\"ahler manifolds. Via this isomorphism, our methods allow computing the index of the Dirac operator. Diagonalization of the relevant spin chains would lead to the determination of its spectrum as well. 

\subsection{Notation and abbreviations.} For the reader's convenience we insert the following table of abbreviations and notations that we use throughout the paper:

\vspace{0.3cm}
\begin{center}
\begin{tabular}{c|c}
   QM  & Quantum Mechanics  \\
   SUSY  &  Supersymmetry\\
   D-model & Spin chain Dolbeault model\\
   K-model & Spin chain K\"ahler -- de Rham model\\
   2b model & $\mathcal{N}=2$ sigma model in 1D\\
   4a model & $\mathcal{N} = 4$ K\"ahler sigma model in 1D\\
   $W$ and $\widetilde{W}$ & Ordinary and equivariant Witten indices\\
   $\mathbf{Z}, \mathbf{\Psi}, \mathbf{\Phi}, \ldots$ &  $\mathcal{N}=2$ superfields\\
   $\mathsf{Z}, \mathsf{\Psi}, \mathsf{\Phi}, \ldots$ &  $\mathcal{N}=1$ superfields\\
   $A,B,C,\ldots$ & Indices numerating sites of the spin chain \\
   $\alpha,\beta,\gamma,\ldots$ & Indices numerating basis vectors of the global vertex of the quiver\\
   $a,b,c,\ldots$ & Indices  numerating basis vectors of local vertices of the quiver\\
   $\circ$ & Scalar product in `ambient space' $\CC^n$: $u\circ v\equiv \sum_{\gamma=1}^n\,u^\gamma v^\gamma$\\
   $\displaystyle{\mathsf{T}^p}$ & $p^{\textrm{th}}$ symmetric power of the defining representation of $\mathfrak{su}_n$\\
   $\upchi_p$ & Character of $\displaystyle{\mathsf{T}^p}$\\
   $\mathds{1}_m$ & Unit $m\times m$ matrix
\end{tabular}
\end{center}

\newpage
\chapter{Quantum \texorpdfstring{$\CP^1$}{Lg} models}

\vspace{0.5cm}
\section{Spin chain Dolbeault model (D-model)} \label{N=2 CP1section}

In this section, we will explain the general idea using the example of the $\CP^1$ model. We wish to start from a simple supersymmetric `oscillator'-type model, which is prototypical for the models studied throughout the paper. Their crucial feature is that the corresponding Hilbert spaces are \emph{finite-dimensional}. As we shall see, they are truncations of 1D sigma models.

We introduce a set of bosonic creation/annihilation operators 
$
z_1^{\alpha},(z_1^\dagger)^\alpha, z_2^{\alpha},(z_2^\dagger)^\alpha \,$ (where $\alpha=1, 2$), 
as well as a pair of fermionic creation/annihilation operators $\psi_{12}, \psi_{12}^\dagger$, satisfying canonical (anti)commutation relations
\bear
\label{CanonicalCommutationRelation}
&&\Bigl[z_A^\alpha, (z_B^\dagger)^{\beta\,}\Bigr]=\delta_{AB}\,\delta^{\alpha\beta}\,,\quad\quad A, B=1, 2\,, \quad \alpha, \beta=1, 2\,,\\ \nonumber 
&& \Bigl\{\psi_{12}, \psi_{12}^\dagger\Bigr\}=1\,.
\eear
All other (anti)commutators vanish. Throughout this paper we will interpret Fock spaces of such oscillators as Hilbert spaces of certain rudimentary spin chains. These spin chains will, in general, have $\mathsf{SU}(n)$ symmetry.
In case of bosonic operators the  capital Latin indices will be numerating sites of the spin chain, whereas Greek indices transform in the ($n$-dimensional) defining representation of $\mathsf{SU}(n)$. The circle $\circ$ will indicate contraction w.r.t.~Greek indices, i.e. it stands for the scalar product in $\CC^n$. 
In turn, fermions determine  interactions between sites of the spin chain, so their  indices refer to pairs of sites they are connecting. For example, in this section we are considering a spin chain with two sites and $\mathsf{SU}(2)$ symmetry.

The model may be defined by the supercharge\footnote{Notice that if one would replace the quadratic combination of the $z$-fields by a linear combination, the corresponding system would be a version of the SUSY harmonic oscillator studied as early as in~\cite{Nicolai76}.}
\bea\label{Q12supercharge}
\mathcal{Q}=\upalpha_{12}\,\psi_{12}\,\big(z_1^\dagger \circ z_2\big)\,,
\eea
where $\upalpha_{12}$ is a real coupling parameter. 
Clearly, $\mathcal{Q}^2=0$, and we define the Hamiltonian by the standard formula
\bea\label{HamDef}
\mathcal{H}=\Bigl\{\mathcal{Q}, \mathcal{Q}^\dagger\Bigr\}\,.
\eea
This defines a QM system with $\mathcal{N}=2$ SUSY. This nomenclature means that there are two real supercharges, which we have packaged in a single complex one.

Classically, the above supercharges (and hence the Hamiltonian) are invariant w.r.t.~the $\mathsf{U}(1)\times \mathsf{U}(1)$ symmetry
\bea\label{CP1symmetry}
z_1 \mapsto e^{i\varphi} z_1\,,\quad\quad z_2 \mapsto e^{i\varphi^\prime} z_2\,,\quad\quad \psi_{12}\mapsto e^{i(\varphi-\varphi^\prime)}\,\psi_{12}\,,\qquad \varphi,\varphi^\prime\in\mathbb{R}\,.
\eea

\noindent
We may readily write out the quantum operators generating these transformations\footnote{Here we choose normal ordering for the oscillators. A change of ordering corresponds to a shift of the values of $p$ and $q$. We will discuss this in more detail in Section~\ref{DDsec}.}:
\bear\label{C1}
 &&\mathcal{C}_1=z^{\dagger}_1 \circ z_1+ \psi_{12}^\dagger \psi_{12} - p_1\,,\\  \nonumber
&&\mathcal{C}_2=z^{\dagger}_2 \circ z_2- \psi_{12}^\dagger \psi_{12} -p_2\,,
\eear
where we have added the constants $p_1:=p$ and $p_2:=p+q$. Since $\mathcal{C}_1$ and $\mathcal{C}_2$ commute with~$\mathcal{Q}$, we may pick a subspace of the full Fock space
\bea\label{C1C2constr}
\mathscr{H}(p_1, p_2):=\Bigl\{\;\mathcal{C}_1=\mathcal{C}_2=0\;\Bigr\}
\eea
without breaking SUSY. These are constraints on the occupation numbers of the oscillators, so they only make sense if $p, q \;\in\; \mathbb{Z}\,.$
We will always assume in the foregoing that the values of these `charges' is such that the constraints allow for a non-empty set of states (concretely, $p > \max\{0,-q\}$).

One easily sees that the space $\mathscr{H}(p_1, p_2)$ is a \emph{finite-dimensional} subspace of the full Fock space. For the reason that will be clear later we call this model \textit{the truncated $\CP^1$  Dolbeault model} (or \textit{the D-model}, in short).

\subsubsection{Hamiltonian and  harmonics.} \label{CP1 Hamiltonian and harmonics}

Let us now examine in more detail the Hamiltonian arising from the SUSY system just described. In particular, we may diagonalize the Hamiltonian and explicitly see the spherical harmonics emerging, which will elucidate  the relation  with the $\CP^1$ sigma model. Very explicitly, the Hamiltonian~(\ref{HamDef})~is
\bea
\mathcal{H}=
\upalpha_{12}^2\left(\big(z_1^\dagger \circ z_2\big)\;\big(z_2^\dagger \circ z_1\big)+(q+2)\psi_{12}^\dagger \psi_{12}\right)\,, \label{HamiltonianDmodel}
\eea
where we have made use of the constraints (\ref{C1C2constr}).

For simplicity, let us analyze the purely bosonic sector, i.e. we set $\psi_{12}^\dagger \psi_{12}=0$. The constraints then imply that ($F$ stands for fermion number)
\bea\label{F0constr}
z^{\dagger}_1 \circ z_1 = p,\,\quad  z^{\dagger}_2 \circ z_2= p+q\,,\quad\quad\quad\quad (F=0)\,,
\eea
so that there are $p$ $z_1$-oscillators and $(p+q)$ $z_2$-oscillators. The most general state has the form
\bea\label{pqwavefunc}
|\Phi\rangle=\Phi_{\alpha_1\cdots \alpha_p|\beta_1\cdots \beta_{p+q}}(z_1^\dagger)^{\alpha_1}\cdots (z_1^\dagger)^{\alpha_p} (z_{2}^\dagger)^{\beta_1}\cdots (z_{2}^\dagger)^{\beta_{p+q}}|0\rangle \,.
\eea
To provide a representation-theoretic description of the wave function we introduce~$\displaystyle{\T^p}$ -- the $p^{\textrm{th}}$  symmetric tensor power of the fundamental representation. In terms of Young diagrams this corresponds to the diagram
\bea
\T^p:=\underbrace{
    \begin{ytableau}
        ~ & ~ & \dots & ~ & ~ \\
    \end{ytableau}
    }_{p}
    \eea
In the present section we are dealing with representation theory of $\mathfrak{su}_2$, but we will preserve this notation for Young diagrams of the form above in the case of $\mathfrak{su}_n$ as well\footnote{For a review of $\SU(n)$ representation theory in terms of bosonic oscillators see~\cite{Affleck_2022}.}.
The wave function~(\ref{pqwavefunc}) belongs therefore to the tensor product $\T^{p}\otimes \T^{p+q}$ of $\SU(2)$ representations. The Hamiltonian acting in this space may also be written in a more familiar form:
\bea\label{Spin12Ham}
\mathcal{H}_{\mathrm{bos}}=\upalpha_{12}^2\,(S_1, S_2)+\mathrm{const}\,,\quad\quad (S_1, S_2)=\sum\limits_{a=1}^3\,S_1^a S_2^a\,,
\eea
where $S_1^a$ and $S_2^b$ are the $\mathfrak{su}_2$ spin operators (which we assume orthonormal w.r.t.~the Killing metric) acting in the space just described. They can be written very explicitly in terms of the oscillators:
\bea
S_1^a=z_1^\dagger\, \sigma^a\, z_1\,,\quad\quad S_2^a=z_2^\dagger\, \sigma^a\, z_2\,,
\eea
with $\sigma^a$ the Pauli matrices. This is the well-known Schwinger-Wigner oscillator representation~\cite{Schwinger} (see~\cite{Bykov_2013, Affleck_2022} for $\mathfrak{su}_n$ generalizations).

Standard $\mathsf{SU}(2)$ representation theory (`addition of spins') tells that $\displaystyle\T^p\otimes \T^{p+q}$ may be decomposed as $\oplus_{k=0}^p \T^{2k+q}$. Accordingly, $|\Phi \rangle$ is decomposed in components $|\Phi \rangle_k\in \T^{2k+q}$ as follows:
\bea\label{PsiKstates}
|\Phi \rangle_k=\Phi^{(k)}_{\alpha_1\cdots \alpha_k\beta_1\cdots \beta_{k+q}}(z_1^\dagger)^{\alpha_1}\cdots (z_1^\dagger)^{\alpha_k} (z_{2}^\dagger)^{\beta_1}\cdots (z_{2}^\dagger)^{\beta_{k+q}} \,\left(\epsilon_{\alpha\beta}(z_1^\dagger)^\alpha (z_2^\dagger)^\beta\right)^{p-k}|0\rangle \,,
\eea
where $\Phi^{(k)}$ is now fully symmetric w.r.t.~all indices. 
Using this fact, we may act with $\mathcal{H}$ on the state $|\Phi \rangle_k$ to find that it is an eigenfunction:
\bea\label{eigenfunctruncation}
\mathcal{H}\,|\Phi \rangle_k=\upalpha_{12}^2\,k(k+1+q)\,|\Phi \rangle_k\,,\quad\quad k=0, \ldots\,, p\,.
\eea
Remarkably, for $q=0$ one recognizes in this expression the spectrum of the Laplacian on the sphere $\CP^1\simeq \mathbb{S}^2$, \emph{truncated to first $(p+1)$ spherical harmonics}. If $q>0$, the eigenfunctions are the well-known monopole harmonics (with monopole charge $q$) described in~\cite{Tamm1991,WuYang,Kuwabara} (for a recent exposition see~\cite{BykSmilga}). 

The spectrum~(\ref{eigenfunctruncation}) has the following curious property: if one replaces $p\mapsto p+1$ all that changes is that an additional state with $k=p+1$ is added, whereas the rest of the energy levels remain intact. This can be explained  by 
observing that there is an operator $\mathcal{O}_2$ that commutes with the Hamiltonian:
\bea\label{O2operator}
\Bigl[\mathcal{O}_2, \,\mathcal{H}\Bigr]=0\,,\quad\quad \textrm{where}\quad\quad \mathcal{O}_2:=\epsilon_{\alpha\beta}(z_1^\dagger)^\alpha (z_2^\dagger)^\beta\,.
\eea
This operator does not commute with the constraints~(\ref{C1}), though: $\Bigl[\mathcal{O}_2, \mathcal{C}_1\Bigr]=\mathcal{O}_2$ and $\Bigl[\mathcal{O}_2, \mathcal{C}_2\Bigr]=\mathcal{O}_2$. This means that acting with  $\mathcal{O}_2$ effectively raises the value of $p$ by one: if $|\psi\rangle \in \mathscr{H}(p_1, p_2)$, then $\mathcal{O}_2|\psi\rangle \in \mathscr{H}(p_1+1, p_2+1)$. However, the energy of both states is the same due to~(\ref{O2operator}). As we shall see in Chapter 3, this phenomenon generalizes to systems with $\mathsf{SU}(n)$ symmetry.

Summarizing the results so far, we have found that the oscillator system can be seen as a truncation of the Laplace operator spectral problem to the  first $p$ harmonics. 
The full spectrum of the Laplacian on $\CP^1$ is recovered in the limit $p\to \infty$. Given the logic of our present exposition, where we simply postulated the supercharge~(\ref{Q12supercharge}), this might seem surprising. To shed light on this in Chapter 2 we will provide a direct derivation of the one-dimensional $\CP^1$ sigma model from the supersymmetric spin chain in the limit $p\to \infty$.    Moreover, this phenomenon is inherited by analogous models with flag manifold target spaces. We  conjecture that our construction generalizes to   coadjoint orbits of other classical compact Lie groups.

\subsubsection{Index of the D-model.} 
In the previous section we concentrated on the bosonic sector of the theory. Clearly, the sector with fermion number $F=1$ could be studied quite analogously. Instead of repeating most of the analysis, we will concentrate here on a slightly different problem: that of analyzing the \emph{zero-energy} states and computing the Witten index, which has a well-known topological meaning.

We found in the previous section that the sector with $F=0$ corresponds to the representation $\displaystyle\T^{p} \otimes \T^{p+q}$. In the case of one fermion, we have 
\bea
z^{\dagger}_1 \circ z_1 = p-1\,,\qquad z^{\dagger}_2 \circ z_2 = p+q+1\,, \qquad (F=1)
\eea
so that this part of the Hilbert space is  $\displaystyle\T^{p-1} \otimes \T^{p+q+1}$. The full  Hilbert space of the D-model is given by
\begin{align}
    \mathscr{H}(p_1, p_2) = \left(\T^{p} \otimes \T^{p+q}\right) \oplus \left(\T^{p-1} \otimes \T^{p+q+1}\right)\,. \label{HilbertSpaceDmodel}
\end{align}
The Witten index is thus
\begin{gather}
    W = \mathsf{Sdim}\big(\mathscr{H}(p_1, p_2)\big):=\mathsf{dim}\big(\T^{p} \otimes \T^{p+q}\big) - \mathsf{dim}\big(\T^{p-1} \otimes \T^{p+q+1}\big) = 
    \label{Witten CP1}\\
    = (p+1)(p+q+1) - p(p+q+2) = 1+q\,.\nonumber
\end{gather}
Notice that, rather amusingly, the answer is indepent of $p$! We can as well write out the zero-energy states explicitly. These are the states~(\ref{PsiKstates}) with $k=0$: 
\bea\label{Psi0}
|\Phi\rangle_0=\big(\Phi^{(0)}\big)_{\beta_1\ldots \beta_q} (z_{2}^\dagger)^{\beta_1}\cdots (z_{2}^\dagger)^{\beta_{q}}\,\left(\epsilon_{\alpha\beta}(z_1^\dagger)^\alpha (z_2^\dagger)^\beta\right)^p |0\rangle\,.
\eea

In fact, we can also compute the equivariant Witten index. To this end, one should simply replace the dimensions of representation in (\ref{Witten CP1}) with their characters. Recalling the expression for the $\SU(2)$ characters
\begin{equation}
\upchi\big({\T^{p}}\big):=\upchi_p=\frac{t^{p+1}-t^{-p-1}}{t-t^{-1}}\nonumber
\end{equation}
we arrive at the following result for the equivariant index:
\begin{eqnarray}
  \tilde{W}=  \upchi_{p}\upchi_{p+q}-\upchi_{p-1}\upchi_{p+q+1}=
  \upchi_q\,. \label{EquivWitten CP1}
\end{eqnarray}
Again, the answer is independent of $p$ and correctly reproduces the character of $\T^{q}$ -- the representation of the zero modes~(\ref{Psi0}). Moreover, $\tilde{W}$ coincides with the equivariant index of the Dolbeault operator on $\CP^1$ twisted by the line bundle $\mathcal{O}(q)$ (cf.~\cite[Sec.~9]{Pestun}).  These surprising facts will be explained later on.

\section{Spin chain K\"ahler-de Rham model (K-model)}\label{N4CP1}

In the previous section we  constructed a finite-dimensional QM model for the Dolbeault complex on $\CP^1$. A natural question is whether one can  construct an analogous model for the K\"ahler-de Rham complex. As is well-known (cf. the book~\cite{SmilgaDiffGeom}), this implies $\mathcal{N}=4$ SUSY. To this end, on top of $\psi_{12}$ we will add an extra fermion, called $\phi_{21}$, transforming in the dual representation. We assume that the fermion $\phi_{21}$ and its conjugate $\phi_{21}^\dagger$ satisfy the same canonical commutation relations as $\psi_{12}$ and $\psi_{12}^\dagger$, i.e.~$\Bigl\{\phi_{21},\phi^\dagger_{21}\Bigr\} = 1$. 
Next, we introduce the following two complex supercharges: 
\bear
\mathcal{Q}_1=\upalpha_{12}\,\psi_{12}\,\big(z_1^\dagger \circ z_2\big)\,,\quad\quad 
\mathcal{Q}_2=\upalpha_{12}\,\phi^\dagger_{21}\,\big(z_1^\dagger \circ z_2\big)\,,
\eear
as well as suitable analogues of the constraints~(\ref{C1}):
\bear
 &&\tilde{\mathcal{C}}_1=z^{\dagger}_1 \circ z_1+ \psi_{12}^\dagger \psi_{12} -\phi_{21}^\dagger \phi_{21}- p_1\,,\nonumber\\ \label{CP1constr}
   &&\tilde{\mathcal{C}}_2=z^{\dagger}_2 \circ z_2- \psi_{12}^\dagger \psi_{12}+\phi_{21}^\dagger \phi_{21} - p_2\,.
\eear
As in the previous sections the Hilbert space $\mathscr{H}(p_1,p_2)$ of the K-model  is constructed from the Fock space by imposing $\tilde{\mathcal{C}}_1 = \tilde{\mathcal{C}}_2 =0$ on states.

Clearly, $\mathcal{Q}_1^2=\mathcal{Q}_2^2=\Bigl\{\mathcal{Q}_1, \mathcal{Q}_2\Bigr\}=0$. Computing the anti-commutator of the two supercharges and using the constraints, we get
\bea
\Bigl\{\mathcal{Q}_1, \mathcal{Q}^\dagger_2\Bigr\}=\upalpha_{12}^2\,(p_1-p_2)\,\psi_{12}\,\phi_{21}\,.
\eea
Thus, a necessary condition for the $\mathcal{N}=4$ SUSY algebra is to set\footnote{Notice that here, in contrast to the $\mathcal{N}=2$ case, one could apriori choose an arbitrary ordering in the constraints~(\ref{CP1constr}) -- the corresponding values of $p_1, p_2$ would be determined from the algebra relation $\Bigl\{\mathcal{Q}_1, \mathcal{Q}^\dagger_2\Bigr\}=0$.} $p_1=p_2$. Besides, for the constraints to make sense one should at least require
\bea
\label{p1equalp2}
p_1=p_2:= p\in \mathbb{Z}\,.
\eea
The rest of the SUSY algebra is easily checked, so that it has the final form
\bear\label{SUSYalg}
&&\Bigl\{\mathcal{Q}_\uprho, \mathcal{Q}_\upgamma^\dagger\Bigr\}=\delta_{\uprho\upgamma}\,\mathcal{H}\,,\qquad \uprho,\upgamma=1,2\,,\quad\quad\textrm{where}\\
&&\mathcal{H}=\upalpha_{12}^2\left({1\over 2}\Bigl\{z_1^\dagger \circ z_2, z_2^\dagger \circ z_1\Bigr\}+\psi_{12}^\dagger \psi_{12}+\phi_{21} \phi_{21}^\dagger-2  \phi_{21}^\dagger \phi_{21} \psi_{12}\psi_{12}^\dagger-1\right)\,.\label{HamiltonianKmodelCP1}
\eear
We have written the Hamiltonian in a way, manifestly symmetric w.r.t.~the $\mathsf{U}(2)$ rotations
\bea\label{SU2Rsymm}
\left(\begin{array}{c}
    \psi_{12}   \\
    \phi_{21}^\dagger 
\end{array}\right)\; \mapsto g \cdot \left(\begin{array}{c}
    \psi_{12}   \\
    \phi_{21}^\dagger 
\end{array}\right)\,,\quad\quad g\in \mathsf{U}(2)\,.
\eea
The constraints~(\ref{CP1constr}) are also invariant w.r.t.~this transformation. This symmetry is the $\mathsf{U}(2)$ $R$-symmetry of the SUSY algebra~(\ref{SUSYalg})  that rotates $(\mathcal{Q}_1, \mathcal{Q}_2)$ as a doublet.

Let us explain why we have insisted on keeping the `non-holomorphic' transformation~(\ref{SU2Rsymm}) rather than relabeling the fermions $\phi_{21}^\dagger\equiv \phi_{12}$. The reason is that we wish to adhere to the conventional definition of the Fock space vacuum:
\bea
\label{VacuumKtheory}
z_A^\alpha|0\rangle=\psi_{12}|0\rangle=\phi_{21}|0\rangle=0\,.
\eea
In this case the fermion number operator $F=\psi_{12}^\dagger \psi_{12}+\phi_{21}^\dagger \phi_{21}$ coincides with differential form degree via the standard mapping of fermions to differential forms~\cite{WittenSUSYQM}. One may thus view $\psi_{12}^\dagger$ as $\mathrm{d}z$ and $\phi_{21}^\dagger$ as $\mathrm{d}\smallthickbar{z}$, where $z$ is the complex coordinate on $\CP^1$. The downside of this definition is that the vacuum state so defined is not invariant w.r.t the $\mathsf{U}(2)$ rotations~(\ref{SU2Rsymm}).

\subsubsection{Index of the K-model.} Just like in the $\mathcal{N}=2$ system above, we may compute the Witten index here. Depending on the fermion number, one has the following $\SU(2)$ representations:
\bear
&&F=0:\quad\quad \T^p\otimes \T^p\,,\\
&&F=1:\quad\quad \left(\T^{p-1}\otimes \T^{p+1}\right)\oplus \left(\T^{p+1}\otimes \T^{p-1}\right)\,,\\
&&F=2:\quad\quad \T^p\otimes \T^p\,.
\eear
As a result, the equivariant Witten index takes the form
\bea
\tilde{W}=2\, \upchi_p^2-2\, \upchi_{p-1}\upchi_{p+1}=2=\mathrm{Eu}(\CP^1)\,,
\eea
where $\mathrm{Eu}$ stands for the Euler characteristic. 
Rather remarkably, just like in the $\mathcal{N}=2$ system above, the answer is independent of~$p$ and coincides with the Euler characteristic of $\CP^1$. In fact, there are two SUSY ground states, which may be constructed explicitly as follows:
\bear
&&|\mathrm{vac}_1\rangle =\Big(\epsilon_{\alpha\beta} (z_1^\alpha)^\dagger (z_2^\beta)^\dagger\Big)^p\,|0\rangle\,,\\
&&|\mathrm{vac}_2\rangle =\Big(\epsilon_{\alpha\beta} (z_1^\alpha)^\dagger (z_2^\beta)^\dagger\Big)^p\,\psi_{12}^\dagger \phi_{21}^\dagger|0\rangle\,,
\eear
where $|0\rangle$ is the Fock space vacuum defined in~(\ref{VacuumKtheory}).

The two states form a doublet w.r.t.~the $\SU(2)$-part of~(\ref{SU2Rsymm}). Identifying the space of vacua with $H^\ast (\CP^1)$, this $\SU(2)$ is easily seen to coincide with the Lefschetz $\SU(2)$ acting in the cohomology ring of K\"ahler manifolds (see~\cite{GriffithsHarris} for the definition or~\cite{Cecotti} for a discussion in the context SUSY QM models). The remaining $\mathsf{U}(1)\subset \mathsf{U}(2)$ generated by ${F'=\psi_{12}^\dagger \psi_{12}-\phi_{21}^\dagger \phi_{21}}$ acts on forms of Hodge type $(m_L, m_R)$ as multiplication by $m_L- m_R$. 

\section{Quiver formulation}\label{QuiverSection}
It turns out that the content of both models discussed above may be succinctly described in terms of framed colored quiver diagrams:  
\bea\label{12quivdiag}
\begin{overpic}[scale=0.6,unit=0.75mm]{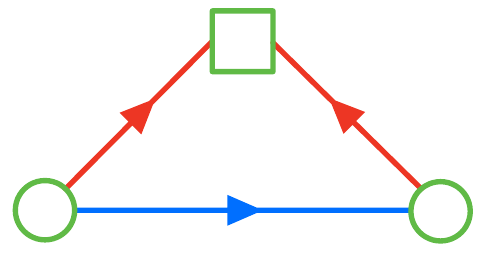}
\put(4.3,5.5){\footnotesize $\CC$}
\put(30.5,28.5){{\footnotesize $\CC^2$}}
\put(11,20){$z_1$}
\put(57.9,5.5){\footnotesize $\CC$}
\put(51,20){$z_2$}
\put(30.5,11){$\psi_{12}$}
\put(24,-7.5){D-model}
\end{overpic}
\quad\quad\quad 
\raisebox{0.02\height}{
\begin{overpic}[scale=0.6,unit=0.75mm]{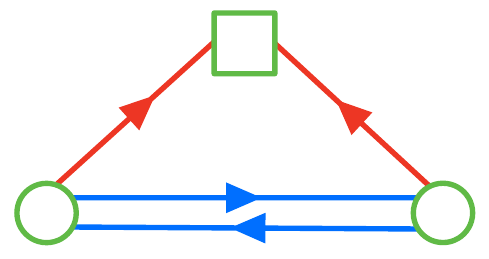}
\put(4.5,5){\footnotesize $\CC$}
\put(30.8,28){\footnotesize $\CC^2$}
\put(11,20){$z_1$}
\put(58.3,5){\footnotesize $\CC$}
\put(51,20){$z_2$}
\put(30.5,12){$\psi_{12}$}
\put(30.5,-1){$\phi_{21}$}
\put(24,-9){K-model}
\end{overpic}
}
\eea

\vspace{0.3cm}
Diagrams of this type will be widely used in the generalizations below, so let us describe the procedure of associating to such graphs the Hilbert spaces of our theories as well as their classical counterparts  (i.e.~phase spaces). 

The quivers that we will encounter will have two types of vertices and two types of edges.
The vertices will be represented either by circles or squares.  Circular vertices will be called `local' -- they correspond to the action of a gauge group, whereas square vertices will be called `global' and correspond to the action of a global symmetry group. The red and blue edges will be associated with bosonic and fermionic fields respectively.

We start with the construction of phase spaces, which in this case are called \textit{quiver varieties}. 
To each vertex of the quiver we associate a vector space $\mathds{C}^m$ (for some $m$) and the group $\mathsf{U}(m)$ acting on it. To each oriented edge that goes from $\mathds{C}^m$ to $\mathds{C}^k$ we associate the linear space of $k\times m$-matrices, denoted by  $\mathrm{Hom}\big(\mathds{C}^m,\mathds{C}^k\big)$, equipped with the symplectic form 
\bea
i\,\mathsf{Tr}\big(\mathrm{d}\zeta^\dagger\wedge\mathrm{d}\zeta\big)\,,
\eea
where ${\zeta\in \mathrm{Hom}\big(\mathds{C}^m,\mathds{C}^k\big)}$. If an edge is fermionic, then we additionally assume that the matrix elements are Grassmann (i.e.~anti-commuting) variables. These vector spaces are subject to the $\mathsf{U}(m)$ and $\mathsf{U}(k)$ actions associated with the vertices. Concretely, the $\mathsf{U}(k)\times\mathsf{U}(m)$-action on  $\zeta\in\mathrm{Hom}\big(\mathds{C}^m,\mathds{C}^k\big)$ is given by
\begin{equation}
\zeta\;\mapsto\; g_1 \zeta g_2^\dagger\,,\qquad g_1\in\mathsf{U}(k)\,,\quad g_2\in \mathsf{U}(m)\,.
\end{equation}
To construct the  quiver variety, one takes the direct sum of $\mathrm{Hom}$'s over all oriented edges, which is naturally a symplectic manifold with  the symplectic structure of a direct product, and performs the symplectic quotient w.r.t. the action of all $\mathsf{U}$-groups associated to the local vertices. The resulting  manifold is additionally equipped with the action of the  $\mathsf{U}$-groups in the global vertices.

Hilbert spaces associated to the quivers are constructed via geometric quantization of the quiver varieties. Technically this means that the matrix elements of all $\zeta$'s become quantum oscillators, whereas the moment maps corresponding to the symplectic reduction are imposed as constraints on the Fock space of these oscillators. 

In practice, we will encounter quivers of a very special type. All our quivers have the combinatorial type of a simplex with one global vertex. The  edges that include the global vertex are bosonic and point  into that vertex. All other edges are fermionic and correspond to either one or two (pointed in opposite directions) fermionic lines depending on the model. The simplest examples of such quivers describing the $\CP^1$ D- and K-models are shown in~(\ref{12quivdiag}).

Sometimes, for the ease of visualization, we will hide the global vertex of the quiver, simply indicating the red lines that are meant to flow into it.

\newpage
\chapter{Classical \texorpdfstring{$\CP^1$}{Lg} models in superspace}

\vspace{0.5cm}
\section{Classical D-model}\label{superspacesec}

The goal of the present section is to define a mechanical `classical spin model', which upon quantization leads to the spin chain Dolbeault  model described in Section~\ref{N=2 CP1section}. The classical analogue of the D-model has $\mathcal{N}=2$ supersymmetry, just like its quantum counterpart. This means that we can write the corresponding Lagrangians in three different ways: either by using $\mathcal{N}=2$ superfields, $\mathcal{N}=1$ superfields or in component form. 
We will choose the following strategy: first we postulate an abstract $\mathcal{N}=2$  superfield system, 
subsequently studying its successive reductions to $\mathcal{N}=1$ superfields as well as to components,  
finally making sure that it actually reproduces the classical version of the D-model.

\subsection{$\mathcal{N}=2$ superspace.}

We start by defining the free gauged $\mathcal{N}=2$ superfield system, subsequently adding interactions at the second step. 

In $\mathcal{N}=2$  superspace with coordinates  $\big(t,\theta_c,\smallthickbar{\theta}_c\,\big)$ the free system may be described by two chiral bosonic doublet superfields $\mathbf{Z}_1^\alpha, \mathbf{Z}_2^\alpha$, one chiral fermionic superfield $\mathbf{\Psi}_{12}$ and  two real gauge superfields  $\mathbf{V}_1, \mathbf{V}_2$ of general type.
In order to fix notations we recall that chiral fields satisfy the chirality conditions $\thickbar{D}_c\mathbf{Z}_1^\alpha=\thickbar{D}_c\mathbf{Z}_2^\alpha = \thickbar{D}_c\mathbf{\Psi}_{12}=0$ (together with the complex conjugate expressions), where the superderivatives are\footnote{Here $\thickbar{D}_c$ is Hermitian conjugate to  $D_c$. The conjugation of $D_c\mathbf{B}$ and $D_c\mathbf{F}$ gives $\thickbar{D}_c\thickbar{\mathbf{B}}$ and $-\thickbar{D}_c\thickbar{\mathbf{F}}$ for $\mathbf{B}$  and $\mathbf{F}$ a bosonic and fermionic superfield, respectively.}
\begin{equation}
\label{derivatives}
\thickbar{D}_c = -\frac{\partial}{\partial\smallthickbar{\theta}_c} + i\theta_c\frac{\partial}{\partial t}\,,\qquad D_c = \frac{\partial}{\partial\theta_c} - i\smallthickbar{\theta}_c\frac{\partial}{\partial t}\,,\qquad \thickbar{D}_c^2=D_c^2 = 0\,.
\end{equation}
The free gauged Lagrangian may then be written as follows  (see~\cite{Nicolai76,Nicolai77, HoweTownsend}):
\bea\label{N2freeLagr}
\mathscr{L}_0=\frac{1}{2}\int\,\mathrm{d}^2\theta_c\,\Big(|\mathbf{Z}_1|^2 e^{\mathbf{V}_1}+|\mathbf{Z}_2|^2 e^{\mathbf{V}_2}+|\mathbf{\Psi}_{12}|^2 e^{\mathbf{V}_1-\mathbf{V}_2}+\xi_1 \mathbf{V}_1+\xi_2 \mathbf{V}_2\Big)\,,
\eea
where $\xi_1, \xi_2$ are the (real) Fayet–Iliopoulos (FI) terms. Later on we will match the values of these parameters with the $p_A$'s of  the D-model. By definition, $|\mathbf{Z}_A|^2 = \thickbar{\mathbf{Z}}_A\circ\mathbf{Z}_A$ and $|\mathbf{\Psi}_{12}|^2 = \thickbar{\mathbf{\Psi}}_{12}\mathbf{\Psi}_{12}$.

In order to introduce interactions we do not add extra terms to  (\ref{N2freeLagr}) but rather modify the chirality conditions:
\bear\label{chirconstr1}
&&\thickbar{D}_c\mathbf{Z}_1=\vkappa\, \mathbf{\Psi}_{12} \mathbf{Z}_2\,,\qquad D_c\thickbar{\mathbf{Z}}_1 = \smallthickbar{\vkappa}\,\thickbar{\mathbf{\Psi}}_{12}\thickbar{\mathbf{Z}}_2\,,\qquad \thickbar{D}_c\mathbf{Z}_2=\thickbar{D}_c\mathbf{\Psi}_{12}=0\,.
\eear
Here $\vkappa$ is a coupling constant (again, it will be matched with the coupling of the D-model in the next section). Supersymmetric systems with such constraints\footnote{We would like to thank E.~Ivanov for pointing out this paper and explaining the possibility of having such constraints.} have been studied in~\cite{IvanovToppan}. Our claim is that the model described by~(\ref{N2freeLagr})-(\ref{chirconstr1}) is the classical counterpart of the D-model of Section \ref{N=2 CP1section}. 
Before checking this statement in components we reduce $\mathcal{N}=2$ to $\mathcal{N}=1$ superfields. The resulting form of the Lagrangian will be useful for establishing a relation between the D-model and the supersymmetric $\CP^1$ sigma model.

If the reader is somewhat bewildered by the unusual form of the constraints~(\ref{chirconstr1}), one can as well pass to standard chiral fields, at the expense of making the Lagrangian more complicated. Indeed, we may set $\mathbf{\Psi}_{12}=\thickbar{D}_c \tilde{\mathbf{\Psi}}_{12}$ and solve the nonlinear constraint via $\mathbf{Z}_1=\tilde{\mathbf{Z}}_1+\vkappa \,\tilde{\mathbf{\Psi}}_{12} \mathbf{Z}_2$, where $\tilde{\mathbf{Z}}_1$ is a conventional chiral superfield satisfying $\thickbar{D}_c\tilde{\mathbf{Z}}_1=0$. Substituing these values in the Lagrangian~(\ref{N2freeLagr}), one gets an expression in terms of chiral fields $\tilde{\mathbf{Z}}_1, \mathbf{Z}_2$ and  superfields $\tilde{\mathbf{\Psi}}_{12}, \mathbf{V}_1, \mathbf{V}_2$ of general type.

\subsection{Reduction to $\mathcal{N}=1$ superspace.}\label{n1Dmodel}

In this section we discuss the reduction of the model (\ref{N2freeLagr}) with the deformed chirality conditions (\ref{chirconstr1}) to  $\mathcal{N}=1$ superspace. 
To this end one introduces the `real' coordinates in $\mathcal{N} = 2$ superspace
\begin{equation}
\theta_c = \theta+i\theta^\prime\,,\quad\smallthickbar{\theta}_c = \theta-i\theta^\prime\quad\Rightarrow\quad \theta = \frac{\theta_c+\smallthickbar{\theta}_c}{2}\,,\quad\theta^\prime = \frac{\theta_c-\smallthickbar{\theta}_c}{2i}\
\end{equation}
as well as the real superderivatives
\begin{equation}
\label{RealSuperderivatives}
D = D_c-\thickbar{D}_c = \frac{\partial}{\partial\theta} - 2i\theta\frac{\partial}{\partial t}\,,\qquad D^\prime = i\big(D_c+\thickbar{D}_c\big) = \frac{\partial}{\partial\theta^\prime} -2i\theta^\prime\frac{\partial}{\partial t}\,.
\end{equation}
The measure of integration in superspace may then be written as
$
\int\mathrm{d}^2\theta_c: = i
\int\mathrm{d}\theta\mathrm{d}\theta^\prime\,.
$

Now we want to explicitly integrate over, say, $\theta^\prime$. The standard way of doing this is the following. Since taking the integral over $\theta^\prime$ is the same as applying    
$D^\prime$ 
up to a total derivative (see (\ref{derivatives})), it means that upon applying $D^\prime$ to  the $\mathcal{N} = 2$ superfield Lagrangian, the dependence on $\theta^\prime$ will be a total derivative. Thus, upon differentiating, we may safely put $\theta^\prime = 0$. The remaining Lagrangian gives the desired $\mathcal{N} = 1$ formulation, since everything is expressed in terms of $\theta$. 
Clearly, in the resulting expression all superderivatives must be expressed in terms of $D$: for example,
\begin{equation}
D^\prime \mathbf{Z}_1 = iD\mathbf{Z}_1+2i\vkappa\,\mathbf{\Psi}_{12} \mathbf{Z}_2\,,\quad
D^\prime\,\thickbar{\mathbf{Z}}_1 = -iD\thickbar{\mathbf{Z}}_1+2i\,\smallthickbar{\vkappa}\,\thickbar{\mathbf{\Psi}}_{12} \thickbar{\mathbf{Z}}_2\,.
\end{equation}

In the course of the reduction each chiral, or generalized chiral, superfield is reduced to a single $\mathcal{N}=1$ superfield, for example $\mathbf{Z}_A\mapsto \mathbf{Z}_A|_{\theta^\prime = 0}$. However, a superfield of general type is reduced to \emph{two} $\mathcal{N}=1$ superfields, $\mathbf{V}_A\mapsto \left(\mathbf{V}_A|_{\theta^\prime = 0}, \,D^\prime\mathbf{V}_A|_{\theta^\prime = 0}\right)$. We will introduce the following notation for these $\mathcal{N}=1$ fields:
\begin{align}
\Big\{\mathbf{Z}_A&\big(\theta,\theta^\prime\big),\mathbf{V}_A\big(\theta,\theta^\prime\big), D^\prime\mathbf{V}_A\big(\theta,\theta^\prime\big),\mathbf{\Psi}_{12}\big(\theta,\theta^\prime\big)\Big\}\Big|_{\theta^\prime = 0}:=\nonumber\\&:=\Big\{\mathsf{Z}_A(\theta),\,\mathsf{V}_A(\theta),\,-2i\,\mathsf{\Lambda}_A(\theta),\,\mathsf{\Psi}_{12}(\theta)\Big\}\,.\label{N1Dsuperfields}
\end{align}
Using this approach, we obtain the following expression for the Lagrangian in $\mathcal{N} = 1$ superspace:
\begin{align}
\mathscr{L} = \frac{i}{2}\int\mathrm{d}\theta
\Big[&2i\,\thickbar{\mathsf{Z}}_A\circ D\mathsf{Z}_A-2i\,\thickbar{\mathsf{\Psi}}_{12} D\mathsf{\Psi}_{12}+2i\,\smallthickbar{\vkappa}\,\thickbar{\mathsf{\Psi}}_{12}\thickbar{\mathsf{Z}}_{2}\circ\mathsf{Z}_1 +2i\vkappa\,\mathsf{\Psi}_{12}\thickbar{\mathsf{Z}}_1\circ\mathsf{Z}_2\Big)+\nonumber\\ \label{DmodelN1superfields}
&+\mathsf{\Lambda}_1\Big(|\mathsf{Z}_1|^2+|\mathsf{\Psi}_{12}|^2+\xi_1\Big)+\mathsf{\Lambda}_2\Big(|\mathsf{Z}_2|^2-|\mathsf{\Psi}_{12}|^2+\xi_2\Big)\Big]\,.
\end{align}
Here we have eliminated the exponents of the $\mathsf{V}_A$-fields by doing a `complexified gauge transformation'
\begin{equation}
\mathsf{Z}_1\rightarrow e^{-\frac{1}{2}\mathsf{V}_1}\,\mathsf{Z}_1\,,\qquad\mathsf{Z}_2\rightarrow e^{-\frac{1}{2}\mathsf{V}_2}\,\mathsf{Z}_2\,,\qquad
\mathsf{\Psi}_{12}\rightarrow e^{-\frac{1}{2}(\mathsf{V}_1-\mathsf{V}_2)}\,\mathsf{\Psi}_{12}
\end{equation}
together with a shift of the $\mathsf{\Lambda}_A$'s after differentiation. 
We see that the constraints imposed by the Lagrange multipliers  $\mathsf{\Lambda}_A$ are a superfield incarnation of (\ref{C1}). Therefore we identify 
$\xi_1 = -p_1$ and $\xi_2 = -p_2$. Once we pass to  component fields in the next section, it will be clear that, in order to match the D-model, we should additionally set  $\vkappa = \smallthickbar{\vkappa} = -\upalpha_{12}$, where $\upalpha_{12}$ is the parameter of the quantum D-model.

\subsection{Component formulation and relation with quantum D-model.}

The final step is to rewrite the theory (\ref{DmodelN1superfields}) in components and check that it coincides with the classical version of the D-model. The component expansion of the superfields~(\ref{N1Dsuperfields})~is:
\begin{align}
\mathsf{Z}_A := z_A + i\theta\gamma_A\,,\quad\mathsf{\Psi}_{12} := i\big(\psi_{12} + \theta\zeta_{12}\big)\,,\qquad\mathsf{\Lambda}_A := \eta_A + \theta\lambda_A\,,\quad A = 1,2\,. \label{DmodelSuperfieldsComponents}
\end{align}
In terms of these component fields the Lagrangian (\ref{DmodelN1superfields}) is rather complicated. It can be significantly simplified by noting that the kinetic term has the following expansion:
\begin{align}
&\int\mathrm{d}\theta\,\Big(-\thickbar{\mathsf{Z}}_A\circ D\mathsf{Z}_A\Big) = i\,\smallthickbar{z}_A\circ\dot{z}_A + \smallthickbar{\gamma}_A\circ\gamma_A\,,\\
&\int\mathrm{d}\theta\,\Big(\thickbar{\mathsf{\Psi}}_{12} D\mathsf{\Psi}_{12}\Big) = i\smallthickbar{\psi}_{12}\dot{\psi}_{12}+\smallthickbar{\zeta}_{12}\zeta_{12}\,.
\end{align}
As a result, $\gamma_A$ and $\zeta_{12}$ are non-dynamical degrees of freedom and thus can be eliminated by using the equations of motion. Doing this, we get
\begin{align}
\mathcal{S} = \int \mathrm{d}t\,\Big[i\,\smallthickbar{z}_A\circ&\dot{z}_A+i\smallthickbar{\psi}_{12}\circ\dot{\psi}_{12}-\upalpha_{12}^2\big(\smallthickbar{z}_1\circ z_2\big)\big(\smallthickbar{z}_2\circ z_1\big)-\upalpha_{12}^2\smallthickbar{\psi}_{12}\psi_{12}\big(\smallthickbar{z}_2\circ z_2 - \smallthickbar{z}_1\circ z_1\big)+\nonumber\\
&+\lambda_1\big(\smallthickbar{z}_1\circ z_1 + \smallthickbar{\psi}_{12}\psi_{12} - p_1\big) +\lambda_2\big(\smallthickbar{z}_2\circ z_2 - \smallthickbar{\psi}_{12}\psi_{12} - p_2\big)\Big]\,.\label{ActionInComponentsCP1}
\end{align}
Note that the Lagrange multipliers $\eta_A$ fall out of this expression as well.

It is clear from the above action that the abstract $\mathcal{N} = 2$ theory (\ref{N2freeLagr}) supplemented with the modified chirality conditions (\ref{chirconstr1}) is exactly the classical analogue of the D-model described in Section~\ref{N=2 CP1section}. Indeed, quantization of this theory reproduces\footnote{In order to distinguish the quantum and classical cases, we write $z^\dagger_A$ for the Hermitian conjugate operator in the quantum case, whereas in the classical case  we write $\smallthickbar{z}_A$.} the commutation relations~(\ref{CanonicalCommutationRelation}), the supercharge~(\ref{Q12supercharge}) and the Hamiltonian, 
as well as   the constraints~(\ref{C1}). For details of this see  Appendix~\ref{ClassicalCP1Dapp}.

\section{Classical K-model}

In this section we discuss the classical analogue of the K-model constructed in Section~\ref{N4CP1}. We will be using the strategy from the previous section. Concretely, we postulate an abstract model in $\mathcal{N} = 2$ superspace, then find its reduction to $\mathcal{N} = 1$ superspace and, finally, to component form that can be directly compared to the K-model. The calculations are similar to the case of the D-model, so we will focus on the necessary modifications. 

\subsection{$\mathcal{N} = 2$ superspace.}
To introduce the free system\footnote{See Section \ref{superspacesec} for $\mathcal{N} = 2$ superspace conventions.} we consider the following field content in $\mathcal{N}=2$ superspace: two bosonic chiral doublets $\mathbf{Z}_1^\alpha, \mathbf{Z}_2^\alpha$, two fermionic chiral fields $\mathbf{\Psi}_{12}, \mathbf{\Phi}_{21}$ and two additional  chiral superfields $\mathbf{\Lambda}_1, \mathbf{\Lambda}_2$. 
We postulate the free system of the form
\begin{align}
\mathscr{L}_0 = \frac{1}{2}&\int\mathrm{d}^2\theta\Big(|\mathbf{Z}_1|^2+|\mathbf{Z}_2|^2+|\mathbf{\Psi}_{12}|^2+|\mathbf{\Phi}_{21}|^2\Big)+\nonumber
\\
+&\xi\int\mathrm{d}\theta_c\,\Big(\mathbf{\Lambda}_1+\mathbf{\Lambda}_2\Big)-\xi\int\mathrm{d}\smallthickbar{\theta}_c\,\Big(\thickbar{\mathbf{\Lambda}}_1+\thickbar{\mathbf{\Lambda}}_2\Big)\,,\label{N2lKmodelCP1action}
\end{align}
where $\xi$ is a (real) FI parameter. 
In order to add an interaction in this system we deform the chirality conditions for our fields in the following way:
\begin{align}
\label{ModifiedChirality}
\begin{pmatrix}
\thickbar{D}_c\mathbf{Z}_1 \\
\thickbar{D}_c\mathbf{Z}_2
\end{pmatrix}
=
\underbracket{
\begin{pmatrix}
\mathbf{\Lambda}_1 & \vkappa\,\mathbf{\Psi}_{12} \\ 
\vkappa\,\mathbf{\Phi}_{21} & \mathbf{\Lambda}_2
\end{pmatrix}
}_{\scalebox{1.1}{$:=\mathscr{A}$}}
\begin{pmatrix}
\mathbf{Z}_1 \\
\mathbf{Z}_2
\end{pmatrix}\,,
\end{align}
which involves the definition of a superconnection $\thickbar{D}_c-\mathscr{A}$. Equation~(\ref{ModifiedChirality}) is the statement about the existence of a full basis of covariantly constant sections (in this case corresponding to $\begin{pmatrix}
\mathbf{Z}_1^\alpha \\
\mathbf{Z}_2^\alpha
\end{pmatrix}$ with $\alpha=1, 2$), which generically are linearly independent. The superconnection thus has to be flat\footnote{Such zero curvature constraints in superspace were encountered in the study of 2D $\mathcal{N}=2$ Kac-Moody algebras in~\cite{HullSpence}.}, i.e.
\bea
\thickbar{D}_c\,\mathscr{A}-\mathscr{A}^2=0\,,
\eea
which in terms of matrix elements may be written as
\begin{align} \label{DLambda1eq}
&\thickbar{D}_c \mathbf{\Lambda}_1 = \vkappa\,\mathbf{\Psi}_{12}\mathbf{\Phi}_{21}\,,\qquad \thickbar{D}_c\mathbf{\Psi}_{12} = \mathbf{\Psi}_{12}\big(\mathbf{\Lambda}_2-\mathbf{\Lambda}_1\big)\,,\\ \label{DLambda2eq}
&\thickbar{D}_c \mathbf{\Lambda}_2 = \vkappa\,\mathbf{\Phi}_{21}\mathbf{\Psi}_{12}\,,\qquad 
\thickbar{D}_c \mathbf{\Phi}_{21} =  \boldsymbol{\mathbf{\Phi}}_{21}\big(\mathbf{\Lambda}_1-\mathbf{\Lambda}_2\big)\,. 
\end{align}
Here $\vkappa$ is a (possibly complex) coupling constant which will be related to the coupling constant of the K-model further on. Notice that, as a consequence of the equations, $\mathbf{\Lambda}_1+\mathbf{\Lambda}_2$ is still a chiral superfield in the ordinary sense, so that the superpotential term in~(\ref{N2lKmodelCP1action}) makes sense\footnote{Thus, one can introduce   only a single FI parameter due to the modified chirality constraints~(\ref{ModifiedChirality}). This is parallel to the situation in the K-model, where, as explained in Section~\ref{N4CP1} (see formula~(\ref{p1equalp2})), one can introduce only a single parameter $p$. Below we will see that $\xi=-p$.}.

It is easily verified that equations~(\ref{ModifiedChirality})-(\ref{DLambda2eq}) are consistent in their entirety, and applying the superderivative $\thickbar{D}_c$ to each equation does not give new relations.

\subsection{Comparison with the $D$-model.} 
At first sight the $\mathcal{N} = 2$ formulations of the D- and K-models look somewhat different. However, they  are closely related. To see this, instead of the action~(\ref{N2freeLagr}) featuring explicit gauge field factors, one could start with a `free' action of the form
\begin{align}\label{freeactionDmodel}
\mathscr{L}_0 = \frac{1}{2}&\int\mathrm{d}^2\theta\Big(|\mathbf{Z}_1|^2+|\mathbf{Z}_2|^2+|\mathbf{\Psi}_{12}|^2\Big)+\left(\int\mathrm{d}\theta_c\,\Big(\xi_1 \mathbf{\Lambda}_1+\xi_2 \mathbf{\Lambda}_2\Big) + \mathrm{c.c.}\right)\,,
\end{align}
which resembles~(\ref{N2lKmodelCP1action}). 
The fields $\mathbf{Z}_1, \mathbf{Z}_2$ are not chiral, but rather satisfy a generalized chirality constraint
\begin{align}
\label{ModifiedChiralityD}
\begin{pmatrix}
\thickbar{D}_c\mathbf{Z}_1 \\
\thickbar{D}_c\mathbf{Z}_2
\end{pmatrix}
= 
\underbracket{
\begin{pmatrix}
\mathbf{\Lambda}_1 & \vkappa\,\mathbf{\Psi}_{12} \\ 
0 & \mathbf{\Lambda}_2
\end{pmatrix}
}_{\scalebox{1.1}{$:=\tilde{\mathscr{A}}$}}
\begin{pmatrix}
\mathbf{Z}_1 \\
\mathbf{Z}_2
\end{pmatrix}\,,
\end{align}
In order for this to have solutions, the connection $\thickbar{D}_c-\tilde{\mathscr{A}}$ needs to be flat, just as above. In particular, it follows that
\bea\label{lambdachiral}
\thickbar{D}_c \mathbf{\Lambda}_1=\thickbar{D}_c\mathbf{\Lambda}_2=0\,,
\eea
so that $\mathbf{\Lambda}_1$ and $\mathbf{\Lambda}_2$ are chiral fields in the usual sense. 
To get back to the original formulation~(\ref{N2freeLagr}), one should solve the chirality constraints~(\ref{lambdachiral}) as $\mathbf{\Lambda}_A = \thickbar{D}_c \mathbf{V}_A$, where $\mathbf{V}_A$ are unconstrained superfields. Then, making the change of variables $\mathbf{Z}_A\mapsto e^{\mathbf{V}_A} \mathbf{Z}_A$ and $\mathbf{\Psi}_{12}\mapsto e^{\mathbf{V}_1-\mathbf{V}_2}\mathbf{\Psi}_{12}$, one recovers the action~(\ref{N2freeLagr}) and the constraints~(\ref{chirconstr1}).

Notice an essential difference between the D-model constraints~(\ref{ModifiedChiralityD}) and the K-model constraints~(\ref{ModifiedChirality}). In the D-model case each of the two fields $\mathbf{\Lambda}_1$ and $\mathbf{\Lambda}_2$ is chiral separately. As a result, it is possible to add two FI parameters via the superpotential in~(\ref{freeactionDmodel}). In the K-model, instead, only their sum is chiral, resulting in a single FI parameter. This means that in the K-model one cannot introduce `magnetic charges'~$q$ without explicitly breaking supersymmetry.

\subsection{Reduction to $\mathcal{N} = 1$ superspace.}
The reduction from $\mathcal{N}=2$ to $\mathcal{N} = 1$ superfields repeats almost verbatim the procedure described in Section \ref{n1Dmodel}. As in the case of the D-model we pick $\vkappa =\smallthickbar{\vkappa} = -\upalpha_{12}$ and define $\mathcal{N} = 1$ superfields as follows:
\begin{equation}
\Big\{\mathbf{Z}_A\big(\theta,\theta^\prime\big),\mathbf{\Psi}_{12}\big(\theta,\theta^\prime\big),\mathbf{\Phi}_{21}\big(\theta,\theta^\prime\big)\Big\}\Big|_{\theta^\prime = 0}
:=
\Big\{\mathsf{Z}_A(\theta),\mathsf{\Psi}_{12}(\theta),  \mathsf{\Phi}_{21}(\theta)\Big\}\,.
\end{equation}
In case of the `Lagrange multipliers' we set
\begin{equation}
\Big\{\mathbf{\Lambda}_A\big(\theta,\theta^\prime\big)+\thickbar{\mathbf{\Lambda}}_A\big(\theta,\theta^\prime\big)\Big\}\Big|_{\theta^\prime = 0}
:= \Big\{-\mathsf{\Lambda}_A(\theta)\Big\}\,.
\end{equation}
 Integrating over $\theta^\prime$ explicitly, we get the action in terms of $\mathcal{N} = 1$ superfields:
\begin{align}
&\qquad\qquad\mathcal{S} = \int\mathrm{d}t\,\mathrm{d}\theta
\Big[-\thickbar{\mathsf{Z}}_A\circ D\mathsf{Z}_A+\thickbar{\mathsf{\Psi}}_{12} D\mathsf{\Psi}_{12}+\thickbar{\mathsf{\Phi}}_{21}D\mathsf{\Phi}_{21}+\nonumber\\
&+
\upalpha_{12}\Big(\thickbar{\mathsf{\Psi}}_{12}\,\thickbar{\mathsf{Z}}_{2}\circ\mathsf{Z}_1 + \mathsf{\Psi}_{12}\,\thickbar{\mathsf{Z}}_1\circ\mathsf{Z}_2\Big)+\upalpha_{12}\Big(\thickbar{\mathsf{\Phi}}_{21}\thickbar{\mathsf{Z}}_1\circ\mathsf{Z}_2 + \mathsf{\Phi}_{21}\thickbar{\mathsf{Z}}_2\circ\mathsf{Z}_1\Big)+
\nonumber\\
&+\mathsf{\Lambda}_1\Big(|\mathsf{Z}_1|^2+|\mathsf{\Psi}_{12}|^2-|\mathsf{\Phi}_{21}|^2+\xi\Big)+\mathsf{\Lambda}_2\Big(|\mathsf{Z}_2|^2-|\mathsf{\Psi}_{12}|^2+|\mathsf{\Phi}_{21}|^2+\xi\Big)\Big]\,.
\end{align}
As one can check, this action is in fact the classical incarnation of the K-model of Section~\ref{N4CP1}. To this end one should identify the leading components of $\mathsf{Z}_A, \mathsf{\Psi}_{12}, \mathsf{\Phi}_{21}$ with $z_A, i\psi_{12}, i\phi_{21}$ respectively and eliminate all other non-dynamical components. As in the case of the D-model, only the top component of $\mathsf{\Lambda}_A$ is the true Lagrange multiplier coupling to the constraints (the leading component of $\mathsf{\Lambda}_A$ is an auxiliary field and simply drops out of the action).  The  formulation in terms of components, upon  eliminating the auxiliary fields, gives the constraints (\ref{CP1constr}) and the Hamiltonian~(\ref{HamiltonianKmodelCP1}). It is easy to see that the FI parameter should be identified with   parameter $p$ of the K-model as~$\xi=-p$.

\section{Deriving the sigma model} \label{Derive of CP1}

So far, we have discussed some properties of the D- and K-models, as well as their formulations in superspace. But what is their relation to the $\CP^1$ sigma model? Our next goal is to show how the equivariant Witten index $\Tilde{W}$ and the $g$-twisted partition function $\mathscr{Z}$ 
\big(where $g \in \SU(2)$\big) of the D- and K-models are related to those of the SUSY $\CP^1$ model.

\subsection{From D-model to the $\mathcal{N} = 2$ sigma model.} \label{Derive of CP^1 D}

To start with, let us recall the definition of the Witten index and the  partition function: 
\begin{gather}
    \Tilde{W}(g) = \mathsf{STr}_{\mathscr{H}(p_1, p_2)}\Big(g\cdot e^{-\beta \mathcal{H}}\Big), \quad \mathscr{Z}\left(g\right) = \mathsf{Tr}_{\mathscr{H}(p_1, p_2)}\Big(g\cdot e^{-\beta \mathcal{H}}\Big), 
\end{gather}
where $\beta$ is a parameter. By $\mathscr{H}(p_1, p_2)$  and $\mathcal{H}$ we denote the Hilbert space~(\ref{HilbertSpaceDmodel}) and Hamiltonian (\ref{HamiltonianDmodel}) of the D-model respectively. One can write path integral expressions for $\Tilde{W}(g)$ and $\mathscr{Z}\left(g\right)$. The classical action standing in the exponent in the integrand is the same in both cases and is given by 
\begin{align}
        \mathcal{S}=\int_{0}^{\beta} \mathrm{d}t \int &\mathrm{d}\theta\, \Big[-\thickbar{\mathsf{Z}}_A \circ D \mathsf{Z}_A + \thickbar{\mathsf{\Psi}}_{12} D\mathsf{\Psi}_{12}  + \upalpha_{12} \big( \mathsf{\Psi}_{12} \thickbar{\mathsf{Z}}_1 \circ \mathsf{Z}_2 + \thickbar{\mathsf{\Psi}}_{12} \thickbar{\mathsf{Z}}_2 \circ \mathsf{Z}_1 \big) +  \nonumber\\
        &+\mathsf{\Lambda}_1 \Big(|\mathsf{Z}_1|^2 + |\mathsf{\Psi}_{12}|^2 - p_1 \Big)+\mathsf{\Lambda}_{2} \Big(|\mathsf{Z}_2|^2 - |\mathsf{\Psi}_{12}|^2 - p_2\Big)\Big]. \label{N=1SUSY CP1 path int}
\end{align}
However, in the two cases we should impose different boundary conditions on the bosonic and fermionic components of the superfields (\ref{DmodelSuperfieldsComponents}) (for details see \cite{SmilgaDiffGeom}). In short, these are 
    \begin{align}
        z_A(\beta) &= g\cdot z_A(0)\,, & \gamma_A(\beta) &= \pm g\cdot \gamma_A(0)\,, \label{CP1 boundary conditions} \\
        \psi_{12}(\beta) &= \pm\psi_{12}(0)\,, &  \zeta_{12}(\beta) &= \zeta_{12}(0)\,,\nonumber
    \end{align}
    where `$+$' corresponds to the case of $\Tilde{W}(g)$, whereas  `$-$' corresponds to $\mathscr{Z}\left(g\right)$. 
    
    It is convenient to work with (\ref{N=1SUSY CP1 path int}) in a slightly different form. For this purpose, we introduce the matrix superfield $\mathsf{Z} = \begin{pmatrix}
    \mathsf{Z}_1 & \mathsf{Z}_2   
    \end{pmatrix}$. We will assume that this matrix is non-degenerate\footnote{The set of such matrices is dense in the original integration domain, so the value of the integral will not change if we restrict to this set.}. In this case  
    the polar decomposition theorem  states that the factorization $\mathsf{Z}=\mathsf{U}\mathsf{H}$ is unique, where $\mathsf{U}$ is a unitary matrix-valued bosonic superfield and $\mathsf{H}$ a Hermitian positive-definite  matrix-valued bosonic superfield. Components of these superfields inherit the boundary conditions from the components of $\mathsf{Z}_1$ and $\mathsf{Z}_2$.  
    With these definitions, the action reads:
    \begin{gather}
        \mathcal{S} = \int \mathrm{d}t\, \mathrm{d}\theta\,\Big[\mathsf{Tr}\left(-\mathsf{K} \mathsf{U}^{\dagger} D \mathsf{U} - \mathsf{H} D \mathsf{H}\right) + \thickbar{\mathsf{\Psi}}_{12} D\mathsf{\Psi}_{12} + \upalpha_{12}\left(\mathsf{\Psi}_{12} \mathsf{K}_{12} + \thickbar{\mathsf{\Psi}}_{12} \mathsf{K}_{21}\right)\Big],\label{NewFormOfSUSY CP1}
    \end{gather}
    where $\mathsf{K} := \mathsf{H}^2$ and $\mathsf{K}_{AB}$ are its matrix elements. It is implied that $\mathsf{U}$ is subject to the constraints $\thickbar{\mathsf{U}}_A\circ \mathsf{U}_B = \delta_{AB}$, where $\mathsf{U}_A$ denotes the $A$-th column.  
    It should also be noted that 
    \bea\label{Kdiag}
    \mathsf{K}_{11} = p_1 - |\mathsf{\Psi}_{12}|^2\quad \textrm{and} \quad \mathsf{K}_{22} = p_2 + |\mathsf{\Psi}_{12}|^2
    \eea
    due to the constraints in (\ref{N=1SUSY CP1 path int}).
    
    Next, let us extract the terms $p\,\sum_{A=1}^2 \thickbar{\mathsf{U}}_A D \mathsf{U}_A = p \,D\ln\left(\varepsilon_{\alpha_1 \alpha_2}\mathsf{U}^{\alpha_1}_1 \mathsf{U}^{\alpha_2}_2\right)$ and $\mathsf{Tr}\left(\mathsf{H}D\mathsf{H}\right) = \frac{1}{2}D\,\mathsf{Tr}\left(\mathsf{H}^2\right)$ from the first and second parts of the trace in (\ref{NewFormOfSUSY CP1}), respectively. These terms are total (super)derivatives and, thus, vanish\footnote{Both terms are $\SU(2)$-invariant. Only their bosonic parts could contribute when we integrate over~$\theta$, but due to the boundary conditions (\ref{CP1 boundary conditions}) all bosonic fields are periodic.}. So, we end up with 
    \begin{align}
        \mathcal{S} = \int \mathrm{d}t \,&\mathrm{d}\theta \,\bigg[|\mathsf{\Psi}_{12}|^2  \,\thickbar{\mathsf{U}}_1 \circ D \mathsf{U}_1 - |\mathsf{\Psi}_{12}|^2\, \thickbar{\mathsf{U}}_2 \circ D \mathsf{U}_2- \mathsf{K}_{12} \thickbar{\mathsf{U}}_2 \circ D \mathsf{U}_1- \mathsf{K}_{21} \thickbar{\mathsf{U}}_1 \circ D \mathsf{U}_2 -\nonumber \\ 
        &- q\,\thickbar{\mathsf{U}}_2 \circ D \mathsf{U}_2
        + \frac{1}{2}\big(\,\thickbar{\mathsf{\Psi}}_{12} D \mathsf{\Psi}_{12} + D \thickbar{\mathsf{\Psi}}_{12} \mathsf{\Psi}_{12} \big) + \upalpha_{12} \left(\mathsf{\Psi}_{12} \mathsf{K}_{12} + \thickbar{\mathsf{\Psi}}_{12} \mathsf{K}_{21}\right)\bigg]\,, 
    \end{align}
    where we have additionally symmetrized the term $\thickbar{\mathsf{\Psi}}_{12} D \mathsf{\Psi}_{12}$.
    
    The obvious way to proceed is to integrate over $\mathsf{K}_{12}$ and $\mathsf{K}_{21}$. However, these fields are not arbitrary due to the positive-definiteness of $\mathsf{H}=\sqrt{\mathsf{K}}$, so, in general, there are  restrictions on the integration domain. However, these restrictions are lifted in the limit $p \to \infty$ with $q$ fixed.
    Indeed, in this limit the diagonal elements of $\mathsf{K}$ tend to infinity, as one can see from~(\ref{Kdiag}). Thus, $\mathsf{H}=\sqrt{\mathsf{K}}$ approaches a diagonal matrix with positive elements.
    In this limit the fields $\mathsf{K}_{12}$ and $\mathsf{K}_{21}$ are therefore Lagrange multipliers generating the constraints $\mathsf{\Psi}_{12} = \left(\upalpha_{12} \right)^{-1} \thickbar{\mathsf{U}}_2 \circ D \mathsf{U}_1$ and ${\thickbar{\mathsf{\Psi}}_{12} =  \left(\upalpha_{12} \right)^{-1} \thickbar{\mathsf{U}}_1 \circ D \mathsf{U}_2}$. Substituting these values in the action, we finally arrive at
    \begin{gather}
        \mathcal{S} = \int \mathrm{d}t \,\mathrm{d}\theta \,\bigg[\frac{i}{2\upalpha_{12}^2} \Big( D \thickbar{\mathsf{U}}_1 \circ \mathsf{U}_2 \thickbar{\mathsf{U}}_2 \circ \dot{\mathsf{U}}_1 + \dot{\thickbar{\mathsf{U}}}_1 \circ \mathsf{U}_2 \thickbar{\mathsf{U}}_2 \circ D \mathsf{U}_1\Big) - q\,\thickbar{\mathsf{U}}_2 \circ D \mathsf{U}_2\bigg]\,. \label{homSUSy CP1}
    \end{gather}
    Clearly, this is a supersymmetric  $\CP^1$ sigma model, with an additional magnetic field, written in $\mathcal{N}=1$ superspace and using homogeneous coordinates~\cite{BykSmilga}. As we have seen before, the quantum D-model has $\mathcal{N}=2$ supersymmetry. It is thus natural to expect that the same supersymmetry should persist in the sigma model. As we shall momentarily show,~(\ref{homSUSy CP1}) is indeed the $\mathcal{N}=2b$ model.   Therefore we have proven that the equivariant Witten index and the twisted partition function of the D-model coincide with those of a SUSY $\CP^1$ sigma model in the limit $p\rightarrow\infty$, $q$ fixed.

 Let us now verify that~(\ref{homSUSy CP1}) is indeed the $\mathcal{N}=2b$ model. First, recall  the well-known expression 
 for the latter in $\mathcal{N}=2$ superspace~\cite{IvanovSmilga}:
\begin{gather}
    -i\,\int \mathrm{d}^2\theta\, \mathrm{d}t \left( \frac{\thickbar{D}_c\thickbar{\mathbf{Z}}D_c\mathbf{Z}}{4\upalpha^2_{12} \left(1+\thickbar{\mathbf{Z}}\mathbf{Z}\right)^2} -\frac{q}{2}\ln\left( 1+ \thickbar{\mathbf{Z}}\mathbf{Z} \right) \right), \label{N=2 CP1}
\end{gather}
where $\mathbf{Z}=z+\sqrt{2}\,\theta \psi - i \theta \smallthickbar{\theta}\, \dot{z}$ is a bosonic chiral  superfield.

In order to compare (\ref{homSUSy CP1}) to (\ref{N=2 CP1}), we should explicitly resolve the constraints $\thickbar{\mathsf{U}}_A\circ \mathsf{U}_B = \delta_{AB}$ on the $\mathsf{U}$-fields, that is
\begin{gather}
    \mathsf{U}_1 = \big(1+\thickbar{\mathsf{Z}}\mathsf{Z}\big)^{-1/2} \begin{pmatrix}
        \thickbar{\mathsf{Z}} \\
        -1     
    \end{pmatrix}, \quad \mathsf{U}_2 = \big(1+\thickbar{\mathsf{Z}}\mathsf{Z}\big)^{-1/2} \begin{pmatrix}
        1 \\
        \mathsf{Z}
    \end{pmatrix}, \label{U fields constrains resolution}
\end{gather}
where $\mathsf{Z}=z+i\theta \psi$ is a bosonic superfield. Substituting this in~(\ref{homSUSy CP1}) and further reducing to component fields, we get
\begin{gather}
    \mathcal{S}=\int \mathrm{d}t \Bigg( \frac{\Dot{\smallthickbar{z}}\Dot{z}}{\upalpha^2_{12} \left(1+\smallthickbar{z}z\right)^2} + \frac{i}{2\upalpha^2_{12}}\frac{\smallthickbar{\psi}\nabla\psi+\psi\nabla\smallthickbar{\psi}}{\left(1+\smallthickbar{z}z\right)^2} + q\left(\frac{i\,\smallthickbar{z}\dot{z}+\smallthickbar{\psi}\psi}{1+\smallthickbar{z}z} +  \frac{z \smallthickbar{z}\,\psi\smallthickbar{\psi}  }{\left(1+\smallthickbar{z}z\right)^2}\right)\Bigg),
    \nonumber\\
    \text{where} \quad \nabla\psi := \dot{\psi}-2\left(\frac{\smallthickbar{z}\dot{z}}{1+\smallthickbar{z}z}\right)\psi,\quad \nabla\smallthickbar{\psi} := \dot{\smallthickbar{\psi}}-2\left(\frac{\dot{\smallthickbar{z}}z}{1+\smallthickbar{z}z}\right)\smallthickbar{\psi}.
\end{gather}
This precisely coincides with the component reduction of~(\ref{N=2 CP1}); it is the system studied in~\cite{Akulov, Mezincescu} (as well as in~\cite{IvanovSmilga} for more general target spaces).

In fact, the discovered relation between the D-model and the sigma model on $\CP^1$ allows one to calculate the spectrum of the Dolbeault Laplacian on $\CP^1$, since it is the Hamiltonian of the $\mathcal{N}=2b$ sigma model. This directly follows from the equality of the twisted partition functions. If we choose a fairly large value for $p$ and compute the spectrum of the D-model, then we will actually calculate a part of the Dolbeault Laplacian spectrum. This idea was illustrated in Section \ref{CP1 Hamiltonian and harmonics} using the example of the purely bosonic sector of the theory.

\subsection{From the K-model to the $\mathcal{N} = 4$ sigma model.}
In the previous section we discussed the relation between the D-model  of Section~\ref{N=2 CP1section} and the $\mathcal{N}=2b$ supersymmetric $\mathds{CP}^1$ sigma model. However, we know from Section~\ref{N4CP1} that there also is a rather analogous K-model with $\mathcal{N} = 4$ SUSY. So, is there a similar relation between the K-model and the $\mathcal{N}=4$ supersymmetric $\mathds{CP}^1$ sigma model?

To answer this question, we work  with the path integral expressions for the equivariant Witten index $\Tilde{W}(g)$ and the twisted partition function $\mathscr{Z} \left(g\right)$ of the K-model. The action featuring in the path integral is
\begin{align}
    &\quad\mathcal{S}=\int \mathrm{d}t \,\mathrm{d}\theta\,\Big[-\thickbar{\mathsf{Z}}_A \circ D \mathsf{Z}_A + \thickbar{\mathsf{\Psi}}_{12} D\mathsf{\Psi}_{12} + \thickbar{\mathsf{\Phi}}_{21} D\mathsf{\Phi}_{21} + \nonumber\\
    &+ \upalpha_{12}\left(\mathsf{\Psi}_{12} \thickbar{\mathsf{Z}}_1 \circ \mathsf{Z}_2 + \thickbar{\mathsf{\Psi}}_{12} \thickbar{\mathsf{Z}}_2 \circ \mathsf{Z}_1 + \thickbar{\mathsf{\Phi}}_{21} \thickbar{\mathsf{Z}}_1 \circ \mathsf{Z}_2 + \mathsf{\Phi}_{21} \thickbar{\mathsf{Z}}_2 \circ \mathsf{Z}_1\right) + \label{N=4SUSY CP1} \\
    &+\mathsf{\Lambda}_1 \left(|\mathsf{Z}_1|^2 + |\mathsf{\Psi}_{12}|^2  - |\mathsf{\Phi}_{21}|^2 - p \right) + \mathsf{\Lambda}_{2} \left(|\mathsf{Z}_2|^2 - |\mathsf{\Psi}_{12}|^2  + |\mathsf{\Phi}_{21}|^2 - p\right)\Big], \nonumber
\end{align}
where $\mathsf{Z}_1, \mathsf{Z}_2$ and $\mathsf{\Psi}_{12}, \mathsf{\Phi}_{21}$ are bosonic and fermionic superfields, respectively. As in Section~\ref{Derive of CP^1 D}, we should impose (twisted) periodic boundary conditions on the bosonic components of all superfields, and either (twisted) periodic or antiperiodic boundary conditions on the fermionic components, depending on whether we are working with the equivariant Witten index or the twisted partition function.

To proceed from~(\ref{N=4SUSY CP1}) one can follow the same line as in Section~\ref{Derive of CP^1 D}, getting rid of the non-dynamical degrees of freedom. A small difference  is that here one cannot eliminate both $\mathsf{\Psi}_{12}$ and $\mathsf{\Phi}_{21}$ -- one of them will remain. Our choice is to keep~$\mathsf{\Phi}_{21}$. Thus, we end up with
\begin{align}
    \mathcal{S} = &\int \mathrm{d}t \,\mathrm{d}\theta\bigg[ \frac{i}{2\upalpha_{12}^2} \left( D \thickbar{\mathsf{U}}_1 \circ \mathsf{U}_2 \thickbar{\mathsf{U}}_2 \circ \dot{\mathsf{U}}_1 + \dot{\thickbar{\mathsf{U}}}_1 \circ \mathsf{U}_2 \thickbar{\mathsf{U}}_2 \circ D \mathsf{U}_1\right)
    + \thickbar{\mathsf{\Phi}}_{21} D \mathsf{\Phi}_{21}+ \mathsf{\Phi}_{21} D \thickbar{\mathsf{\Phi}}_{21} + \nonumber\\
    &+ \frac{i}{\upalpha_{12}}\left( \thickbar{\mathsf{\Phi}}_{21} \thickbar{\mathsf{U}}_1 \circ \dot{\mathsf{U}}_2 +  \mathsf{\Phi}_{21} \thickbar{\mathsf{U}}_2 \circ \dot{\mathsf{U}}_1 \right) - 2\, \thickbar{\mathsf{\Phi}}_{21} \mathsf{\Phi}_{21} \left(\,\thickbar{\mathsf{U}}_1 \circ D \mathsf{U}_1 - \thickbar{\mathsf{U}}_2 \circ D \mathsf{U}_2 \right)\bigg]\,, \label{N=4 CP1 Hom}
\end{align}
where again it is assumed that $\mathsf{U}_A$'s are subject to the constraints  ${\thickbar{\mathsf{U}}_A  \circ \mathsf{U}_B = \delta_{AB}}$. We should now compare~(\ref{N=4 CP1 Hom}) with the known action of the $\mathcal{N}=4$ SUSY $\CP^1$ sigma model. To this end, we explicitly resolve the constraints on $\mathsf{U}_A$ following (\ref{U fields constrains resolution}). Finally, one can rewrite the action in terms of component  fields\footnote{The superfield $\mathsf{\Phi}_{21}$ contains a non-dynamical boson, which has also been  eliminated.}:
\begin{align}
    \mathcal{S} = \int \mathrm{d}t \,\Bigg[\frac{\dot{\smallthickbar{z}}\dot{z}}{ 2\upalpha_{12}^2\left(1+\smallthickbar{z}z\right)^2}&+\frac{i}{2\upalpha_{12}^2 }\frac{\smallthickbar{\psi}\nabla \psi + \psi \nabla \smallthickbar{\psi}}{\left(1+\smallthickbar{z}z\right)^2}+i \left(\smallthickbar{\varphi}\, \nabla\varphi + \varphi \nabla\smallthickbar{\varphi} \right) 
    -\\
    &- \frac{ \smallthickbar{\varphi}\,\nabla \psi - \varphi \nabla\smallthickbar{\psi}}{\upalpha_{12}\left(1+\smallthickbar{z}z\right)} -\frac{4\smallthickbar{\varphi}\varphi \smallthickbar{\psi}\psi}{\left(1+\smallthickbar{z}z\right)^2}\Bigg].\nonumber
\end{align}
To bring this expression to standard form, we rescale the fermions  $\sqrt{2}\,\psi \rightarrow \psi$, $2\upalpha_{12}\left(1+\smallthickbar{z}z\right)\varphi \rightarrow \varphi$ and diagonalize the quadratic form of the fermions\footnote{During the diagonalization process, a boundary term will arise, however it vanishes since it is bosonic and therefore periodic.}. At the end we get
\begin{gather}
    \mathcal{S} = \frac{1}{2\upalpha_{12}^2}\int \mathrm{d}t \Bigg(\frac{\dot{\smallthickbar{z}}\dot{z}}{ \left(1+\smallthickbar{z}z\right)^2}+\frac{i}{2}\frac{\smallthickbar{\psi}\nabla \psi +\psi \nabla \smallthickbar{\psi} }{\left(1+\smallthickbar{z}z\right)^2}+\frac{i}{2}\frac{\smallthickbar{\varphi}\,\nabla \varphi + \varphi \nabla \smallthickbar{\varphi}}{\left(1+\smallthickbar{z}z\right)^2}-\frac{2\smallthickbar{\varphi}\varphi \smallthickbar{\psi}\psi}{\left(1+\smallthickbar{z}z\right)^4}\Bigg),
\end{gather}
which is the well-known form of  the $\mathcal{N}=4a$ SUSY $\CP^1$ model~\cite{WittenSUSYbreaking, DavisCP1, SmilgaHowTo, IvanovSmilga}. Therefore we have proven that the K-model gives rise to the $\mathcal{N}=4a$ SUSY $\CP^1$ sigma model in the limit when $p$ tends to infinity.

\newpage
\chapter{Oscillator calculus on \texorpdfstring{$\SU(n)$}{Lg} orbits}

\vspace{1cm}
We started in Section~\ref{N=2 CP1section} by describing a SUSY extension of the spin-spin Hamiltonian~(\ref{Spin12Ham}) acting in the tensor product $\displaystyle{\T^p\otimes \T^{p+q}}$ of $\mathfrak{su}_2$ representations. We then showed that in the limit $p\to\infty$ this system is equivalent to a 1D SUSY sigma model with target space $\CP^1$, whose Hamiltonian is the Laplacian suitably extended to act on differential forms. 
In the present section we wish to extend the above results to the case of more general (co)adjoint orbits\footnote{In the present paper we discuss solely the 1D models. Two-dimensional sigma models with flag manifold target spaces have been studied, for example,  in~\cite{PerelomovChiral,Donagi,BykLagEmb, Bykov_2013,Tanizaki,Ohmori,Amari1,Amari2,Amari3}.}. In the case of $\SU(n)$, all (co)adjoint orbits are of the form
\bea
\mathcal{O}_{\Uplambda}=\Bigl\{\; g \Uplambda  g^{-1}\,,\; g\in \SU(n)\; \Bigr\}\,,\label{coadj orb}
\eea
where $\Uplambda \in \mathfrak{su}_n$ is a traceless Hermitian matrix. In particular, $\mathcal{O}_\Uplambda$ is a homogeneous space
\bea
\mathcal{O}_{\Uplambda}=\frac{\SU(n)}{\mathsf{Stab}_{\Uplambda}}\,,
\eea
where the stabilizer $\mathsf{Stab}_{\Uplambda}$ 
is determined by the eigenvalues of $\Uplambda$.

Since $\Uplambda$ is Hermitian, it can be diagonalized, the corresponding diagonal matrix also belonging to the orbit~(\ref{coadj orb}). Consider the most general diagonal $\Uplambda$ allowing several coincident eigenvalues:
\setlength{\arraycolsep}{1.3pt}
\renewcommand*{\arraystretch}{0.4}
\begin{equation}\label{Lambdamatrix}
  \Uplambda= \begin{pNiceMatrix}[margin]
\Block[borders={bottom,right}]{4-4}{}
\,\uplambda_1& & & & \Block{4-1}{\scriptstyle \, n_1}\\
   &\uplambda_1 & & & \\
   &   & \Ddots & & \\
   \RowStyle[cell-space-bottom-limit=4pt]{}
   &   &      &\uplambda_1& \\
   \RowStyle[cell-space-top-limit=4pt]{}
   &   &      &   &\Block[draw]{4-4}{}
                   \,\uplambda_2 & & & & \Block{4-1}{\scriptstyle n_2} \\
   &   &      &   &   &\uplambda_2 & & &\\
   &   &      &   &   &   &\Ddots & & \\
   \RowStyle[cell-space-bottom-limit=4pt]{}
   &   &      &   &   &   &      &\uplambda_2& \\ 
   &   &      &   &   &   &      &   &{\Ddots}\\
   \RowStyle[cell-space-top-limit=4pt]{}
   &   &      &   &   &   &      &   &      &\Block[draw]{4-4}{}
\,\uplambda_k & & & & \Block{4-1}{\scriptstyle \;n_k} \\
   &   &      &   &   &   &      &   &      &   &\uplambda_k & & &\\
   &   &      &   &   &   &      &   &      &   &   &\Ddots & &\\
   \RowStyle[cell-space-bottom-limit=4pt]{}
   &   &      &   &   &   &      &   &      &   &   &      &\uplambda_k& \\
\end{pNiceMatrix} 
\end{equation}

The corresponding orbit $\mathcal{O}_{\Uplambda}=\mathcal{F}_{n_1, \ldots, n_k}$ is then the partial flag manifold
\begin{gather}\label{flagquotient}
    \mathcal{F}_{n_1, n_2, \ldots,n_k}:=\frac{\SU(n)}{\mathsf{S}\big(\mathsf{U}(n_1)\times \mathsf{U}(n_2) \times \dots\times \mathsf{U}(n_k) \big)}, \quad 
    \text{where}\quad \sum_{A=1}^k n_A = n. 
\end{gather}
It is also often useful to introduce the partial sums $d_B=\sum_{A=1}^B n_A$. In this notation $d_k=n$ and we also set $d_0\equiv 0$.

Let us note that one could as well represent the partial flag manifold as a quotient of  $\mathsf{U}(n)$ by its subgroup $\mathsf{U}(n_1) \times \dots\times \mathsf{U}(n_k)$. It is this representation that is more convenient for us in this paper, although the choice of the definition used will be made depending on the context.

Let us recall two geometric definitions of this manifold. The first one is that the flag manifold $\mathcal{F}_{n_1, n_2, \ldots,n_k}$ may be viewed as the space of ordered $k$-tuples of orthonormal planes of dimensions $n_1, \ldots, n_k$. To enumerate the vectors spanning each plane, we introduce the sets of labels:
\bea\label{IAdef}
I_A:=\Bigl\{\;m\;\textrm{integer}\;,\quad d_{A-1}< m\leq d_{A}\;\Bigr\}\,,\quad\quad A=1, \ldots, k\,.
\eea
One can then construct an orthonormal basis $(u_1, \ldots, u_{n_A})$, such that the vectors $\{u_i\}_{i\in I_A}$ span the $A$-th plane. Moreover, the basis within one plane $\{u_i\}_{i\in I_A}$ is defined up to an action of $\mathsf{U}(n_A)$.  
This is essentially a rephrasing of what the quotient space~(\ref{flagquotient}) is.

The second definition is that $\mathcal{F}_{n_1, n_2, \ldots,n_k}$ is a $k$-tuple of linear subspaces
\bea\label{flagcompldef}
L_A\subset \CC^n \quad \textrm{such that}\quad L_A\subset L_{A+1}\quad \textrm{and}\quad \mathrm{dim}\,L_A=d_A\quad (A=1, \ldots, k)\,.
\eea
In order to pass from~(\ref{flagquotient}) to the latter definition, one should choose a complex structure on the quotient space. One can prove (cf.~the review~\cite{Affleck_2022} or~\cite{BorelHirzebruch, AlekseevskyPerelomov}) that complex structures on $\mathcal{F}_{n_1, n_2, \ldots,n_k}$ are in one-to-one correspondence with total orderings on the set $I_1, \ldots, I_k$. 
Given an ordering, one can define the space $L_A$ as the linear span of vectors with labels in the first $A$ sets. For example, the definition~(\ref{flagcompldef}) corresponds to the standard (lexicographic) ordering $I_1<I_2< \cdots$.

It is often convenient to describe total orderings in terms of \emph{acyclic} tournament diagrams: these are graphs with $k$ vertices, such that every pair of vertices is connected by an arrow indicating the ordering between the nodes; besides, the graph is required to contain no cycles. For an example with $k=3$ see diagram~(\ref{tournament}). This description in terms of tournament diagrams (quivers) will be useful for constructing SUSY extensions below.
\bea\label{tournament}
\begin{overpic}[scale=0.5,unit=1mm]{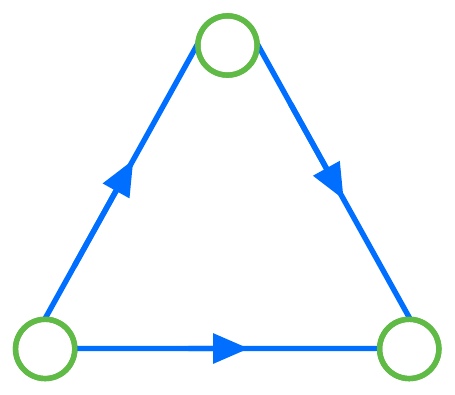}
\end{overpic}
\eea

\section{Oscillator calculus on complete flags}\label{completeflagsec}

 A `typical' orbit corresponds to the case when all eigenvalues of~(\ref{Lambdamatrix}) are distinct. It is called the manifold of complete flags and may be written as follows:
\bea\label{Fndef}
\mathcal{F}_n:=\frac{\SU(n)}{\mathsf{S}(\mathsf{U}(1)^n)} = \frac{\mathsf{U}(n)}{\mathsf{U}(1)^n}\,.
\eea
It is this orbit that we will be studying in the present section. In Section~\ref{CPnsec} we will consider the opposite case of complex projective space $\CP^{n-1}$, which corresponds to the situation when the maximal number of eigenvalues of $\Uplambda$ coincide.

In the foregoing discussion we will exclusively concentrate on the `spin chain' part of the story, leaving apart the detailed derivation of the sigma model, which could in principle be recovered along the lines of our analysis  of the $\CP^1$ model in Section~\ref{Derive of CP1}. The bosonic cores of the Hamiltonians that we will be considering are straightforward generalizations of~(\ref{Spin12Ham}):
\bea
\mathcal{H}_{\mathrm{bos}}=\sum\limits_{A<B}\,\upalpha_{AB}^2\,(S_A, S_B) + \mathrm{const} \label{Spin Ham 123}
\eea
acting in a tensor product $\displaystyle{\T^{p_1}\otimes \T^{p_2}\otimes \cdots \otimes \T^{p_n}}$. In the purely bosonic setup it was shown in~\cite{Bykov_2024} that in the limit $p\to\infty$ such Hamiltonian gives rise to a sigma model on $\mathcal{F}_n$ with the following metric:
\bea\label{F1nmetr}
\mathrm{d}s^2=\sum\limits_{A<B}\,{1\over\upalpha_{AB}^2}\,|\smallthickbar{u}_A\circ \mathrm{d} u_B|^2\,,\qquad u_A\circ u_B = \delta_{AB}\,.
\eea
The spectrum of the Laplacian for the normal metric -- when all $\upalpha_{AB}$ are equal -- was computed in~\cite{Yamaguchi}. Another  interesting special case, where the spectrum in general is not known, is when the metric is K\"ahler\footnote{In case of K\"ahler metrics there are natural sigma models with $\mathcal{N}=4$ SUSY that may be obtained by dimensional reduction of $\mathcal{N}=1$ models in 4D~\cite{ZUMINO1979203}.}. To find the corresponding constraint on the parameters, assume that the complex structure has been chosen according to the ordering $1<2<\ldots<n$. In this case the one-forms $\smallthickbar{u}_A\circ \mathrm{d} u_B$ for $A<B$ are  of type~$(1, 0)$. It is then an elementary exercise to extract the fundamental Hermitian form from the metric~(\ref{F1nmetr}) and to check that it is closed if and only if the following condition holds 
(for more details on this see the review~\cite{Affleck_2022}):
\bea\label{Kahconstr}
{1\over\upalpha_{AC}^2}={1\over\upalpha_{AB}^2}+{1\over\upalpha_{BC}^2}\quad\quad \textrm{for all}\quad\quad A<B<C\,.
\eea
In fact, these constraints may be explicitly solved by
\bea
{1\over \upalpha_{AB}^2}=r_A-r_B\,,
\eea
where $\{r_A\}_{A=1, \ldots, n}$ is a decreasing sequence of real numbers.

\subsection{D-model for $\mathcal{F}_3$.}\label{FlagDolbeaultIndex}

We start with the case $n=3$ to describe some of the salient new features as compared to the case of $\CP^1\simeq \mathcal{F}_2$. The field content of this model is summarized as follows:
\bea
z_A^\alpha\,,\quad\quad A,\, \alpha=1, 2, 3,\quad\quad \psi_{12}, \psi_{23}, \psi_{13}\,,
\eea
together with the corresponding (conjugate) creation operators. We adopt the following ansatz for the supercharge:
\bea\label{F123Q}
\mathcal{Q}=\upalpha_{12}\,\psi_{12}\big(z_1^\dagger\circ z_2\big)+\upalpha_{23}\,\psi_{23} \big(z_2^\dagger\circ z_3\big)+\upalpha_{13}\,\psi_{13} \big(z_1^\dagger\circ z_3\big)+\upbeta \,\psi_{12}\psi_{23}\psi_{13}^\dagger 
\eea
and assume that, at least generically, all $\upalpha_{AB}$ are non-vanishing. The crucial new feature is the appearance of a term cubic in the fermions. It is inevitable in order for the supercharge to be nilpotent. We find
\bear\label{betavalue}
\mathcal{Q}^2=(\upalpha_{12}\upalpha_{23}+\upbeta \upalpha_{13})\,\psi_{12}\psi_{23}\big(z_1^\dagger\circ z_3\big)=0\quad 
\Rightarrow\quad \upbeta=-\frac{\upalpha_{12}\upalpha_{23}}{\upalpha_{13}}\,.
\eear
The supercharge is invariant w.r.t.~$\mathsf{U}(1)^3$ global symmetry. 
As in the $\CP^1$ case, we wish to gauge this $\mathsf{U}(1)^3$ symmetry. To this end, we introduce the constraint operators:
\bear
&&\mathcal{C}_1=
z_1^\dagger \circ z_1+\psi_{12}^\dagger\psi_{12}+\psi_{13}^\dagger\psi_{13}-p_1 \,,\label{Constraints F123}\\
&&\mathcal{C}_2=
z_2^\dagger \circ z_2-\psi_{12}^\dagger\psi_{12}+\psi_{23}^\dagger\psi_{23}-p_2\,, \label{Constraints F123-2}\\
&&\mathcal{C}_3=
z_3^\dagger \circ z_3-\psi_{23}^\dagger\psi_{23}-\psi_{13}^\dagger\psi_{13}-p_3\,. \label{Constraints F123-3}
\eear
They commute with $\mathcal{Q}$ by construction. As before, we will require that $p_1, p_2, p_3 \in \mathbb{Z}$. The corresponding Hilbert space -- a truncation of the oscillator Fock space -- is defined as follows:
\bea
\mathscr{H}(p_1, p_2, p_3)=\Bigl\{\, \mathcal{C}_1=\mathcal{C}_2=\mathcal{C}_3=0 \,\Bigr\}\,.
\eea
The structure of the Hilbert space may be summarized by the following quiver (see  Section~\ref{QuiverSection}):
\bea\label{F123pic}
\begin{overpic}[scale=0.7,unit=1mm]{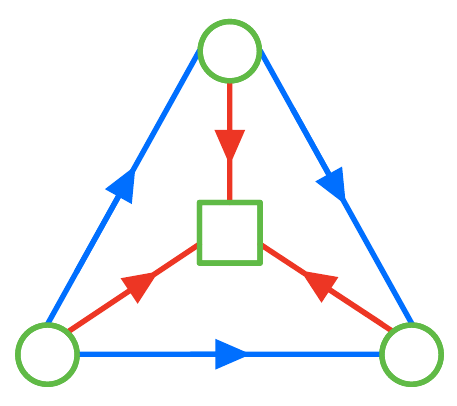}
\put(4.3,5){\footnotesize $\CC$}
\put(15,9.5){$z_1$}
\put(47.5,5){\footnotesize $\CC$}
\put(37,9.5){$z_3$}
\put(25,1){$\psi_{13}$}
\put(25.3,19.3){\footnotesize $\CC^3$}
\put(25.8,41){\footnotesize $\CC$}
\put(22,31){$z_2$}
\put(6.2,25){$\psi_{12}$}
\put(41.8,25){$\psi_{23}$}
\end{overpic}
\eea

The Hamiltonian of the model is defined simply as $\mathcal{H}=\Bigl\{\mathcal{Q}, \mathcal{Q}^\dagger\Bigr\}$ and may be computed from~(\ref{F123Q}). Some general remarks that we would like to make will not require its explicit expression. First of all, we wish to compute the $\SU(3)$-equivariant Witten index which can be written in the form 
\bea
\tilde{W}=\mathsf{Tr}_{\mathscr{H}(p_1, p_2, p_3)}\left((-1)^F \prod\limits_{\alpha=1}^3\,t_\alpha^{-\mathcal{J}_\alpha}\,e^{-\beta \mathcal{H}}\right)\,.
\eea
Here $\mathcal{J}_\alpha := \sum_{A=1}^{3} \big(z^{\dagger}_A\circ z_A\big)$ may be thought of as the Cartan elements of $\mathfrak{u}_3$, and since we are computing $\SU(3)$-characters we additionally require $\prod\limits_{i=\alpha}^3\,t_\alpha=1$. As the index is independent of~$\beta$, we may set $\beta=0$, which allows recasting  $\tilde{W}$ as a supercharacter:
\bea\label{WF123supercharacter}
\tilde{W}=\mathsf{STr}_{\mathscr{H}(p_1, p_2, p_3)}\left( \prod\limits_{\alpha=1}^3\,t_\alpha^{-\mathcal{J}_\alpha}\right)\,.
\eea
To compute this, we will use the following trick. We start by evaluating a modified  supercharacter, which we call $\mathcal{Z}$,  in the unconstrained Fock space:
\begin{align}
&\mathcal{Z}(t|s)=\mathsf{STr}\left(\prod\limits_{\alpha=1}^3\,t_\alpha^{-\mathcal{J}_\alpha}\,\prod\limits_{A=1}^3\,s_A^{\mathcal{C}_A}\right)=\nonumber\\
=&\prod\limits_{\alpha, A=1}^3\,\frac{1}{1-s_A t_{\alpha}^{-1}}\,\prod\limits_{B<C}^3\,\left(1-\frac{s_B}{s_C}\right)\,\frac{1}{s_1^{p_1}s_2^{p_2}s_3^{p_3}}\,.\label{partfuncF123}
\end{align}
Note that the minus sign in the bracket comes from the presence of the $(-1)^F$ in the definition of the index. 
We may then reduce to the constraint surface by taking residues\footnote{Which is tantamount to averaging w.r.t.~the gauge group $\mathsf{U}(1)^3$. This interpretation will be important for non-Abelian generalizations (the case of partial flags considered below).} w.r.t.~$s_1, s_2, s_3$:
\bea
\tilde{W}(t)=\frac{1}{(2\pi i)^3}\,\oint {\mathrm{d}s_1\over s_1}\,\oint {\mathrm{d}s_2\over s_2}\,\oint {\mathrm{d}s_3\over s_3}\;\mathcal{Z}(t|s)\,.
\eea
Initially, all contour integrals are assumed to be over small circles around the origin. A glance at~(\ref{partfuncF123}) reveals, though, that $\mathcal{Z}(t|s)$, as a function of the $s$-variables, has poles at $s_A=0$, as well as at $s_A= t_\alpha$ for all combinations of $A$ and $\alpha$. We may thus wish to close the contours at infinity, picking up these latter poles in the process. This seems to suggest that there will be $3^3 = 27$ contributions. However, in a configuration where two of the $s_A$'s take the same value the numerator in~(\ref{partfuncF123}) vanishes. As a result, the only non-zero contributions will come from configurations where $(s_1, s_2, s_3)=(t_1, t_2, t_3)$ and permutations thereof, resulting in only $3! = 6$ possibilities. We arrive at the following\footnote{Here and in what follows $\mathsf{S}_m$ is the permutation group of $m$ elements.} (using the fact that $t_1t_2t_3=1$):
\bea \label{Witten index for full flag}
\tilde{W}(t)=\sum\limits_{\mathrm{\sigma \in \mathsf{S}_3}}\,\frac{t_{\sigma(1)}^{-p_1}\,t_{\sigma(2)}^{-p_2}\,t_{\sigma(3)}^{-p_3}}{\prod\limits_{k<l}\,\left(1-\frac{t_{\sigma(l)}}{t_{\sigma(k)}}\right)}\,.
\eea

\subsubsection{$p\,$-independence.}\label{pindepsec}

Assume that $p_3 \geq p_2 \geq p_1$. 
In this case~(\ref{Witten index for full flag}) is simply the Weyl character formula \cite{WeylTheClassicalGroups, Zhelobenko, KOIKE1987466} (see Appendix~\ref{Weylcharacterapp}) for the representation, whose Young diagram has rows of lengths\footnote{Our definition (\ref{WF123supercharacter}) of the supercharacter is related to the usual one by a formal substitution $t\rightarrow t^{-1}$. Thus, we should interpret (\ref{Witten index for full flag}) as a character of the representation dual to that with the Young diagram $(N-p_1, N-p_2, N-p_3)$, where $N$ is a fairly large number. The same comment holds true in the general case.}  $(p_3, p_2, p_1)$. 
For general $p_A$'s, $\tilde{W}(t)$ is the character of a virtual representation. We present the red Young diagram for the representation and the blue Young diagram, which is drawn upside down, for its dual\\
\vspace{0.3cm}
\begin{equation}\label{Youngdiag}
\underbrace{
\begin{overpic}[scale=0.4,unit=1mm]{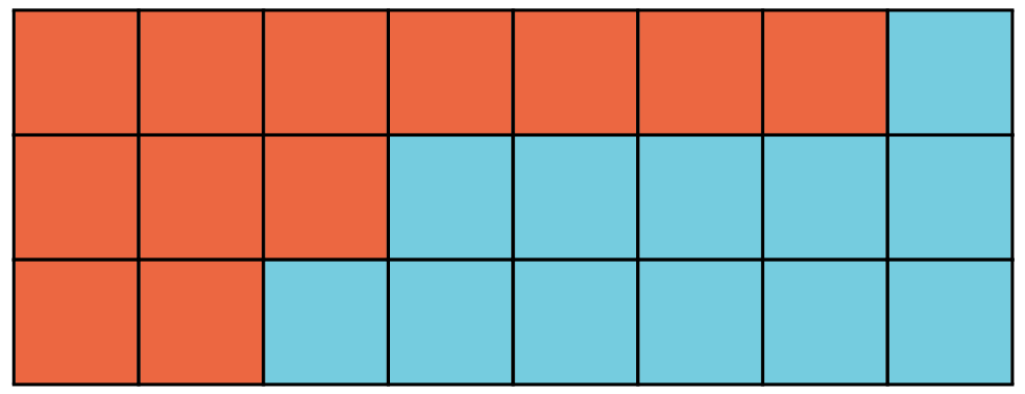}
    \put(1,19){$\xleftrightarrow{\makebox[5.7cm]{}}$}
    \put(21,22){$p_3$}
    \put(1,10.8){$\xleftrightarrow{\makebox[2.31cm]{}}$}
    \put(11.5,13.8){$p_2$}
    \put(1,2.3){$\xleftrightarrow{\makebox[1.47cm]{}}$}
    \put(4,5.3){$p_1$}
\end{overpic}}_{N}
\end{equation}

\vspace{0.3cm}
\noindent
In fact the dual representation will appear for a D-system similar to that described in Section \ref{FlagDolbeaultIndex}, with the difference being that the fermion arrows in the quiver (\ref{F123pic}) point in the opposite direction.

Looking at~(\ref{Witten index for full flag}), notice that the shift $p_A \mapsto p_A+1$ for all~$A$ produces a factor of $t_1t_2t_3=1$, so it has no effect on $\tilde{W}(t)$. This is in accord with the fact that adding a full column to a $\SU(n)$ Young diagram has no effect on the representation. For the foregoing it will be useful to parametrize $p_A=p+q_A$; then one can say that the index depends only on $q_A$ but not on $p$.

Curiously, there is a related `$p$-independence' property of the full Hamiltonian. More precisely, we will now prove that the spectrum of $\mathcal{H}_p$ is contained in the spectrum of~$\mathcal{H}_{p+1}$ (where the index now indicates the relevant representation). To this end we introduce the operator
\bea
\mathcal{O}_3:=
\epsilon_{\alpha\beta\gamma}(z_1^\dagger)^\alpha (z_2^\dagger)^\beta (z_3^\dagger)^\gamma
\eea
-- the straightforward generalization of~(\ref{O2operator}). It is easy to see that it has the following commutation properties:
\bea
\Bigl[z_A^\dagger \circ z_B, \;\mathcal{O}_3\Bigr]=\delta_{AB}\,\mathcal{O}_3\,.
\eea
As a result, it commutes with the supercharges $\Bigl[\mathcal{Q}, \mathcal{O}_3\Bigr]=\Bigl[\mathcal{Q}^\dagger, \mathcal{O}_3\Bigr]=0$ and, therefore, with the Hamiltonian. Besides, $\Bigl[\mathcal{C}_A, \mathcal{O}_3\Bigr]=\mathcal{O}_3$, so that 
\bea
|\psi\rangle\in \mathscr{H}(p_1, p_2, p_3)\quad \Rightarrow \quad |\psi'\rangle:=\mathcal{O}_3|\psi\rangle\in \mathscr{H}(p_1+1, p_2+1, p_3+1)\,.
\eea
The latter Hilbert space corresponds to the shift $p\mapsto p+1$. Moreover, if $|\psi\rangle$ is an eigenstate of the Hamiltonian, $\mathcal{H}|\psi\rangle=E|\psi\rangle$, then so is $|\psi'\rangle$: $\mathcal{H}|\psi'\rangle=E|\psi'\rangle$. This proves that $\mathrm{Spec}\,\mathcal{H}_p\subset \mathrm{Spec}\,\mathcal{H}_{p+1}$.

\subsubsection{Higher-$n$ generalizations 
of the D-model.}
\label{nN2}
One can now easily generalize to the case of complete flag manifolds $\mathcal{F}_n$ for arbitrary $n$: what remains is to write an expression for the corresponding nilpotent supercharge $\mathcal{Q}$.

Assume that we have a tournament with $n$ vertices, together with a total ordering on this set of vertices (i.e.~an acyclic tournament, as we explained earlier). Without loss of generality, we may set $1<2<\cdots<n$. Then the supercharge has the form\footnote{The form of the supercharge strongly resembles that of the BRST charge. Perhaps a relation could be found along the lines of~\cite{Frenkel}.}
\bea\label{QN2gen}
\mathcal{Q}=\sum\limits_{  A<B}\,\upalpha_{AB}\,\psi_{AB}\big(z_A^\dagger \circ z_B\big)-\sum\limits_{  A<B<C}\,\frac{\upalpha_{AB}\upalpha_{BC}}{\upalpha_{AC}}\,\psi_{AB}\,\psi_{BC}\,\psi^\dagger_{AC}\,.
\eea
Let us prove that it is nilpotent. First of all, when squaring the term linear in $\psi$'s, one  gets  the sum $\sum_{  A<B<C}\,\upalpha_{AB}\upalpha_{BC}\,\psi_{AB}\psi_{BC}\big(z_A^\dagger \circ z_C\big)$, but this is easily seen to cancel against the cross-terms arising from the anti-commutators of $\psi$ with $\psi^\dagger$, as in the $n=3$ example before. However, at first sight one will have additional terms arising from squaring the second piece of the supercharge~(\ref{QN2gen}). These may appear when one of the fermions $\psi$ enters two triangles in opposite directions. The only two such possibilities are depicted in the following diagram:
\bea
\begin{overpic}[scale=0.5,unit=0.7mm]{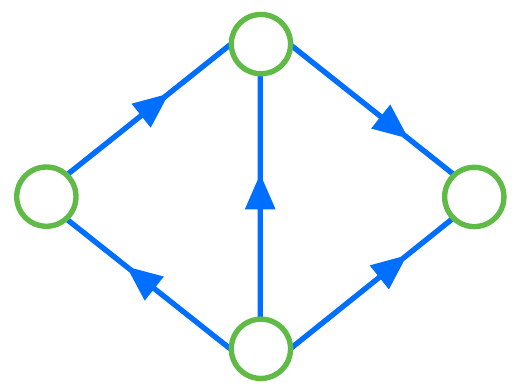}
\put(34,23){$\psi$}
\end{overpic} \hspace{1.5cm}
\begin{overpic}[scale=0.5,unit=0.7mm]{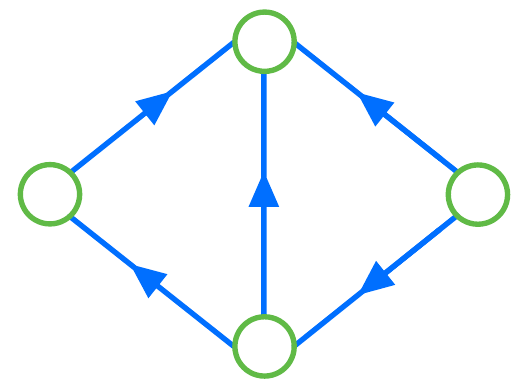}
\put(34,23){$\psi^\prime$}
\end{overpic}
\eea
However, these two diagrams always come in pairs. To show this, consider the left diagram. Since any two vertices must be connected by an arrow, let us draw the arrow between the left-most and right-most vertices. Only the direction from left to right is allowed, as the opposite one would lead to an upper triangle being cyclic. Thus, we arrive at the following situation pictured by a simplex:
\bea\label{4vertfig}
\begin{overpic}[ scale=0.5,unit=0.66mm]{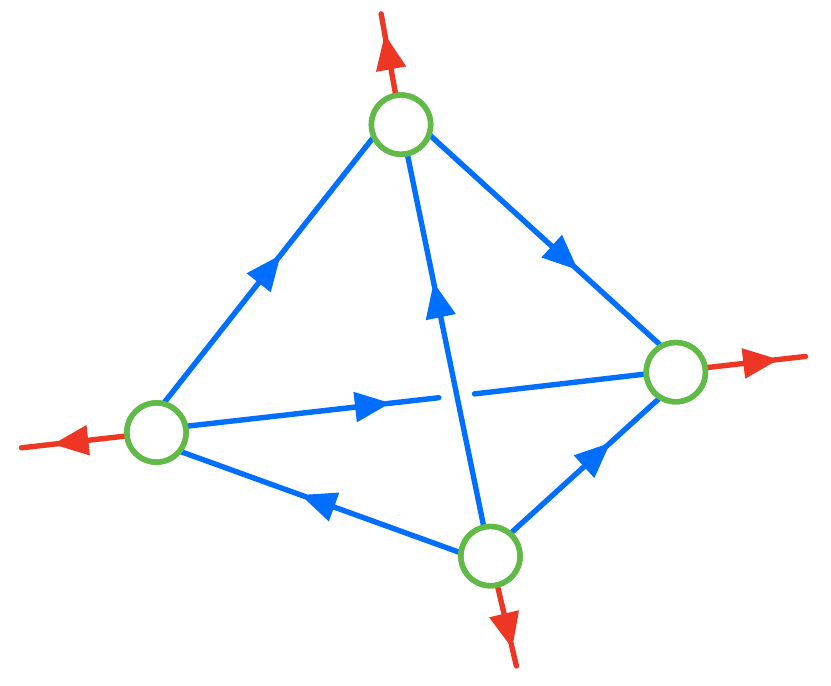}
\put(57.3,50){$\psi$}
\put(44,29.5){$\psi^\prime$}
\put(18.5,30.8){\footnotesize $2$}
\put(85.3,38.3){\footnotesize $4$}
\put(61.5,15){\footnotesize $1$}
\put(50,70.3){\footnotesize $3$}
\end{overpic}
\eea
Here we have numbered the vertices as $1, 2, 3 ,4$, although these could be any numbers from $1$ to $n$ with the same ordering (determined by the arrows). 
It is now clear that interchanging $\psi$ with $\psi'$ is precisely like interchanging the two diagrams above. Therefore it suffices to consider the diagram~(\ref{4vertfig}). Its contribution to the cubic part of the supercharge  has the form 
\bea
\mathcal{Q}_{(3)}=\mathcal{Q}_{123}+\mathcal{Q}_{134}+\mathcal{Q}_{234}+\mathcal{Q}_{124}\,,\quad \textrm{where}\quad \mathcal{Q}_{ABC}=\frac{\upalpha_{AB}\upalpha_{BC}}{\upalpha_{AC}}\,\psi_{AB}\psi_{BC}\psi^\dagger_{AC}\,.
\eea
A direct calculation shows that $\mathcal{Q}_{(3)}^2=\Bigl\{\mathcal{Q}_{123}, \mathcal{Q}_{134}\Bigr\}+\Bigl\{\mathcal{Q}_{234}, \mathcal{Q}_{124}\Bigr\}=0$ proving that the supercharge, as defined in~(\ref{QN2gen}), is nilpotent.

Just like in the $n=3$ case above, here we could compute the equivariant Witten index $\tilde{W}$, estabilishing its independence of the cutoff $p$. Although one can guess the answer by extrapolating~(\ref{Witten index for full flag}), we postpone the calculation to Section~\ref{partialflagsec}, where it will be done in greater generality. However, in the ample case $p_n\geq p_{n-1}\geq \cdots \geq p_1$ one can easily construct the zero-energy states in the bosonic sector ($F=0$) explicitly, generalizing formula~(\ref{Psi0}) of $n=2$. First of all,  the straightforward generalization of the constraints~(\ref{Constraints F123})-(\ref{Constraints F123-3}) reduces in this sector to
\bea
\left(z_A^\dagger \circ z_A-p_A\right)|\Phi\rangle_{F=0}=0\,,\quad\quad A=1,\,\cdots, n\,.
\eea
This means that $|\Phi\rangle_{F=0}$ contains exactly $p_A$ oscillators of type $(z_A^\alpha)^\dagger$.  
Besides, any such state
is annihilated by $\psi_{AB}$, so that  $\mathcal{Q}|\Phi\rangle_{F=0}=0$. Similarly, $\mathcal{Q}^\dagger|\Phi\rangle_{F=0}=0$
leads to the set of constraints
\bea
\big(z_B^\dagger \circ z_A\big)|\Phi\rangle_{F=0}=0\,,\quad\quad \textrm{for}\quad A< B\,,
\eea
which imply that the oscillators $(z_A^\alpha)^\dagger$ enter in \emph{skew-symmetric} combinations with all $(z_B^\beta)^\dagger$ for $B>A$ (for more on this see~\cite{Bykov_2013, Affleck_2022}). Recall that, due to the ample property, the number of $z_A$-oscillators is no less than the number of $z_B$-oscillators for $B>A$, so that the latter requirement makes sense. One concludes that the zero-modes $|\Phi\rangle_{F=0}$ furnish an irreducible representation, whose Young diagram has rows of lengths $(p_n,\, \cdots, p_1)$. Although we will not do this here, one should be able to prove that  there are no other zero-energy states for $F>0$, so that the equivariant Witten index $\tilde{W}(t)$ is the character of this representation, in agreement with the results~(\ref{Witten index for full flag})-(\ref{Youngdiag}) for the case $n=3$.  

Finally, we notice that the spectra of the Hamiltonians have the property that $\mathrm{Spec}\,\mathcal{H}_p\subset \mathrm{Spec}\,\mathcal{H}_{p+1}$.  This can be proven by introducing the operator
\bea
\mathcal{O}_n:=
\epsilon_{\alpha_1\cdots \alpha_n}(z_1^{\alpha_1})^\dagger \cdots (z_n^{\alpha_n})^\dagger
\eea
and following the same steps as in Section~\ref{pindepsec}. In fact, this argument applies to all of the models considered in the present paper (including the purely bosonic Hamiltonian~(\ref{Spin Ham 123})).

\subsection{K-model for $\mathcal{F}_3$.}\label{KmodelF123}

Here we wish to extend the above constructions to $\mathcal{N}=4$ SUSY and describe the K\"ahler-de Rham complex for the flag manifold, in a way similar to the treatment of the $\CP^1$ model in Section~\ref{N4CP1}.

Just as in the $n=2$ case, we double the number of fermions, and propose the following quiver diagram:
\bea\label{F3Kdiag}
\begin{overpic}[scale=0.8,unit=0.7mm]{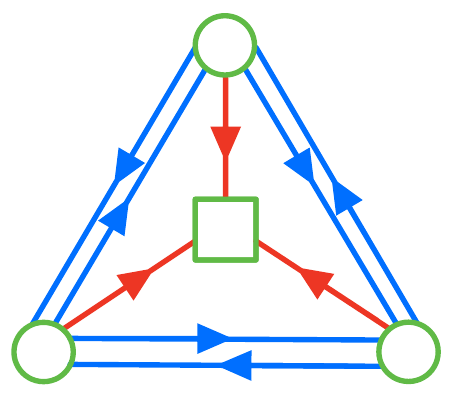}
\put(6.5,8){\footnotesize $\CC$}
\put(28,20){$z_1$}
\put(77,8){\footnotesize $\CC$}
\put(52.7,21){$z_3$}
\put(35,48.5){$z_2$}
\put(41.5,67){\footnotesize $\CC$}
\put(25,34){$\psi_{12}$}
\put(53.5,36){$\psi_{23}$}
\put(41,31){\footnotesize $\CC^3$}
\put(39,16){$\psi_{13}$}
\put(12.5,42.5){$\phi_{21}$}
\put(68,41){$\phi_{32}$}
\put(39,0.8){$\phi_{31}$}
\end{overpic}
\eea
We make the following ansatz for the two complex supercharges:
\bear\label{F123Q1Q2}
&&\mathcal{Q}_1=\upalpha_{12}\,\psi_{12} \big( z_1^\dagger\circ z_2 \big)+\upalpha_{23}\,\psi_{23} \big( z_2^\dagger\circ z_3 \big) +\upalpha_{13}\,\psi_{13} \big( z_1^\dagger\circ z_3 \big) +\\ \nonumber &&\hspace{1cm}+\upbeta \,\psi_{12}\psi_{23}\psi_{13}^\dagger +\upgamma_1\,\psi_{12}\phi_{32}^{\dagger}\phi_{31}+\updelta_1 \,\phi_{21}^{\dagger}\psi_{23}\phi_{31}\,,\\ \label{F123Q1Q2sec}
&&\mathcal{Q}_2=\upalpha_{12}\,\phi_{21}^\dagger \big( z_1^\dagger\circ z_2 \big) +\upalpha_{23}\,\phi_{32}^\dagger \big( z_2^\dagger\circ z_3 \big) +\upalpha_{13}\,\phi_{31}^\dagger \big( z_1^\dagger\circ z_3 \big)-\\ \nonumber &&\hspace{1cm}-\upbeta \,\phi_{32}^\dagger\phi_{21}^\dagger\phi_{31} +\upgamma_2\,\phi_{21}^\dagger\psi_{23}\psi_{13}^\dagger+\updelta_2 \,\psi_{12}\phi_{32}^\dagger\psi_{13}^\dagger\,.
\eear
As we explain below, the extra terms are necessary for the algebra of extended supersymmetry to hold. The supercharges~(\ref{F123Q1Q2})-(\ref{F123Q1Q2sec}) still admit the $\mathsf{U}(1)^3$ symmetry. The corresponding charges may be read off from the diagram~(\ref{F3Kdiag}) and are expressed as follows:
\bear
&&\tilde{\mathcal{C}}_1=
z_1^\dagger \circ z_1+\psi_{12}^\dagger\psi_{12}+\psi_{13}^\dagger\psi_{13}-\phi_{21}^\dagger\phi_{21}-\phi_{31}^\dagger\phi_{31}-p_1\,,\\
&&\tilde{\mathcal{C}}_2=
z_2^\dagger \circ z_2-\psi_{12}^\dagger\psi_{12}+\psi_{23}^\dagger\psi_{23}+\phi_{21}^\dagger\phi_{21}-\phi_{32}^\dagger\phi_{32}-p_2\,,\\
&&\tilde{\mathcal{C}}_3=
z_3^\dagger \circ z_3-\psi_{23}^\dagger\psi_{23}-\psi_{13}^\dagger\psi_{13}+\phi_{32}^\dagger\phi_{32}+\phi_{31}^\dagger\phi_{31}-p_3\,.
\eear

Given the ansatz~(\ref{F123Q1Q2})-(\ref{F123Q1Q2sec}), one should check that the $\mathcal{N}=4$ superalgebra holds. This involves the following steps:
\begin{itemize}
    \item $\mathcal{Q}_1^2=\mathcal{Q}_2^2=0$. This requires $\upbeta=-\dfrac{\upalpha_{12}\upalpha_{23}}{\upalpha_{13}}$, just as before in~(\ref{betavalue}).
    \item $\Bigl\{\mathcal{Q}_1, \mathcal{Q}^\dagger_2\Bigr\}=0$. In the case of $\CP^1$ (see Section~\ref{N4CP1}) this constraint only led to the condition that the charges $p_A$ be equal. Here instead we arrive at a set of conditions that are solved as follows:
\bear
&&p_A=p_B := p\quad \textrm{for all}\quad A, B\,,
\\
&&\upgamma_1=\upgamma_2=-\dfrac{\upalpha_{12}\upalpha_{13}}{\upalpha_{23}}\equiv \upgamma\,,\quad\quad \updelta_1=\updelta_2=-\dfrac{\upalpha_{23}\upalpha_{13}}{\upalpha_{12}}\equiv \updelta\,.
\eear
\item $\Bigl\{\mathcal{Q}_1, \mathcal{Q}_2\Bigr\}=0$. It turns out that this condition constrains the coefficients $\upalpha_{AB}$:
\bea\label{Kahcond}
{1\over \upalpha_{13}^2}={1\over \upalpha_{12}^2}+{1\over \upalpha_{23}^2}\,.
\eea
This is nothing but the condition~(\ref{Kahconstr}) that the resulting metric on $\mathcal{F}_3$ is K\"ahler.
\item $\Bigl\{\mathcal{Q}_1, \mathcal{Q}^\dagger_1\Bigr\}=\Bigl\{\mathcal{Q}_2, \mathcal{Q}^\dagger_2\Bigr\}\equiv\mathcal{H}$. This does not lead to any further constraints.
\end{itemize}

The upshot is that, in the $\mathcal{N}=4$ case, the SUSY algebra determines the coefficients $\upbeta, \upgamma_i, \updelta_i$ ($i=1,2$) in the supercharges~(\ref{F123Q1Q2})-(\ref{F123Q1Q2sec}) and, moreover, leads to the K\"ahler constraint~(\ref{Kahcond}) on the parameters of the metric. As a result, from a differential-geometric standpoint, the supercharges determine the differentials in the K\"ahler-de Rham complex. 

Our next goal is the calculation of the Witten index of the K-model. Computing the partition function without constraints, in this case we get
\bear\label{partfuncF123KahDR}
\mathcal{Z}(t|s)=
\prod\limits_{\alpha, A=1}^3\,\frac{1}{1- s_A t^{-1}_{\alpha}}\,\prod\limits_{B\neq C}^3\,\left(1-\frac{s_B}{s_C}\right)\,\frac{1}{(s_1s_2s_3)^{p}}\,.
\eear
Note the difference with~(\ref{partfuncF123}) due to the doubling of fermions in this model. Taking  residues w.r.t.~$s_1, s_2, s_3$, here we find that the numerator cancels against the denominator at each pole 
$(s_1, s_2, s_3)=\big({t_{\sigma(1)}}, {t_{\sigma(2)}}, {t_{\sigma(3)}}\big)$,
so that we are left with
\bea
\tilde{W}(t)=\frac{1}{(2\pi i)^3}\,\oint {\mathrm{d}s_1\over s_1}\,\oint {\mathrm{d}s_2\over s_2}\,\oint {\mathrm{d}s_3\over s_3}\;\mathcal{Z}(t|s)=\sum\limits_{\sigma\in \mathsf{S}_3}\,1=3!=6=\mathrm{Eu}\left(\mathcal{F}_3\right)\,.
\eea
The result is independent of $t$ and is equal to the Euler characteristic of the flag manifold. As is well-known, each SUSY vacuum corresponds to an element of $H^\ast (\mathcal{F}_3)$, and here each such element may be represented by a  $\mathsf{SU}(3)$-invariant form.

For the purpose of higher-$n$ generalizations it is useful to rewrite the supercharges in a manifestly $\mathsf{U}(2)$-symmetric form. To this end we introduce the $\mathsf{U}(2)$-doublets
\bea
\Psi_{AB}=\left(\begin{array}{c}
    \psi_{AB}   \\
    ~\\
      \phi_{BA}^\dagger
\end{array}\right)\,,\quad\quad A<B\,.
\eea
In this notation the doublet of supercharges takes the following form (taking into account that $\upbeta=\upgamma+\updelta$):
\bear\label{QSU2inv}
&&\mathcal{Q}=\upalpha_{12}\,\Psi_{12} \big( z_1^\dagger\circ z_2 \big) +\upalpha_{23}\,\Psi_{23} \big( z_2^\dagger\circ z_3 \big)+\upalpha_{13}\,\Psi_{13} \big( z_1^\dagger\circ z_3 \big)+\\ \nonumber
&&\quad\quad +\frac{\upalpha_{12}\upalpha_{13}}{\upalpha_{23}} \,\Psi_{12}\big(\Psi_{13}^\dagger\Psi_{23}\big)-\frac{\upalpha_{23}\upalpha_{13}}{\upalpha_{12}}\, \Psi_{23}\big(\Psi_{13}^\dagger\Psi_{12}\big)\,.
\eear

\subsubsection{Higher-$n$ generalizations ($\mathcal{N}=4$).} 
Again, the above theory is easily generalizable to higher values of $n$. Without going into details, let us  write out the expression for the supercharge in the general case -- this  is a straightforward extension of~(\ref{QSU2inv}):
\begin{align}
\mathcal{Q}=\sum\limits_{  A<B<C}\,\bigg(\frac{\upalpha_{AB}\upalpha_{AC}}{\upalpha_{BC}}\,&\Psi_{AB}\left(\Psi_{AC}^\dagger\Psi_{BC}\right)-\frac{\upalpha_{BC}\upalpha_{AC}}{\upalpha_{AB}}\,\Psi_{BC}\left(\Psi_{AC}^\dagger\Psi_{AB}\right)\bigg)+\nonumber\\
&+\sum\limits_{  A<B}\,\upalpha_{AB}\,\Psi_{AB}\big( z_A^\dagger \circ z_B\big)
\end{align}
with the K\"ahler condition~(\ref{Kahconstr}):
\bea\label{Kahproperty}
{1\over \upalpha_{AC}^2}={1\over \upalpha_{AB}^2}+{1\over \upalpha_{BC}^2}\quad\quad \textrm{for}\quad\quad A<B<C\,.
\eea

The proof that\footnote{Here we denote the two components of the doublet of supercharges $\mathcal{Q}$ by $\mathcal{Q}_\uprho$ for $\uprho=1,2$.} $\Bigl\{\mathcal{Q}_\uprho, \mathcal{Q}_\upgamma\Bigr\}=0$ is similar to the one of Section~\ref{nN2}. To start with, we split the supercharge in a term linear in the fermions plus a term cubic in the fermions: ${\mathcal{Q}=\mathcal{Q}^{(1)}+\mathcal{Q}^{(3)}}$. One easily calculates
\bea
\Bigl\{\mathcal{Q}^{(1)}_\uprho, \mathcal{Q}^{(1)}_\upgamma\Bigr\}=\sum\limits_{A<B<C}\,\upalpha_{AB}\upalpha_{BC} \,\big(z_A^\dagger\circ z_C\big)\Big[(\Psi_{AB})_\uprho (\Psi_{BC})_\upgamma+(\Psi_{AB})_\upgamma (\Psi_{BC})_\uprho\Big]\,.
\eea
Direct calculation shows (with the use of the K\"ahler property~(\ref{Kahproperty})) that this cancels exactly against an analogous commutator $\Bigl\{\mathcal{Q}^{(1)}_\uprho, \mathcal{Q}^{(3)}_\upgamma\Bigr\}+\Bigl\{\mathcal{Q}^{(1)}_\upgamma, 
\mathcal{Q}^{(3)}_\uprho\Bigr\}$. Thus, it remains to check that $\Bigl\{\mathcal{Q}^{(3)}_\uprho, \mathcal{Q}^{(3)}_\upgamma\Bigr\}=0$. Possible unwanted terms come from the same diagram~(\ref{4vertfig}), as in the $\mathcal{N}=2$ case earlier. Here one has
\bear
 \nonumber\Bigl\{\mathcal{Q}^{(3)}_\uprho, \mathcal{Q}^{(3)}_\upgamma\Bigr\}&&=\Bigl\{(\mathcal{Q}_{123})_\uprho, (\mathcal{Q}_{134})_\upgamma\Bigr\}+\Bigl\{(\mathcal{Q}_{123})_\upgamma, (\mathcal{Q}_{134})_\uprho\Bigr\}+\\&&+\;\Bigl\{(\mathcal{Q}_{234})_\uprho, (\mathcal{Q}_{124})_\upgamma\Bigr\}+\Bigl\{(\mathcal{Q}_{234})_\upgamma, (\mathcal{Q}_{124})_\uprho\Bigr\}=0\,,
\eear
where again one needs to use the K\"ahler condition. 

\section{Oscillator calculus on \texorpdfstring{$\CP^{n-1}$}{Lg}}\label{CPnsec}

In this section, building upon the previously studied case of $\CP^1$, we will describe the oscillator variables and the respective Hilbert spaces corresponding to truncations of the $\CP^{n-1}$ model. 

\subsection{D-model.}\label{D model CPn} We start with the definition of the desired spin chain type system. 
The field content and the structure of the Hilbert space can be read off from the following quiver diagram:
\bea\label{CPnquivdiag}
\begin{overpic}[scale=0.8,unit=0.8mm]{Flag12new2.pdf}
\put(5.5,7.3){\footnotesize $\CC$}
\put(38.5,36){\footnotesize $\CC^{n}$}
\put(15.5,25){$z_1$}
\put(70.4,6.7){\footnotesize $\CC^{n\!-\!1}$}
\put(62,25){$z_2$}
\put(37.3,13){$\psi_{12}$}
\end{overpic}
\eea

\vspace{-0.3cm}\noindent
Following Section \ref{QuiverSection}, we introduce the matrices
\begin{align}
    \psi_{12} \in \mathrm{Hom}(\CC,\CC^{n-1}),\quad z_1 \in \mathrm{Hom}(\CC,\CC^n),\quad z_2 \in \mathrm{Hom}(\CC^{n-1},\CC^n). \label{CPn field content}
\end{align}
Recall that every element of these matrices is either a fermionic  ($\psi_{12}$) or bosonic ($z_1, z_2$) annihilation operator.

In order to define the model, we introduce the manifestly nilpotent supercharge\footnote{Starting from this section, we drop the $\circ$ notation, since most fields will no longer be vectors in~$\CC^n$ but rather matrices of various shapes.} 
\begin{gather}\label{QCPn}
    \mathcal{Q} := z_1^\dagger\, z_2\, \psi_{12}\,.
\end{gather}
It leads to the Hamiltonian $\mathcal{H} = \Bigl\{\mathcal{Q}, \mathcal{Q}^\dagger \Bigr\}$ of the model. 
The model classically possesses invariance under the following $\mathsf{U}(1)\times \mathsf{U}(n-1)$ global transformations, which is a generalization of the Abelian $\mathsf{U}(1)\times \mathsf{U}(1)$ symmetry  (\ref{CP1symmetry}) in the case of $\CP^1$:
\begin{gather}
    z_1 \mapsto  z_1\,e^{i\varphi}\,, \quad z_2 \mapsto z_2\, g\,,\quad \psi_{12} \mapsto g^{-1}\, \psi_{12}\,e^{i\varphi}\,,\qquad g\in \mathsf{U}(n-1)\,,\quad\varphi\in\mathbb{R}\,.
\end{gather}
We wish to gauge this symmetry. Thus, we impose the `scalar' constraint $\mathcal{C}_1=0$ and matrix  constraint\footnote{In what follows  $\mathcal{C}_{2}^{ab}\,(a,b=2,\dots,n)$ denotes the matrix elements of $\mathcal{C}_{2}$.} $\mathcal{C}_{2}=0$, where
\bear\label{CPnconstr1}
  &&  \mathcal{C}_1 := z_1^{\dagger}\, z_1 + \psi^{\dagger}_{12} \psi_{12} - p_1, \\ \label{CPnconstr2}
  &&  \mathcal{C}_{2} := z_2^{\dagger}\, z_2 + \N\psi_{12} \psi_{12}^{\dagger}\N - p_2\,\mathds{1}_{n-1}, 
\eear
and $\N\dots\N$ stands for normal ordering. It is defined by the standard convention 
\begin{align}
    \N z z^{\dagger}\N = z^{\dagger} z, \quad \N\psi \psi^{\dagger}\N = - \psi^{\dagger} \psi
\end{align}
for $z^{\dagger}$ a bosonic creation operator and $\psi^{\dagger}$ a fermionic one. One easily checks that the constraints (\ref{CPnconstr1}) and~(\ref{CPnconstr2}) furnish the Lie algebra of $\mathfrak{u}_1\oplus \mathfrak{u}_{n-1}$. 

To compute the Witten index, we will follow the method of Section~\ref{FlagDolbeaultIndex}. To start with, we compute the twisted partition function of the free  system without constraints, i.e.~the super-character of the Fock space generated by $z_1, z_2, \psi_{12}$. The supercharge~(\ref{QCPn}), when acting in the full Fock space, has $\SU(n)\times \mathsf{U}(1)\times \mathsf{U}(n-1)$ as its symmetry group, so the partition function will be twisted by elements of this group:
\bea\label{partfuncZF3}
\mathcal{Z}(g_L|h_R)=\mathsf{STr}_{\textrm{Fock space}}\left(g_L\,h_R\right)\,,\quad\quad g_L\in \SU(n)\,,\;h_R\in \mathsf{U}(1)\times \mathsf{U}(n-1)\,.
\eea
The subscripts $L$ and $R$ refer to left and right action in  the following sense. If we package all $z_A$'s in a single matrix $Z:=(z_1, z_2)$, then the action of the two groups is described by $Z\mapsto g_L \,Z\,h_R^{-1}$. In particular, the two actions commute.

To compute the Witten index, we need to take the trace over the subsector of states, invariant w.r.t.~the gauge group $\mathsf{U}(1)\times \mathsf{U}(n-1)$. This is equivalent to averaging w.r.t.~this group:
\bea\label{Windexintegral}
\tilde{W}(g_L)=\int\,\mathrm{d}h_R\,\mathcal{Z}(g_L|h_R)\,.
\eea
To simplify the integral, we use the fact that the partition function~(\ref{partfuncZF3}) is invariant w.r.t.~the adjoint action, i.e.~$\mathcal{Z}(g_0g_Lg_0^{-1}|h_0h_Rh_0^{-1})=\mathcal{Z}(g_L|h_R)$. Therefore it depends only on the eigenvalues of $g_L$ and $h_R$. In fact, we already used this fact implicitly in~(\ref{WF123supercharacter}), writing the Witten index in terms of exponents of the Cartan generators of~$\mathfrak{su}_3$. In the integral~(\ref{Windexintegral}) we may as well pass to integration over the eigenvalues of~$h_R$ by using the Weyl integration formula\footnote{In the case of $G=\mathsf{U}(n)$ the Weyl integration formula reads (cf.~\cite{Zhelobenko}):
\bea
\int\,\mathrm{d}g\,f(g)={1\over n!}\int\,\prod\limits_{l=1}^n\frac{\mathrm{d}z_l}{2\pi i\,z_l}\,\prod\limits_{j\neq k}\,\left(1-\frac{z_j}{z_k}\right)\,f(z_1, \ldots, z_n)\,,
\eea
where integration is over unit circles around the origin. Alternatively, one can make the change of variables $z_l=e^{i\varphi_l}$ and integrate over $\varphi_l\in [0, 2\pi)$. The more algebraic-looking expression above is useful when the function $f(g)$ admits an analytic continuation in the $z_l$-variables, just as in the cases we are considering here. 
}. As a result, we get:
\bear \label{WittenIndexIntegral}&&
\tilde{W}(t_1,\,\ldots\,,t_n)={1\over (n-1)!}\oint\,\frac{\mathrm{d}s_1}{2\pi i s_1}\,\oint\,\prod\limits_{d=2}^{n}\frac{\mathrm{d}s_{(2,d)}}{2\pi i s_{(2,d)}}\,\prod\limits_{e\neq f}\,\left(1-\frac{s_{(2,e)}}{s_{(2,f)}}\right)\times\\ && \hspace{5cm} \nonumber \times\; \mathsf{Str}_{\textrm{Fock space}}\left(\prod\limits_{\alpha=1}^{n} t_\alpha^{-\mathcal{J}_\alpha}\,s_1^{\mathcal{C}_1} \prod\limits_{a=2}^{n} \left(s_{(2,a)}\right)^{\mathcal{C}_{2}^{aa}}\right)\,.
\eear
Notice that the operator under the supertrace includes only the diagonal constraints out of the full set~(\ref{CPnconstr1})-(\ref{CPnconstr2}). The Cartan generators $\mathcal{J}_\alpha$ of the global symmetry group~$\mathsf{U}(n)$ are defined as follows:
\bea\label{Jglobcharge}
\mathcal{J}_\alpha=\N\left(Z\,Z^{\dagger}\right)^{\alpha \alpha}\N\,,\quad\quad \alpha=1\,, \ldots, n\,.
\eea
Since the symmetry group is really\footnote{More precisely, it is $\mathsf{PSU}(n)$ that acts faithfully.} $\SU(n)$, we will impose the additional constraint $t_1 \cdots t_n=1$. Further details of the calculation of the integral~(\ref{WittenIndexIntegral}) are presented in Appendix~\ref{Indexaapp}, the final answer being
\bea\label{CompleteFlagWittenIndexD}
\tilde{W}(t_1,\,\ldots\,,t_n)=\sum\limits_{\alpha=1}^n\,\frac{t_\alpha^{q}}{\prod\limits_{\beta\neq \alpha}^n\left(1- {t_\beta\over t_\alpha}\right)}\,,\quad\quad \textrm{where}\quad\quad q=p_2-p_1\,.
\eea
This is the expression for the character of the representation dual to $\T^q$. In other words, it is the representation with a rectangular Young diagram of height $n-1$ and width $q$. Just as in all of our earlier examples, the answer is independent of $p$.

\subsection{K-model.}

In this subsection we perform the  $\mathcal{N}=4$ extension of the above construction, using the $\CP^1$ example as our main reference.

Essentially  we should  double the number of fermions. Therefore, in addition to the fields~(\ref{CPn field content}) we introduce one more matrix $\phi_{21} \in \mathrm{Hom}(\CC^{n-1}, \CC)$ of fermionic annihilation operators. The quiver~(\ref{CPnquivdiag}) is now replaced by
\bea
\begin{overpic}[scale=0.75,unit=0.75mm]{Flag12doubledNew2.pdf}
\put(6,6){\footnotesize $\CC$}
\put(38.5,35.5){\footnotesize $\CC^n$}
\put(15,25){$z_1$}
\put(70.4,6){\footnotesize $\CC^{n\!-\!1}$}
\put(63.5,25){$z_2$}
\put(37.3,15){$\psi_{12}$}
\put(37.3,0){$\phi_{21}$}
\end{overpic}
\eea
The two manifestly nilpotent supercharges of the system are
\bear
  &&  \mathcal{Q}_1 = z_1^{\dagger}\, z_2 \, \psi_{12}\,, \qquad \mathcal{Q}_2 = \phi_{21}\, z^{\dagger}_2\, z_1\,,
\eear
whereas the constraints defining the Hilbert space take the form
\bear
  &&\tilde{\mathcal{C}}_1 = z_1^{\dagger}\, z_1 + \psi^{\dagger}_{12} \psi_{12} + \N\phi_{21}\phi^{\dagger}_{21}\N - p_1 = 0\,, \\
  &&\tilde{\mathcal{C}}_{2} = z_2^{\dagger}\,z_2 + \N\psi_{12}\psi_{12}^{\dagger}\N + \phi_{21}^{\dagger}\phi_{21} - p_2 \mathds{1}_{n-1}= 0\,. \nonumber
\eear
One can check that the $\mathcal{N}=4$ supersymmetry algebra
\begin{gather}
\Bigl\{\mathcal{Q}_\upmu, \mathcal{Q}_\upnu\Bigr\}=0\,,\quad\quad \Bigl\{\mathcal{Q}_\upmu, \mathcal{Q}^\dagger_\upnu\Bigr\}=\delta_{\upmu\upnu} \mathcal{H}
\end{gather}
requires $p_1=p_2$, by analogy with Sections \ref{N4CP1} and \ref{KmodelF123}. Similarly to the previous examples we calculate the equivariant Witten index of the model:
\begin{gather}
    \Tilde{W}(t) = n = \mathrm{Eu}(\CP^{n-1})\,. 
\end{gather}

\section{Oscillator calculus on partial flags}\label{partialflagsec}

In this section we generalize to the case of partial flag manifolds and calculate the respective Witten indices. The definition  of partial flag manifolds as well as their conventional  parametrizations  were reviewed at the beginning of this Chapter.

\subsection{D-model.}
The D-model corresponding to the partial flag manifold $\mathcal{F}_{n_1, n_2, \ldots,n_k}$ is encrypted by a quiver of the following type (here we present the example of~$\mathcal{F}_{n_1,n_2,n_3,n_4}$) \bea
\begin{overpic}[ scale=0.6,unit=0.6mm]{flag1234-tetraNewNew.pdf}
\put(22,41){\footnotesize $\CC^{n_{2}}$}
\put(109.8,51){\footnotesize $\CC^{n_{4}}$}
\put(78.5,20){\footnotesize $\CC^{n_1}$}
\put(63.3,93){\footnotesize $\CC^{n_3}$}
\end{overpic}\label{DFlag}
\eea
where with red color we have indicated   the bosonic lines flowing to the unique global vertex corresponding to the vector space $\CC^{n}$, where $n=n_1+\dots+n_k$. The full quiver is a simplex with $k+1$ vertices, $k$ of them  local (with $\CC^{n_A}$ vector spaces sitting in them) and one global,  and acyclic faces. 
Thus, the field content consists of the matrices of bosonic and fermionic annihilation operators
\begin{align}\label{fieldContentPartialFlags}
    z_A \in \mathrm{Hom}(\CC^{n_A}, \CC^n)\,,\quad \psi_{AB} \in \mathrm{Hom}(\CC^{n_A}, \CC^{n_B})\,,\quad A,B =1,2,\dots,k\,;\, A<B\,.
\end{align}

We propose the following expressions for the supercharge and constraints: 
\begin{align}
    &\mathcal{Q} = \sum\limits_{A<B}\,\upalpha_{AB} \mathsf{Tr}\left(z_A^{\dagger}z_B \,\psi_{AB}\right) + \sum\limits_{A<B<C}\frac{\upalpha_{AB}\upalpha_{BC}}{\upalpha_{AC}}\mathsf{Tr}\left(\psi^{\dagger}_{AC}\, \psi_{BC}\, \psi_{AB}\right),\\
    &\mathcal{C}_A=z_A^{\dagger}z_A + \sum\limits_{B\,<\,A}\N\psi_{BA}\psi_{BA}^{\dagger}\N+ \sum\limits_{A\,<\,C}\psi_{AC}^{\dagger}\psi_{AC} - p_A \mathds{1}_{n_A} = 0\,,\label{constraintspartialFlags}
\end{align}
where $A=1,2,\dots,k$. The matrix elements of the constraints, denoted by $\mathcal{C}_{A}^{ab}$, are indexed by lowercase Latin letters $a,b$, where $a, b \in I_A$ (see~(\ref{IAdef}) for the definition of~$I_A$).

In our calculation of the equivariant Witten index we will be closely following  the method of Section~\ref{D model CPn}. Specifically, we express it as the average of the full Fock space partition function $\mathcal{Z}(g_L|h_R):=\mathsf{Str}_{\textrm{Fock space}}\left(g_L\,h_R\right)$ w.r.t.~the gauge group:
\begin{align}
    \tilde{W}(g_L)=\int\,\mathrm{d}h_R\,\mathcal{Z}(g_L|h_R)&\,,
\end{align}
where $g_L\in \SU(n)\,, h_R\in \mathsf{U}(n_1)\times \dots \times \mathsf{U}(n_k)$. Following the same arguments as in Section~\ref{D model CPn}, we arrive at 
\begin{align}
    \tilde{W}(t_1,\dots,t_n)=\prod_{A=1}^k \frac{1}{n_A!}\prod_{d\in I_A}\left(\oint\frac{\mathrm{d} s_{(A,d)}}{2\pi i s_{(A,d)}} \right) \prod_{\substack{e,f \, \in I_A \\ e\neq f}}\left(1-\frac{s_{(A,e)}}{s_{(A,f)}}\right) \times
    \\
    \times \mathsf{Str}_{\textrm{Fock space}}\left(\prod_{\alpha=1}^n t_\alpha^{-\mathcal{J}_\alpha} \times \prod_{A=1}^k\prod_{a\in I_A} \left(s_{(A,a)}\right)^{C_{A}^{aa}}\right), 
\end{align}
where $\mathcal{C}_{A}^{aa}$ are the diagonal constraints that may be read off from~(\ref{constraintspartialFlags}), and $\mathcal{J}_\alpha$ are the Cartan generators of the global $\mathsf{U}(n)$ symmetry that have the same form as in\footnote{The matrix $Z$ now has the form $Z = (z_1,\, z_2,\dots,\, z_k)$.}~(\ref{Jglobcharge}). 
Just as before, due to the fact that the diagonal $\mathsf{U}(1)\subset \mathsf{U}(n)$ is really part of the gauge group, it will be convenient to impose an additional constrain $t_1\cdots t_n=1$, thus explicitly reducing the global symmetry group to $\SU(n)$.

An explicit calculation of the Fock space partition function gives
\begin{align}
    \mathsf{Str}_{\textrm{Fock space}}\left(\prod_{\alpha=1}^n t_\alpha^{-\mathcal{J}_\alpha} \times \prod_{A=1}^k\prod_{a \in I_A} \left(s_{(A,a)}\right)^{\mathcal{C}_{A}^{aa}}\right) = \prod_{C=1}^k \prod_{c\,\in I_C}\prod_{\alpha=1}^n \frac{1}{1- s_{(C,c)} t^{-1}_\alpha } \times
    \\
    \times\prod_{A=1}^k \prod_{a\,\in I_A}\left(\frac{1}{s_{(A,a)}^{p_A}} \prod_{B<A}\prod_{b\,\in I_B}\left(1-\frac{s_{(B,b)}}{s_{(A,a)}}\right)\right).\nonumber
\end{align}
Now, following the Appendix \ref{Indexaapp}, we can easily see that the equivariant Witten index is given by
\begin{align}
    \Tilde{W}(t) = \prod_{A=1}^k \frac{1}{n_A !} \times \sum_{\sigma \in \mathsf{S}_n} \prod_{A=1}^k \prod_{a\in I_A} \frac{ t^{-p_A}_{\sigma (a)}}{\prod_{B < A} \prod_{b\,\in I_B} \left(1-\frac{t_{\sigma(a)}}{t_{\sigma(b)}}\right)} \,.\label{Witten index for partial flags}
\end{align}
As one can notice, this formula reproduces the results obtained earlier for the cases of complete flag manifolds and projective spaces. The formula~(\ref{Witten index for partial flags}) expresses  the equivariant index of the Dolbeault operator on $\mathcal{F}_{n_1,n_2,\dots,n_k}$ twisted by a line bundle, whose   first Chern class is given by the twists (in a natural basis of ${H^2(\mathcal{F}_{n_1,n_2,\dots,n_k}, \mathbb{Z})\simeq \mathbb{Z}^{k-1}}$, see Section~\ref{DDsec} for more details)
\begin{align}
q_A = p_A-p_{A-1}\,, \quad A=2,\dots,k. \label{twists}
\end{align}

\subsection{K-model.}
Our final step in this section is to describe the K-model for partial flag manifolds. As always, the quiver that defines the model is obtained by doubling the fermionic edges of (\ref{DFlag}), i.e.
\bea
\begin{overpic}[ scale=0.6,unit=0.6mm]{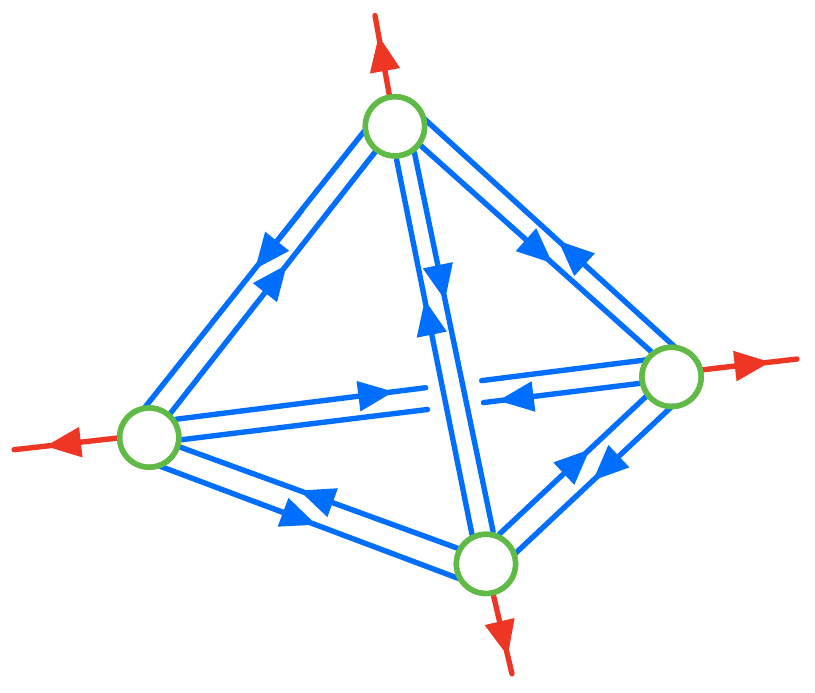}
\put(20.7,41){\footnotesize $\CC^{n_2}$}
\put(109,51.3){\footnotesize $\CC^{n_4}$}
\put(77.8,20){\footnotesize $\CC^{n_1}$}
\put(62.3,94){\footnotesize $\CC^{n_3}$}
\end{overpic}
\eea
It means that, in addition to the field content of the D-model (\ref{fieldContentPartialFlags}), we should introduce the following matrices of fermionic annihilation operators:
\begin{align}
    \phi_{BA} \in \mathrm{Hom}\left(\CC^{n_B}, \CC^{n_A}\right), \quad A,B = 1,2,\dots,k;\,\, A < B.
\end{align}
As in the case of complete flags, we combine fermions into doublets
\begin{align}
    \Psi_{AB} = \begin{pmatrix}
        \psi_{AB} \\
        ~\\
        \phi_{BA}^{\dagger}
    \end{pmatrix},\quad \text{where}\,\, A,B = 1,\dots,k,\,\,A<B. 
\end{align}
Naturally, the doublets have an inner $\CC^2$-index. We denote contraction w.r.t.~this index by  $\wick{\c\Psi_{AC}^\dagger\c\Psi_{BC}} := \sum_{\uprho =1}^2\big(\Psi_{AC}^\dagger\big)^\uprho \,\Psi_{BC}^{\uprho}$.

\noindent
In this notation the doublet of supercharges reads:
\begin{align}
\mathcal{Q}=\sum\limits_{  A<B<C}\,\bigg(\frac{\upalpha_{AB}\upalpha_{AC}}{\upalpha_{BC}}\,&\mathsf{Tr}\left[\wick{\c\Psi_{AC}^\dagger\c\Psi_{BC}\Psi_{AB}}\right]+\frac{\upalpha_{BC}\upalpha_{AC}}{\upalpha_{AB}}\,\mathsf{Tr}\left[\wick{\c\Psi_{AC}^\dagger\Psi_{BC}\c\Psi_{AB}}\right]\bigg)+\nonumber\\
&+\sum\limits_{  A<B}\,\upalpha_{AB}\,\mathsf{Tr}\left[\big(z_A^\dagger \circ z_B\big) \Psi_{AB}\right]
\end{align}
The constraints have the form 
\begin{align}
\mathcal{C}_A&=z_A^{\dagger}z_A + \sum\limits_{B\,<\,A}\N \wick{\c\Psi_{BA}\c\Psi_{BA}^{\dagger}}\N +\sum\limits_{A\,<\,C}\N\wick{\c\Psi_{AC}^{\dagger}\c\Psi_{AC}}\N - p_A \mathds{1}_{n_A} = 0\,,
\end{align}
where $A=1,2,\dots,k$. In order to have the $\mathcal{N}=4$ SUSY algebra we must require that all $p_A$'s be equal and impose the K\"ahler conditions~(\ref{Kahproperty}) on the parameters of the metric.

The calculation of the equivariant Witten index is similar to that of the D-model case. The difference is that, due to the fact that all 
 $p_A$'s are now equal and all fermions are doubled, the denominators cancel against the numerators at each pole. As a result, we end up with
\begin{gather}
    \Tilde{W}(t) = \prod_{A=1}^k \frac{1}{n_A !} \times \sum_{\sigma \in \mathsf{S}_n} 1 = \frac{n!}{\prod\limits_{A=1}^k n_A !} = \mathrm{Eu}\left( \mathcal{F}_{n_1,\dots,n_k}\right)\,.
\end{gather}

\section{Relation to the Dirac complex}
\label{DDsec}

As we discussed, the D-model allows one to calculate the index of the  Dolbeault operator. In the case of K\"ahler manifolds, there is a well-known isomorphism between the twisted Dolbeault and Dirac complexes~\cite{HitchinDD} (more details can be found in~\cite{Michelsohn} or in the books~\cite{SpinGeometry, Berline}; in the supersymmetric context this is discussed in~\cite{IvanovSmilga, SmilgaDiffGeom}). Our results therefore also allow addressing the spectral problem for the Dirac operator and, in particular, finding its index. To start with, let us recall several ways of looking at the isomorphism.

First of all, the supercharge $\mathcal{Q}_{\mathcal{N}=1}$ in a $\mathcal{N}=1$ sigma model coincides with the Dirac operator  acting on Dirac spinors~(cf.~\cite{Alvarez}). The left- and right-handed Weyl components of the Dirac spinor are distinguished by the fermion parity operator ${\gamma_5=(-1)^F}$, which anticommutes with $\mathcal{Q}_{\mathcal{N}=1}$, as it should.
When the target space is K\"ahler, $\mathcal{N}=1$ supersymmetry of the sigma model is automatically upgraded to  $\mathcal{N}=2b$. Indeed, the $\mathcal{N}=2b$ sigma model may be formulated in $\mathcal{N}=1$ language as follows~\cite{Hull}:
\bea\label{genN1lagr}
\mathscr{L}=\int \mathrm{d}\theta \left(i\,g_{\upmu\upnu}D\mathsf{X}^{\upmu} \dot{\mathsf{X}}^{\upnu}+c_{\upmu\upnu\uprho}D\mathsf{X}^{\upmu} D\mathsf{X}^{\upnu} D\mathsf{X}^{\uprho}\right)\,.
\eea
where the metric $g$ is assumed Hermitian w.r.t.~a complex structure $J$, and the 3-form $c$ is constructed from the fundamental Hermitian form $\omega=gJ$ via $c\sim \mathrm{d}\omega$. On a K\"ahler manifold the form $\omega$ is closed, so that $c=0$ and the second term in~(\ref{genN1lagr}) vanishes. As a result, one is left with the minimal $\mathcal{N}=1$ system. 
The (complex) supercharge $\mathcal{Q}$ of the $\mathcal{N}=2b$ model satisfying $\mathcal{Q}^2=0$ is the Dolbeault operator and is related to the $\mathcal{N}=1$ supercharge by $\mathcal{Q}_{\mathcal{N}=1}=\mathcal{Q}+\mathcal{Q}^\dagger$.

\subsection{Clifford algebra in $\mathbb{R}^{2N}$.}

The second way of looking at the isomorphism is by analyzing the relevant Clifford algebra. Let us start from the simplest example. Consider the Clifford algebra corresponding to Euclidean  $\mathbb{R}^{2N}$: ${\{\upgamma^a, \upgamma^b\}=2\delta^{ab}}$, where $a, b= 1, \ldots, 2N$. We may pass to the complex basis 
\bea
{\BbbGamma}^m={1\over 2}\left(\upgamma^{m}+i \,\upgamma^{N+m}\right)\quad \textrm{and} \quad \BbbGamma^{\smallthickbar{m}}={1\over 2}\left(\upgamma^{m}-i\, \upgamma^{N+m}\right)\,, \quad m=1, \ldots, N\,.
\eea
The only non-vanishing anti-commutator is 
$\{\BbbGamma^m, \BbbGamma^{\smallthickbar{m}'}\}=\delta^{m\smallthickbar{m}'}$, which means that $\BbbGamma^m, \BbbGamma^{\smallthickbar{m}}$ may be thought of as fermionic creation/annihilation operators. The standard Dirac spinor representation may then be identified with the corresponding Fock space of dimension $2^N$. In other words, define the vacuum state $|0\rangle$ by the conditions $\BbbGamma^m  |0\rangle=0$. Then an arbitrary state is obtained by applying `raising' operators~$\BbbGamma^{\smallthickbar{m}}$:
\bea
\BbbGamma^{\smallthickbar{m}_1}\cdots \BbbGamma^{\smallthickbar{m}_\ell} |0\rangle\,.
\eea
Next, we introduce the complex coordinates $z^1, \ldots, z^N$ on~$\mathbb{R}^{2N}$, together with their conjugates, and consider a  form of type $(0, \ell)$,  
\bea\label{vform}
v:=v_{\smallthickbar{m}_1\cdots \smallthickbar{m}_{\ell}}(z, \smallthickbar{z}\,)\; \mathrm{d}\smallthickbar{z}^{\,\smallthickbar{m}_1}\wedge \cdots \wedge \mathrm{d}\smallthickbar{z}^{\,\smallthickbar{m}_\ell}\,.
\eea
One can map any such form to the Dirac spinor as follows (recall that $\BbbGamma^{\smallthickbar{m}}$'s anti-commute):
\bea
v\quad \mapsto \quad v_{\smallthickbar{m}_1\cdots \smallthickbar{m}_{\ell}}(z, \smallthickbar{z}\,) \,\BbbGamma^{\smallthickbar{m}_1}\cdots \BbbGamma^{\smallthickbar{m}_\ell}|0\rangle\,.
\eea
At each point $(z, \smallthickbar{z})$ this map identifies the space of all such forms with the Dirac representation; if all $z$'s are taken into account at once, this is an isomorphism  of the bundle of all forms of various ranks with the spinor bundle that we call $S$:
\bea\label{spinbundleforms}
S\simeq \bigoplus\limits_{\ell=0}^N \,\Omega^{(0, \ell)}\,. 
\eea
The corresponding Dirac operator
\bea
\upgamma^a \dd_a=\BbbGamma^{\smallthickbar{m}} \dd_{\smallthickbar{m}}+\BbbGamma^m  \dd_m=\smallthickbar{\dd}+\smallthickbar{\dd}^\ast
\eea
is then expressed via the Dolbeault operator $\smallthickbar{\dd}$ and its Hermitian conjugate\footnote{Hermitian conjugation is w.r.t.~the $L^2$-norm. Concretely, for an $\ell$-form $\theta=\theta_{\smallthickbar{\alpha}_1 \cdots \smallthickbar{\alpha}_\ell} \mathrm{d}z^{\smallthickbar{\alpha}_1}\wedge \cdots \wedge \mathrm{d}z^{\smallthickbar{\alpha}_\ell}$ one has $\smallthickbar{\dd}^\ast \theta\sim \dfrac{\dd}{\dd z^\beta}\theta_{\smallthickbar{\beta} \smallthickbar{\alpha}_2 \cdots \smallthickbar{\alpha}_\ell} \mathrm{d}z^{\smallthickbar{\alpha}_2}\wedge \cdots \wedge \mathrm{d}z^{\smallthickbar{\alpha}_\ell}$, so that $\smallthickbar{\dd}^\ast$ is the holomorphic analogue of the divergence operator.} $\smallthickbar{\dd}^\ast$  acting on such forms\footnote{In  an arbitrary number of dimensions $M$ one may identify the Clifford algebra, as a vector space, with the space of all forms~\cite{AtiyahVector}. Besides, it is a module over itself of dimension $2^M>2^{[{M\over 2}]}$ and splits into several Dirac spinors. The corresponding `Dirac' equation then maps to the Ivanenko-Landau-K\"ahler equation (cf.~\cite{BennTucker, Obukhov}).}.

\subsection{Dirac equation on a K\"ahler manifold.}

This construction may be generalized, in a more subtle way, to the case when $\mathbb{R}^{2N}$ is replaced by a K\"ahler manifold. First, we introduce the vielbein for the K\"ahler metric:
\bea
g_{\alpha\smallthickbar{\beta}}=\sum\limits_{m=1}^N\,E_{\alpha}^m E_{\,\smallthickbar{\beta}}^{\,\smallthickbar{m}}\,.
\eea
The gamma matrices may then be written as
$
\upgamma^\alpha=E^{\alpha}_m \BbbGamma^m\,,\; \upgamma^{\smallthickbar{\alpha}}=E^{\,\smallthickbar{\alpha}}_{\,\smallthickbar{m}} \BbbGamma^{\smallthickbar{m}}\,,
$ 
where $\BbbGamma$'s are the flat space gamma matrices defined above. Given a form of type $(0, \ell)$, we may again construct the Dirac spinor
\bea\label{vformDirac}
v_{\smallthickbar{\alpha}_1\cdots \smallthickbar{\alpha}_{\ell}}(z, \smallthickbar{z}\,) E^{\smallthickbar{\alpha}_1}_{\smallthickbar{m}_1}\,\cdots E^{\smallthickbar{\alpha}_\ell}_{\smallthickbar{m}_\ell}\,\BbbGamma^{\smallthickbar{m}_1}\cdots \BbbGamma^{\smallthickbar{m}_\ell}|0\rangle\,.
\eea
For simplicity consider the anti-holomorphic part of the Dirac operator, this time involving the spin connection:
\begin{align}
&D_{\thickbar{\mathrm{hol}}}:=\gamma^{\smallthickbar{\alpha}} \left(\dd_{\,\smallthickbar{\alpha}}+{1\over 2} (\omega_{\,\smallthickbar{\alpha}})_{m\smallthickbar{n}} \,[\BbbGamma^m, \BbbGamma^{\smallthickbar{n}}]\right)=\nonumber\\ \label{spinconn}
&=\gamma^{\smallthickbar{\alpha}} \left(\dd_{\,\smallthickbar{\alpha}}+{1\over 2} (\omega_{\,\smallthickbar{\alpha}})_{m\smallthickbar{m}}- (\omega_{\,\smallthickbar{\alpha}})_{m\smallthickbar{n}} \,\BbbGamma^{\smallthickbar{n}}\BbbGamma^m  \right).
\end{align}
Here we have used that, on a K\"ahler manifold, the spin connection is valued in $\mathfrak{u}_N$, and therefore is represented by a Hermitian matrix. Acting on~(\ref{vformDirac}), we get 
\begin{align}\label{Dholaction}
&\left[\left(\smallthickbar{\dd}+{1\over 2}\,\smallthickbar{\omega}^{(0)}\right)\wedge v\right]_{\smallthickbar{\alpha}_0\smallthickbar{\alpha}_1\cdots \smallthickbar{\alpha}_\ell} \upgamma^{\smallthickbar{\alpha}_0}\cdots \upgamma^{\smallthickbar{\alpha}_{\ell}}|0\rangle\;+\\ \nonumber &\quad\quad\quad +\ell\,v_{\smallthickbar{\alpha}_1\cdots \smallthickbar{\alpha}_{\ell}}(z, \smallthickbar{z}\,) E^{\smallthickbar{\alpha}}_{\smallthickbar{m}_0}\nabla_{\smallthickbar{\alpha}} \,E^{\smallthickbar{\alpha}_1}_{\smallthickbar{m}_1}\,\cdots E^{\smallthickbar{\alpha}_\ell}_{\smallthickbar{m}_\ell}\,\BbbGamma^{\smallthickbar{m}_0}\,\BbbGamma^{\smallthickbar{m}_1}\cdots \BbbGamma^{\smallthickbar{m}_\ell}|0\rangle\,,
\end{align}
where  $\smallthickbar{\omega}^{(0)}$ is the $\mathfrak{u}_1$-part of the connection in~(\ref{spinconn}). The covariant derivative 
$\nabla$, at face value, involves only the spin connection. However, one can replace it with the full covariant derivative, since the term with the Christoffel symbols is proportional to $E^{\smallthickbar{\alpha}}_{\smallthickbar{m}_0} \Gamma_{\smallthickbar{\alpha}\smallthickbar{\beta}}^{\smallthickbar{\alpha}_1} E^{\smallthickbar{\beta}}_{\smallthickbar{m}_1}$ and therefore vanishes when contracted with the skew-symmetric combination $\BbbGamma^{\smallthickbar{m}_0}\BbbGamma^{\smallthickbar{m}_1}$. Thus, $\nabla$ is really the covariant derivative w.r.t~both the spin and affine connection, hence $\nabla_{\smallthickbar{\alpha}} E^{\smallthickbar{\alpha}_1}_{\smallthickbar{m}_1}=0$ by definition of the spin connection.

It follows that only the first line in~(\ref{Dholaction}) remains, so that the action of $D_{\thickbar{\mathrm{hol}}}$ on the spinor~(\ref{vformDirac}) is isomorphic to the action of the twisted Dolbeault operator on the form~$v$ of type $(0, \ell)$. The holomorphic part of the Dirac operator may be dealt with in a similar manner.

Finally, to analyze the twist, we need an explicit expression for the spin connection on a K\"ahler manifold~\cite{IvanovSmilga}: $(\omega_{\smallthickbar{\alpha}})_{m\smallthickbar{n}}=\dd_{\smallthickbar{\alpha}} E^{\,n}_{\,\beta} \,E_{\,m}^{\,\beta}  $, so that $\smallthickbar{\omega}^{(0)}=\smallthickbar{\dd} \log{\det E}$. The twisted Dolbeault operator in~(\ref{Dholaction}) may thus be written as
\bea\label{twistedDolb}
\left(\smallthickbar{\dd}+{1\over 2}\,\smallthickbar{\omega}^{(0)}\right)\wedge v={1\over (\det E)^{1/2}}\,\smallthickbar{\dd}\left((\det E)^{1/2}\,v\right)
\eea
Since $E^m$ ($m=1, \ldots, N$) are one-forms of type $(1, 0)$, their wedge product ${E^1\wedge \cdots \wedge E^N= (\det E)\,dz^1\wedge \cdots \wedge dz^N}$ is a section of the canonical bundle $\mathcal{K}$. Accordingly, $(\det E)^{1/2}$ should be thought of as a section of the line bundle $\mathcal{K}^{1/2}$, which exists whenever the manifold is spin. 
Otherwise one should use the spin$^c$ structure instead:  this is tantamount to adding an extra gauge field defined by line bundle~$\mathcal{L}$, chosen so that the square root $(\mathcal{K}\otimes \mathcal{L})^{1/2}$ exists.

It  follows that the $\smallthickbar{\dd}$-operator in~(\ref{twistedDolb}) acts on a section of $\mathcal{K}^{1/2}\otimes \Omega^{(0, \ell)}$. As a result,  on a K\"ahler manifold the identification~(\ref{spinbundleforms}) is replaced with $S\simeq \mathcal{K}^{1/2}\otimes \bigoplus\limits_{m=0}^N \,\Omega^{(0,m)}$ (see~\cite[Section~3.4]{Friedrich}).

\subsection{Shifts of the monopole charges.}

So far we have discussed the Dirac operator in the absence of magnetic fields. 
 Throughout the paper we have allowed topologically nontrivial magnetic fields characterized by the monopole charges~$q_A$~(\ref{twists}), so the goal of this section is to explain how these may be taken into account in the Dirac operator.
 
 The Dirac and Dolbeault operators, when coupled to monopole magnetic fields, act on sections of certain line bundles $\mathcal{L}_q$ over the target space $\mathcal{M}\equiv \mathcal{F}_{n_1,n_2,\dots,n_k}$. From this perspective, the magnetic charges are the expansion coefficients of the first Chern class $c_1(\mathcal{L}_q)\in H^2(\mathcal{M}, \mathbb{Z})$ of the line bundle w.r.t.~a basis in $H^2(\mathcal{M}, \mathbb{Z})$. Tensoring with the square root of the canonical class $\mathcal{K}^{1/2}$, as in~(\ref{twistedDolb}), means we pass to the `shifted' line bundle\footnote{Here we use the fact that for the flag manifolds studied in this paper $H^2(\mathcal{M}, \mathbb{Z})=\mathbb{Z}^{k-1}$ is a free abelian group.} 
 \bea\label{qshiftbundle}
 \mathcal{L}_{q'}\otimes \mathcal{K}^{1/2}=\mathcal{L}_{q}\,,
 \eea
 so that passing from the Dolbeault operator to the Dirac operator effectively amounts to a shift  $q\mapsto q'$ of the magnetic charges.
 
 To calculate the shift it thus suffices to know the expansion of the three line bundles in~(\ref{qshiftbundle}) in the same basis of $H^2(\mathcal{M}, \mathbb{Z})$. 
 To this end recall\footnote{We refer to \cite{AchmedZade2019RicciFlatMO} or \cite{AlekseevskyPerelomov} for details.} 
 that there is a natural forgetful projection (here $d_A=\sum_{B=1}^A n_B$)
 \bea
\pi_A:\,\mathcal{F}_{n_1,n_2,\dots,n_k} \mapsto \mathsf{Gr}(d_A,n) := \mathcal{F}_{d_A, n - d_A}
\eea
of the flag manifold to the Grassmannian. Denote the pullback of the tautological bundle over the Grassmannian $\mathsf{Gr}(d_A,n)$ as $U_A$. Using this notation, the magnetic line bundles $\mathcal{L}_q$ and the 
canonical bundle $\mathcal{K}_{\mathcal{M}}$ of $\mathcal{F}_{n_1,n_2,\dots,n_k}$ are expanded as follows\footnote{The sign in the first equality follows from the fact that we've defined the line bundles $\mathcal{L}_q$ to be ample for $q_A>0$. For example, in the D-model the zero-energy states with fermion number zero are  the holomorphic sections of the corresponding line bundle. Comparing with the reference case of  $\mathcal{M}=\CP^1$, one sees from~(\ref{Psi0}) that these may be identified with sections of $\mathcal{O}(q)$.}: 
 \begin{gather}
 c_1 \left(\mathcal{L}_q\right)=-\sum_{A=2}^{k} q_A\, c_1 (U_{A-1})\,,
 \\ \label{flagcanonical}
     c_1 \left(\mathcal{K}_{\mathcal{M}}\right) = \sum_{A=2}^{k} (n_{A} + n_{A-1})\, c_1 (U_{A-1})\,.
 \end{gather}
 Thus,~(\ref{qshiftbundle}) implies the following relation between $q$ and $q'$:
 \bea\label{qshift1}
 q_A'=q_A+{1\over 2}\left(n_A+n_{A-1}\right)\,,\quad A=2, \dots, k
 \eea

Curiously, the relevant shift may be easily implemented by a change of ordering of the oscillators. It follows from Chapter 1 that the only ordering ambiguity that we encounter  is in the constraints of the D-model. In order to arrive at the interpretation of the supercharge as the Dolbeault operator we have used the normal ordering prescription. It turns out that to obtain the Dirac operator instead one should use the Weyl (symmetric) ordering prescription for all  oscillators. For the bosonic and fermionic variables it is defined as follows:  
\bea
\mathsf{W}\left[z^{\dagger} \circ z\right]=  {1\over 2}\left(z^{\dagger} \circ z + z \circ z^{\dagger}\right)\,,\quad\quad \mathsf{W}\left[\psi^{\dagger} \psi\right]=  {1\over 2}\left(\psi^{\dagger} \psi - \psi \psi^{\dagger}\right)\,.
\eea
To explain why Weyl ordering naturally appears here, recall that normal ordering in the D-model was directly related to the choice of complex structure $J$ on the target space. Indeed,  the creation/annihilation operators of the fermions were distinguished by the flow of arrows in the corresponding quivers~(\ref{12quivdiag}), (\ref{F123pic}). The choice of Weyl ordering, instead, makes the construction symmetric (in the absence of magnetic charges, $q_A=0$) w.r.t.~the swap $J \to -J$ to the opposite complex structure. At the same time, when constructing the mapping between forms and Dirac spinors in~(\ref{vformDirac}) we could have considered the space of forms of type $(\bullet, 0)$ instead of $(0, \bullet)$. 
Thus, it is natural to expect that the construction applicable to the Dirac operator should be symmetric under the swap $J\to -J$.

So, what happens if we change the ordering prescription for the constraints from normal to Weyl? It is clear that the only change will occur in the diagonal constraints $\mathcal{C}_{A}^{aa}$  (see~(\ref{constraintspartialFlags})), which define all the twists $q_A=p_A-p_{A-1}$ ($A=2, 3, \dots ,k$). One may summarize the change by a shift in the $q_A$'s:
\bea\label{qchargeshift}
    q_A^{\textrm{Weyl}} = q_A^{\textrm{normal}} +\frac{1}{2}\left(n_A + n_{A-1}\right)\,, 
    \quad A=2, \dots, k\,. 
\eea
Identifying $q^{\textrm{normal}}\equiv q$ and $q^{\textrm{Weyl}}\equiv q'$, we find that this exactly coincides with the shift~(\ref{qshift1}).

One consequence of~(\ref{qchargeshift}) has to do with the existence of spin structure on the flag manifold. In order for the theory to be well-defined, it is necessary to require that $q_A^\prime$'s be integer. Thus, if some of the $\frac{1}{2}\left(n_A+n_{A-1}\right)$'s are not integer, the corresponding $q_A$'s cannot be zero. This is related to the fact that in those cases the manifold is not spin, so that one cannot consistently define the Dirac operator in the absence of a twist.

As an example, consider the case\footnote{The index and zero modes of the Dirac operator on $\CP^{n-1}$ were studied in~\cite{Kirchberg, IvanovSmilga}. The full spectrum in the absence of gauge fields was obtained in~\cite{Semmelmann} and generalized to non-vanishing mononopole charges in~\cite{DolanDirac, BykSmilga}.} of $\CP^{n-1}$. Here we have a single charge $q$ and the shift is~$n/2$. This is consistent with the well-known fact that $\CP^{n-1}$ is a spin manifold if and only if $n$ is even~\cite{Friedrich}.

\newpage
\conclbig{Conclusion and outlook}

\vspace{1cm}
In the present paper we  considered peculiar models of SUSY  quantum mechanics. These are systems of spin chain type so that, in particular, they have finite-dimensional Hilbert spaces. Rather strikingly, their Hamiltonians turn out to be equivalent to Laplace operators on $\SU(n)$ coadjoint orbits, truncated to a finite number of harmonics. The truncation is controlled by an integer $p\geq 0$, which one can think of as the spin, i.e. the representation at a site of the chain.  The spectra of the spin chain  Hamiltonians $\mathcal{H}_p$ have the nested  property that $\mathrm{Spec}\,\mathcal{H}_p\subset \mathrm{Spec}\,\mathcal{H}_{p+1}$. Thus, a subset of eigenvalues of the Laplacian may be inferred by diagonalizing the spin chains for fixed values of $p$. In the `large-spin' limit $p\to \infty$ the full Hilbert space  $L_2(\textrm{orbit})$ is recovered, together with the Laplacian acting on it.

We considered two classes of models: the ones with $\mathcal{N}=2$ SUSY (which we called D-models) and the ones with $\mathcal{N}=4$ SUSY (K-models). The Laplace operators are respectively the Laplace-Dolbeault and Laplace-de Rham operators acting on differential forms. As an application, we have calculated the Witten indices of the models showing that they are independent of the truncation $p$  and that they reproduce the Witten indices of the (twisted) Dolbeault and de Rham operators on $\SU(n)$ coadjoint orbits. 
As is well-known, the twisted Dolbeault complex on a K\"ahler manifold is equivalent to the Dirac complex. Thus, the same calculation also gives the index of the twisted Dirac operator.

It turned out that our models have a rather amusing description in superspace. The corresponding actions superficially look as free actions, however the fields entering these actions satisfy non-linear chirality constraints. These constraints coincide with the standard conditions for chiral superfields only in the limit of vanishing coupling, when the theory is indeed free. 

We envision various possible extensions of our work. First of all, as the title suggests, we expect that our methods may be generalized to the coadjoint orbits of other classical groups. Besides, our SUSY setup might allow constructing interesting (possibly integrable) SUSY spin chains. Another potential direction is in considering coadjoint orbits of infinite-dimensional groups, such as loop groups. The corresponding models would then be models of 2D field theory.

\vspace{0.5cm}
\textbf{Acknowledgments.} Sections 0-2 were written with the support of the Foundation for the Advancement of Theoretical Physics and Mathematics ``BASIS''. Sections 3-9 were supported by the Russian Science Foundation grant № 22-72-10122 (\href{https://rscf.ru/en/project/22-72-10122/}{\emph{https://rscf.ru/en/project/22-72-10122/}}). We would like to thank M.~Grigoriev, E.~Ivanov and A.~Smilga for discussions as well as E.~Ivanov and A.~Smilga for comments on the manuscript.

\appendix

\vspace{2cm}
\appendixbig{}

\setcounter{section}{10}
\newcounter{appcounter}
\setcounter{appcounter}{1}
\renewcommand{\thesection}{\Alph{appcounter}}

\section{Details on the classical \texorpdfstring{$\CP^1$}{Lg} D-model}\label{ClassicalCP1Dapp}

\subsection{Supersymmetry transformations for component fields.}\label{ClassicalQuantumFermions}

In this Appendix we discuss the classical supersymmetry of the $\CP^1$ D-model (\ref{ActionInComponentsCP1}).
We define the supersymmetry transformations (w.r.t.~$\mathcal{Q}$) for bosonic operators $B$ and  fermionic operators $F$ in the standard way:
\begin{equation}
\delta B := \varepsilon\,\Bigl[B,\mathcal{Q}\Bigr]\,,\qquad\delta F :=\Bigl[F,\varepsilon\,\mathcal{Q}\Bigr] = -\varepsilon\, \Bigl\{F,\mathcal{Q}\Bigr\}\,.
\end{equation}
Similarly, we define the SUSY transformations with respect to $\mathcal{Q}^\dagger$ as
\begin{equation}
\smallthickbar{\delta} B := \smallthickbar{\varepsilon}\,\Bigl[B,\mathcal{Q}^\dagger\Bigr]\,,\qquad\smallthickbar{\delta} F := -\smallthickbar{\varepsilon}\, \Bigl\{F,\mathcal{Q}^\dagger\Bigr\}.
\end{equation}
Here we assume that $\varepsilon$ and $\smallthickbar{\varepsilon}$ are two independent fermionic parameters. The explicit nonzero $\mathcal{N}=2$ SUSY transformations are
\begin{align}
\label{ExplicitSUSYCP1_1}
\delta z_1 = \varepsilon\,\psi_{12} z_2\,,
\quad \delta z_2^\dagger &= -\varepsilon\psi_{12} z_1^\dagger\,,\quad\delta\psi_{12}^\dagger = -\varepsilon\,z_1^\dagger\circ z_2\,,\\
\smallthickbar{\delta} z_2 = \smallthickbar{\varepsilon}\,\psi_{12}^\dagger z_1\,,
\quad \smallthickbar{\delta} z_1^\dagger &= -\smallthickbar{\varepsilon}\,\psi_{12}^\dagger z_2^\dagger\,,\quad\smallthickbar{\delta}\psi_{12} = -\smallthickbar{\varepsilon}\, z_2^\dagger\circ z_1\,,
\label{ExplicitSUSYCP1_3}
\end{align}

As one can verify by an explicit calculation, the action (\ref{ActionInComponentsCP1}) is invariant under the SUSY transformations (\ref{ExplicitSUSYCP1_1})-(\ref{ExplicitSUSYCP1_3}), where we should replace $z_A^\dagger\rightarrow \smallthickbar{z}_A$ and $\psi_{12}^\dagger\rightarrow\smallthickbar{\psi}_{12}$.
Let us check this, for example, for the $\delta$-variation of the classical Hamiltonian. The purely bosonic part transforms as 
\begin{equation}
\delta\big(\smallthickbar{z}_1\circ z_2\big)\big(\smallthickbar{z}_2\circ z_1\big) = \varepsilon\,\psi_{12}\big(\smallthickbar{z}_1\circ z_2\big)\big(\smallthickbar{z}_2\circ z_2 - \smallthickbar{z}_1\circ z_1\big)\,.
\end{equation}
The term that includes fermions transforms in the same way but with the opposite sign,
\begin{equation}
\delta\Big(\smallthickbar{\psi}_{12}\psi_{12}\big(\smallthickbar{z}_2\circ z_2 - \smallthickbar{z}_1\circ z_1\big)\Big) = -\varepsilon\,\psi_{12}\big(\smallthickbar{z}_1\circ z_2\big)\big(\smallthickbar{z}_2\circ z_2 - \smallthickbar{z}_1\circ z_1\big)\,,
\end{equation}
where we used the nilpotency of $\psi_{12}$ and $\smallthickbar{\psi}_{12}$. 
It is easy to see that these variations cancel each other. 

One could as well `reverse engineer' and derive the supercharges using Noether's theorem from the known form of the super-transformations~(\ref{ExplicitSUSYCP1_1})-(\ref{ExplicitSUSYCP1_3}). 

\subsection{Classical and quantum fermions.}\label{AppCCF} Let us mention a minor subtlety with the `dequantization' of our quantum oscillators into a  classical system of the type (\ref{ActionInComponentsCP1}). For simplicity we set the `monopole charge' to zero,  $q=0$. By using the constraints one can eliminate the part of the Hamiltonian that contains fermions,
\begin{align}
\smallthickbar{\psi}_{12}\psi_{12}\big(\smallthickbar{z}_2\circ z_2 - &\smallthickbar{z}_1\circ z_1\big) = \Bigl[\text{constraints $\mathcal{C}_1=0$ and $\mathcal{C}_2=0$}\Bigr] = \nonumber\\ &= -2\big(\smallthickbar{\psi}_{12}\psi_{12}\big)^2 = 0\,,\label{EliminationOfFermionsCP1}
\end{align}
where we have used that  $(\smallthickbar{\psi}_{12})^2 = 0$. After such elimination  supersymmetry (\ref{ExplicitSUSYCP1_1})-(\ref{ExplicitSUSYCP1_3}) of the action (\ref{ActionInComponentsCP1})  still holds  but this time only up to the constraints. 

However, at the quantum level one cannot completely eliminate this term. Concretely, by using the same argument as in (\ref{EliminationOfFermionsCP1}), we get
\begin{equation}
{\psi}^\dagger_{12}\psi_{12}\big({z}^\dagger_2\circ z_2 - {z}^\dagger_1\circ z_1\big) = -2\big({\psi}^\dagger_{12}\psi_{12}\big)^2 = -2\,{\psi}^\dagger_{12}\psi_{12}\,,
\end{equation}
where in passing to the last equality we have used the canonical (anti)commutation relation $\Bigl\{\psi_{12},\psi^\dagger_{12}\Bigr\} = 1$. Thus, there seemingly are two different `dequantization' maps:
\begin{equation}
\label{DequantizationAmbiguity}
\big({\psi}^\dagger_{12}\psi_{12}\big)^2\rightarrow\big(\smallthickbar{\psi}_{12}\psi_{12}\big)^2 = 0\qquad\text{or}\qquad \big({\psi}^\dagger_{12}\psi_{12}\big)^2={\psi}^\dagger_{12}\psi_{12}\rightarrow\smallthickbar{\psi}_{12}\psi_{12}\neq 0\,, 
\end{equation}
which manifests the ordering ambiguity. 
As one sees from (\ref{ActionInComponentsCP1}), the  consistent way is to pick the first prescription in (\ref{DequantizationAmbiguity}). Another reason for such choice is that the second prescription  evidently breaks  supersymmetry of the classical action. 

\setcounter{appcounter}{2}
\section{D-model on \texorpdfstring{$\CP^{n-1}$}{Lg}: calculation of the index}\label{Indexaapp}

In this Appendix we present  details of the calculation of the integral~(\ref{Windexintegral}). First of all, we compute the supertrace over the Fock space. 
An elementary calculation shows that
\bear
&&\mathsf{Str}_{\textrm{Fock space}}\left(\prod\limits_{\alpha=1}^{n} t_{\alpha}^{-\mathcal{J}_\alpha}\,s_1^{\mathcal{C}_1} \prod\limits_{a=2}^{n} \left(s_{(2,a)}\right)^{\mathcal{C}^{aa}_2}\right)=\\ \nonumber &&=\prod\limits_{\alpha=1}^n\,\frac{1}{1-s_1 t_\alpha^{-1}}\times \prod\limits_{\beta=1}^n\,\prod\limits_{a=2}^n\,\frac{1}{1-s_{(2,a)} t_\beta^{-1}}\times\prod\limits_{b=2}^n\,\left(1-\frac{s_1}{s_{(2,b)}}\right)\times\frac{1}{s_1^{p_1}}\,\frac{1}{\left(\prod\limits_{c=2}^n\,s_{(2,c)}\right)^{p_2}}\,.
\eear
Next,  let us compute the integrals over $s_1$ and $s_{(2,a)}$ in~(\ref{WittenIndexIntegral}). 
We will first perform the integral over $s_1$, picking the poles at $s_1={t_\alpha}$. As a result, we get the following:
\bear \nonumber
&&\tilde{W}(t_1,\,\ldots\,,t_n)={1\over (n-1)!}\sum\limits_{\alpha=1}^n\,\oint\,\prod\limits_{d=2}^{n}\frac{\mathrm{d}s_{(2,d)}}{2\pi i s_{(2,d)}}\,\prod\limits_{e\neq f}\,\left(1-\frac{s_{(2,e)}}{s_{(2,f)}}\right)\times\\ \nonumber
&&\times \prod\limits_{\alpha \neq \beta}^n\,\frac{1}{1- {t_\alpha\over t_\beta}}\times \prod\limits_{\gamma=1}^n\,\prod\limits_{a=2}^n\,\frac{1}{1-s_{(2,a)} t_\gamma^{-1}}\times\prod\limits_{b=2}^n\,\left(1-\frac{t_{\alpha}}{s_{(2,b)}}\right)\times \,\frac{1}{t_\alpha^{p_1}\left(\prod\limits_{c=2}^n\,s_{(2,c)}\right)^{p_2}}=\\ \nonumber
&&=\bigg[\textrm{canceling the factors}\, \prod\limits_{b=2}^n\,\left(1-\frac{t_{\alpha}}{s_{(2,b)}}\right)\bigg]={(-1)^{n-1}\over (n-1)!}\sum\limits_{\alpha=1}^n\,\prod\limits_{\alpha\neq \beta}^n\,\frac{1}{1- {t_\alpha\over t_\beta}}\times\,\\ \label{WittenIndexIntegralLong}
&&\times \oint\,\prod\limits_{d=2}^{n}\frac{\mathrm{d}s_{(2,d)}}{2\pi i s_{(2,d)}}\,\prod\limits_{e\neq f}\,\left(1-\frac{s_{(2,e)}}{s_{(2,f)}}\right)\times 
 \prod\limits_{\gamma\neq \alpha}^n\,\prod\limits_{a=2}^n\,\frac{1}{1-s_{(2,a)} t_\gamma^{-1}}\times \prod_{b=2}^n \frac{t_\alpha}{s_{(2,b)}}\times\nonumber
 \\
 &&\times \frac{1}{t_\alpha^{p_1}\left(\prod\limits_{c=2}^n\,s_{(2,c)}\right)^{p_2}}\,.
\eear
Finally, we take the remaining $n-1$ integrals over $s_{(2,a)}$. The poles are at $s_{(2,a)}={t_\gamma}$ for $\gamma \neq \alpha$, but not all of them contribute. Indeed, if two of the $s_{(2,a)}$'s coincide, the numerator vanishes, so that the residue is zero. Therefore the relevant configurations~are
\bear\label{stpoles}
&&s_{(2,a)}={t_{\gamma_a}}\,,\quad\quad a=2, \ldots , n\,,\\&&\textrm{where}\quad\quad (t_{\gamma_1}, \ldots t_{\gamma_a})=\textrm{permutation of}\;(t_1, \ldots, \hat{t}_\alpha,\ldots t_n)\,.
\eear
For any such configuration, the first two factors in the integrand of~(\ref{WittenIndexIntegralLong}) cancel each other, and $\prod\limits_{c=2}^n\,s_{(2,c)}=\prod\limits_{c=2}^n\, t_{\gamma_c}=t^{-1}_\alpha$ due to the $\SU(n)$ constraint $\prod\limits_{\gamma=1}^n\, t_{\gamma}=1$. Thus, we see that each of the configurations~(\ref{stpoles}) gives the same contribution, so that in a sum over all poles we simply get an overall factor of $(n-1)!$\,. Ultimately we arrive at the  expression~(\ref{CompleteFlagWittenIndexD}).

\setcounter{appcounter}{3}
\section{Spectrum of the Laplacian on \texorpdfstring{$\mathcal{F}_{3}$}{Lg}} 
Let us illustrate the recipe for calculating the spectrum of the Dolbeault Laplacian on $\mathcal{F}_{3}$, restricting for simplicity to the bosonic sector of the theory\footnote{The strategy  for the K-model is the same.  Moreover, the bosonic sector of the K-model is the same as that of the D-model, with the additional restriction that all monopole charges vanish.}. 
The general method for finding the spectrum is based on a relation with the sigma model, which can be derived in full analogy with the $\CP^1$ case (see Section
\ref{Derive of CP1}). To summarize, one should compute the spectrum of the D-model of Section~\ref{FlagDolbeaultIndex} for fairly large values of~$p_A$'s (these feature in the constraints~(\ref{Constraints F123})-(\ref{Constraints F123-3})). The result will exactly match part of the spectrum of the Dolbeault Laplacian. The entire spectrum can be obtained in the limit  $p_A \rightarrow \infty$ with the differences $p_A - p_B$ fixed. The latter are interpreted as the monopole charges.

In the case of the bosonic sector and $p_1 = p_2 = p_3 = p$, the Hamiltonian is just the  spin chain Hamiltonian (\ref{Spin Ham 123}), where the additive constant is defined so that its minimal eigenvalue is zero\footnote{This is one of the remarkable differences between SUSY QM and ordinary QM. As discussed in~\cite{Bykov_2024}, in  ordinary QM this constant needs to be tuned manually. In contrast to that, in SUSY QM it is fixed automatically.}. Let us now consider specific values of $p$ and calculate the spectrum explicitly. As it turns out, all eigenvalues are solutions to polynomial equations of prescribed order. For a given irreducible representation, the order is equal to the multiplicity of this representation in the spectrum (the spectrum is \emph{not} multiplicity-free).  Thus, we will write out the equations, the corresponding representations as well as the dimensions of these representations (the multiplicities of the eigenvalues).

In the case of $p=1$ and $p=2$  one finds:
\begin{align*}
&&& \textrm{Eigenvalue}\, &&&&\textrm{Representation} && \textrm{Dimension}\nonumber\\
    & p = 1\,:\nonumber\\
    &&&\lambda = 0 &&&&Trivial && 1 \nonumber \\
    &&&\lambda - 2 a =0 &&&&{\tiny \begin{ytableau}
        ~ & ~ & ~
    \end{ytableau}} && 10 \nonumber \\
    &&&\lambda^2 - 2 a \lambda + 3b = 0  &&&&{\tiny \begin{ytableau}
        ~ & ~ \\
        ~ 
    \end{ytableau}} && 8 \nonumber \\
    & p=2\,:\nonumber\\
    &&&\lambda = 0  &&&&Trivial && 1 \nonumber \\
    &&&\lambda - 2 a =0  &&&&{\tiny \begin{ytableau}
        ~ & ~ & ~
    \end{ytableau}} && 10 \nonumber \\
    &&&\lambda - 2 a =0  &&&&{\tiny \begin{ytableau}
        ~ & ~ & ~\\
        ~ & ~ & ~
    \end{ytableau}} && 10 \nonumber \\
    &&&\lambda - 6 a =0  &&&&{\tiny \begin{ytableau}
        ~ & ~ & ~& ~& ~& ~
    \end{ytableau}} && 28 \nonumber \\
    &&&\lambda^2 - 2 a \lambda + 3b = 0  &&&&{\tiny \begin{ytableau}
        ~ & ~ \\
        ~ 
    \end{ytableau}} && 8 \nonumber \\
    &&&\lambda^2 - 8 a \lambda + 12\left(a^2 + b\right) = 0  &&&&{\tiny \begin{ytableau}
        ~ & ~& ~ &~ & ~ \\
        ~ 
    \end{ytableau}} && 35 \nonumber \\
    &&&\lambda^3 - 8 a \lambda^2 + \left(12 a^2 + 28 b\right)\lambda - 48 ab - 80 c = 0 &&&&{\tiny \begin{ytableau}
        ~ & ~& ~ &~  \\
        ~ & ~
    \end{ytableau}} && 27 \nonumber
\end{align*}
where $a := \upalpha^2_{12} + \upalpha^2_{23} + \upalpha^2_{13},\, b:= \left(\upalpha_{12}\upalpha_{13}\right)^2 + \left(\upalpha_{12}\upalpha_{23}\right)^2 + \left(\upalpha_{13}\upalpha_{23}\right)^2$ and $c := \left(\upalpha_{12}\upalpha_{23}\upalpha_{13}\right)^2$ are the elementary symmetric polynomials of $\upalpha_{12}^2, \upalpha_{23}^2, \upalpha_{13}^2$. One easily checks that the sum of dimensions (taking into account the multiplicities) matches $3^3=27$ for $p=1$ and $6^3=216$ for $p=2$.

Notice the important property (proven in Section~\ref{pindepsec}) that the spectrum at $p = 2$ contains the spectrum at~$p = 1$.

\setcounter{appcounter}{4} 
\section{Comparison with the Weyl character formula for \texorpdfstring{$\SU(n)$}{Lg}}\label{Weylcharacterapp}

Here we will explain how our results for the equivariant Witten indices of the D-models coincide with the Weyl character formula.
The Weyl character formula for $\SU(n)$ reads~\cite{WeylTheClassicalGroups}:
\begin{align} \label{Weyl character}
    \chi_{\boldsymbol{\lambda}} (t_1,t_2,\dots, t_n) = \frac{ \mathrm{det} \begin{pmatrix}
        t_1^{\lambda_1+n-1} & t_1^{\lambda_2+n-2} & \dots & t_1^{\lambda_n} \\
        & & &\\
        t_2^{\lambda_1+n-1} & t_2^{\lambda_2+n-2} & \dots & t_2^{\lambda_n} \\
        \vdots & & \ddots & \vdots \\
        t_n^{\lambda_1+n-1} & t_n^{\lambda_2+n-2} & \dots & t_n^{\lambda_n}
    \end{pmatrix}}{\mathrm{det} \begin{pmatrix}
        t_1^{n-1} & t_1^{n-2} & \dots & 1 \\
        & & &\\
        t_2^{n-1} & t_2^{n-2} & \dots & 1 \\
        \vdots & & \ddots & \vdots \\
        t_n^{n-1} & t_n^{n-2} & \dots & 1 \\
    \end{pmatrix}},
\end{align}
where $\boldsymbol{\lambda} = (\lambda_1,\lambda_2, \dots, \lambda_n)$ is a non-increasing sequence of  natural numbers. In fact, $\boldsymbol{\lambda}$ defines the representation, whose Young diagram has rows of lengths $\lambda_\alpha$'s. 

For the numerator we use the standard formula for calculating the determinant by using permutations, that is
\begin{align}
    \sum\limits_{\sigma \in \mathsf{S}_n} (-1)^{\mathrm{sign}(\sigma)} \,\,t_{\sigma(1)}^{\lambda_1 + n - 1} \,t_{\sigma(2)}^{\lambda_2 + n - 2} \dots \,t_{\sigma(n)}^{\lambda_n}\,.
\end{align}
The denominator in~(\ref{Weyl character}) is the Vandermonde determinant. A minor simplification then leads to the following expression for the character: 
\begin{align} \label{character1}
    \chi_{\boldsymbol{\lambda}} (t_1,t_2,\dots, t_n) &= \sum\limits_{\sigma \in \mathsf{S}_n} \frac{t_{\sigma(1)}^{\lambda_1 + n - 1} \,t_{\sigma(2)}^{\lambda_2 + n - 2} \dots \,t_{\sigma(n)}^{\lambda_n}}{\prod\limits_{k < l}\left(t_{\sigma(k)} - t_{\sigma(l)}\right)} =  \sum\limits_{\sigma \in \mathsf{S}_n} \frac{t_{\sigma(1)}^{\lambda_1} \,t_{\sigma(2)}^{\lambda_2} \dots \,t_{\sigma(n)}^{\lambda_n}}{\prod\limits_{k < l}\left(1 - \frac{t_{\sigma(l)}}{t_{\sigma(k)}}\right)}\,. 
\end{align}
The latter form is useful for comparing with the equivariant Witten index~(\ref{Witten index for full flag}) for complete flag manifolds. 

It is more interesting to compare the character $\chi_{\boldsymbol{\lambda}}$ with the equivariant Witten index (\ref{Witten index for partial flags}) for partial flag manifolds. Again, we will be using (\ref{character1}), but in the case of partial flags some of the $\lambda_\alpha$'s are the same. For simplicity, let us assume that $\lambda_1=\ldots =\lambda_m=\Lambda$, then one can rewrite (\ref{character1}) as follows:
\begin{align}
    \sum\limits_{\sigma \in \mathsf{S}_n} \frac{\left(t_{\sigma(1)}\dots t_{\sigma(m)}\right)^{\Lambda + n - m} \cdot t_{\sigma(1)}^{m-1} t_{\sigma(2)}^{m-2} \dots t_{\sigma(m)}^{0} \cdot t_{\sigma(m+1)}^{\lambda_{m+1}+n-m-1} \dots t_{\sigma(n)}^{\lambda_n}}{V_1 \times V_2}\,, \label{CharacterSum}
\end{align}
where we have split the Vandermonde determinant in the denominator into two parts:  $V_1$ contains only $(t_{\sigma(k)} - t_{\sigma(l)})$ with $1 \leq k < l \leq m$ whereas $V_2$ contains all remaining terms. Note that if one replaces $\sigma$ with $\sigma \circ \Tilde{\sigma}$, where $\Tilde{\sigma}$ is a permutation of $\{1,2,\dots,m\}$, the sum (\ref{CharacterSum}) remains the same. However, in each term exactly two  changes will occur: 
\begin{align}
    t_{\sigma(1)}^{m-1}\, t_{\sigma(2)}^{m-2}\, \dots\, t_{\sigma(m)}^{0} &\rightarrow t_{\sigma(\Tilde{\sigma}(1))}^{m-1}\, t_{\sigma(\Tilde{\sigma}(2))}^{m-2}\, \dots\, t_{\sigma(\Tilde{\sigma}(m))}^{0},\\
    V_1 &\rightarrow (-1)^{\mathrm{sign}(\Tilde{\sigma})}\,\, V_1.  \nonumber
\end{align}
Now, if we sum over all possible $\Tilde{\sigma}$'s, then on the one hand we simply get $m!$ times  $\chi_{\boldsymbol{\lambda}}$. On the other hand, swapping the order of summation over $\sigma$ and $\Tilde{\sigma}$,  we get  
\begin{align}
    \sum\limits_{\Tilde{\sigma} \in \mathsf{S}_m} \, (-1)^{\mathrm{sign}(\Tilde{\sigma})} t_{\sigma(\Tilde{\sigma}(1))}^{m-1}\, t_{\sigma(\Tilde{\sigma}(2))}^{m-2}\, \dots\, t_{\sigma(\Tilde{\sigma}(m))}^{0} = V_1.
\end{align}
Thus, we end up with 
\begin{align}
    \chi_{\boldsymbol{\lambda}} (t_1,t_2,\dots, t_n) = \frac{1}{m!} \sum\limits_{\sigma \in \mathsf{S}_n} \frac{\left(t_{\sigma(1)}\dots t_{\sigma(m)}\right)^{\Lambda + n - m} \cdot t_{\sigma(m+1)}^{\lambda_{m+1}+n-m-1} \dots t_{\sigma(n)}^{\lambda_n}}{V_2}\,,
\end{align}
which matches~(\ref{Witten index for partial flags}) for the special case of the flag manifold~$\mathcal{F}_{m,1,1,\dots,1}$. It is clear that, in a similar fashion, one can prove the equivalence of~(\ref{Witten index for partial flags}) and~(\ref{character1}) for an arbitrary partial flag.

\vspace{1cm}    
    \setstretch{0.8}
    \setlength\bibitemsep{5pt}
    \printbibliography
\end{document}